\def\spacingNumerator{5}
\def\spacingDenominator{4}
%%%
%%%   jfmacros.tex
%%%   Version of 12/08/02
%%%   Joel Feldman  feldman@math.ubc.ca
%%%

\def\ifundefined#1{\expandafter\ifx\csname#1\endcsname\relax}
\ifundefined{ftmagnification}  \def\ftmagnification{1200} \fi
\ifundefined{spacingNumerator}  \def\spacingNumerator{5} \fi
\ifundefined{spacingDenominator}  \def\spacingDenominator{4} \fi

%=====================================================================
%======================= PRIMARY MACROS ==============================
%=====================================================================

%%%%%%% VARIOUS SETTINGS
\magnification\ftmagnification
\tolerance=10000
\hsize=17truecm\vsize=23truecm

\parindent=40pt
\mathsurround=0pt
     \multiply\baselineskip by \spacingNumerator
     \divide \baselineskip by \spacingDenominator 

%
%     HEADERS
%
\def\today{\ifcase\month\or January\or February\or March\or April\or
     May\or June\or July\or August\or September\or October\or November\or
     December\fi\space\number\day, \number\year}
%
%     STYLE SHORT FORMS
%
\def\dst{\displaystyle}
\def\sst{\scriptstyle}
\def\tst{\textstyle}
\def\ssst{\scriptscriptstyle}
%
%       EQUATIONS
%     
\def\frac#1#2{\dst {#1\over#2}}     % fractions in displaystyle
\def\sfrac#1#2{{\tst{#1\over#2}}}   % fractions in textstyle    

\def\deqalign#1{\vcenter{\openup1\jot \mathsurround=0pt \ialign{
                \strut\hfil$\displaystyle{##}$&&$\displaystyle{{}##}$\hfil
                \crcr
                #1\crcr}}}         %  double eqalign 

\def\meqalign#1{\vcenter{\openup1\jot \mathsurround=0pt \ialign{
                &\strut\hfil$\displaystyle{##}$&$\displaystyle{{}##}$\hfil&
                \quad$##$\crcr
                #1\crcr}}}         %  multiple eqalign

%
%      GREEK LETTERS
%
\def\al{\alpha}
\def\be{\beta}
\def\ga{\gamma}
\def\de{\delta}
\def\ep{\epsilon}
\def\ze{\zeta}

\def\ka{\kappa}
\def\la{\lambda}

\def\si{\sigma}

\def\ch{\chi}

\def\om{\omega}

\def\De{\Delta}

\def\La{\Lambda}
\def\Si{\Sigma}

\def\Om{\Omega}   
%
%        BOLD FACE, SCRIPT and ROMAN CHARACTERS
%
\def\pmb#1{\setbox0=\hbox{#1}       % generate bold face
     \kern-.025em\copy0\kern-\wd0
     \kern.05em\copy0\kern-\wd0
     \kern-.025em\box0}             %Knuth puts in \raise.0433em before box0 
\def\0{{\bf 0}}

\def\k{{\bf k}}

\def\x{{\bf x}}
\def\y{{\bf y}}

\def\cB{{\cal B}}

\def\cF{{\cal F}}
\def\cG{{\cal G}}
\def\cH{{\cal H}}

%
%      FONT FAMILIES
%
\font\tenfrak                 = eufm10
\font\sevenfrak               = eufm7
\font\fivefrak                = eufb5
\newfam\frakfam
     \textfont\frakfam=\tenfrak
     \scriptfont\frakfam=\sevenfrak   
     \scriptscriptfont\frakfam=\fivefrak
\def\frak{\fam\frakfam\tenfrak}
\font \tensans                = cmss10
\font \fivesans               = cmss10 at 5pt
\font \sevensans              = cmss10 at 7pt
\newfam\sansfam
     \textfont\sansfam=\tensans
     \scriptfont\sansfam=\sevensans
     \scriptscriptfont\sansfam=\fivesans
\def\sans{\fam\sansfam\tensans}
%
%      Blackboard characters
%
\def\bbbr{{\rm I\!R}}  
\def\bbbn{{\rm I\!N}}

\def\bbbc{{\mathchoice {\setbox0=\hbox{$\displaystyle\rm C$}\hbox{\hbox %\bbbc
to0pt{\kern0.4\wd0\vrule height0.9\ht0\hss}\box0}}
{\setbox0=\hbox{$\textstyle\rm C$}\hbox{\hbox
to0pt{\kern0.4\wd0\vrule height0.9\ht0\hss}\box0}}
{\setbox0=\hbox{$\scriptstyle\rm C$}\hbox{\hbox
to0pt{\kern0.4\wd0\vrule height0.9\ht0\hss}\box0}}
{\setbox0=\hbox{$\scriptscriptstyle\rm C$}\hbox{\hbox
to0pt{\kern0.4\wd0\vrule height0.9\ht0\hss}\box0}}}}
\def\bbbq{{\mathchoice {\setbox0=\hbox{$\displaystyle\rm               %\bbbq
Q$}\hbox{\raise
0.15\ht0\hbox to0pt{\kern0.4\wd0\vrule height0.8\ht0\hss}\box0}}
{\setbox0=\hbox{$\textstyle\rm Q$}\hbox{\raise
0.15\ht0\hbox to0pt{\kern0.4\wd0\vrule height0.8\ht0\hss}\box0}}
{\setbox0=\hbox{$\scriptstyle\rm Q$}\hbox{\raise
0.15\ht0\hbox to0pt{\kern0.4\wd0\vrule height0.7\ht0\hss}\box0}}
{\setbox0=\hbox{$\scriptscriptstyle\rm Q$}\hbox{\raise
0.15\ht0\hbox to0pt{\kern0.4\wd0\vrule height0.7\ht0\hss}\box0}}}}
\def\bbbz{{\mathchoice {\hbox{$\sans\textstyle Z\kern-0.4em Z$}}       %\bbbz
{\hbox{$\sans\textstyle Z\kern-0.4em Z$}}
{\hbox{$\sans\scriptstyle Z\kern-0.3em Z$}}
{\hbox{$\sans\scriptscriptstyle Z\kern-0.2em Z$}}}}
%
%      SYMBOLS
%
\def\const{{\rm const}\,}
\def\sgn{{\rm sgn}}
\def\half{\sfrac{1}{2}}

\def\optbar#1{\vbox{\ialign{##\crcr\hfil${\scriptscriptstyle(}\mkern -1mu
         \vrule height 1.2pt width 3pt depth -.8pt
         {\scriptscriptstyle)}$\hfil\crcr
          \noalign{\kern-1pt\nointerlineskip}$\hfil\displaystyle{#1}\hfil$\crcr}}}
\def\<{\left<}
\def\>{\right>}

\def\smprod{\mathop{\textstyle\prod}}
\def\smsum{\mathop{\textstyle\sum}}
\def\set#1#2{\big\{ \ #1\ \big|\ #2\ \big\}}
\def\eval#1{\big|\lower4pt\hbox{$\displaystyle\sst #1$}}
%
%       SECTION TITLES
%
\font \tafontt                = cmbx10 scaled\magstep2
\font \tbfontt                = cmbx10 scaled\magstep1
\def\titlea#1{\centerline{\tafontt #1 }\vskip.5truein}
\def\titleb#1{\removelastskip\vskip.3truein%
\noindent{\tbfontt #1 }\vskip.25truein}
\def\titlec#1{\removelastskip\vskip.15truein\noindent{\bf #1 }\vskip.1truein}

%
%          THEOREM etc.
%
\def\newenvironment#1#2#3#4{\long\def#1##1##2{%
\removelastskip\penalty-100\vskip\baselineskip%
\noindent{#3#2\if!##1!.\else\unskip\ \ignorespaces
##1\unskip\fi\ }{#4\ignorespaces##2\vskip\baselineskip}}}
\newenvironment\lemma{Lemma}{\bf}{\it}
\newenvironment\proposition{Proposition}{\bf}{\it}
\newenvironment\theorem{Theorem}{\bf}{\it}
\newenvironment\corollary{Corollary}{\bf}{\it}
\newenvironment\example{Example}{\bf}{\rm}
\newenvironment\problem{Problem}{\bf}{\rm}
\newenvironment\definition{Definition}{\bf}{\rm}
\newenvironment\remark{Remark}{\bf}{\rm}
\newenvironment\hypothesis{Hypothesis}{\bf}{\it}
\newenvironment\convention{Convention}{\bf}{\it}

\def\Item{\vskip.1in\noindent}

%
%         PROOF, QED  
%
\long\def\proof#1{\removelastskip\penalty-100\vskip\baselineskip\noindent{\bf
            Proof\if!#1!\else\ \ignorespaces#1\fi:\ }\ \ \ignorespaces}
\long\def\prf{\removelastskip\penalty-100\vskip\baselineskip\noindent{\bf
            Proof:\ }\ \ \ignorespaces}
\def\endproof{\hfill\vrule height .6em width .6em depth 0pt\goodbreak\vskip.25in }

%===========================================================================
%===================== Automatic numbering (Frozen) ========================
%===========================================================================
\ifundefined{warnForwardRef}  \def\warnForwardRef{n} \fi
\newcount\chapno
\newcount\sectno
\newcount\equano
\newcount\theono
\newcount\probno

\def\IgNoRe#1{}

\chapno=0
\sectno=0
\equano=0
\theono=0
\probno=0
\def\eqhead{}
\def\frefwarning{\if\warnForwardRef y\immediate\write16{   Forward reference on line \the\inputlineno}\fi}
\def\qqqrefwarning{\immediate\write16{   ??? reference on line \the\inputlineno}}

\def\chap#1{\equano=0\sectno=0\theono=0\probno=0\global\advance\chapno by 1%
\def\eqhead{\ifcase\chapno\or I\or II\or III\or IV\or V\or VI\or VII\or
VIII\or IX\or X\or XI\or XII\or XIII\or XIV\or XV\or XVI\or XVII\or XVIII\or
XIX\or XX\or XXI\or XXII\or XXIII\or XXIV\or XXV\or XXVI\or XXVII\or XXVIII\or XXIX\or XXX\or XXXI\or XXXII\or XXXIII\or XXXIV\or XXXV\or XXXVI\or XXXVII\or XXXVIII\or XXXIX\fi.}%
\titlea{\eqhead \hglue 5pt #1}%
}

\def\sect#1{\global\advance\sectno by 1%
\titleb{\eqhead\number\sectno  \hglue 5pt #1}%
}%

\def\appendix#1#2{\equano=0\sectno=0\theono=0\probno=0\def\eqhead{#1.}
\titlea{Appendix #1: #2}%
}

\def\:#1{\def\temp{\expandafter\IgNoRe\string#1}%
\expandafter\ifx\csname\temp\endcsname\relax%
\expandafter\gdef#1{\qqqrefwarning ???}\fi#1}

\def\Eqn{{\hbox{\global\advance\equano by 1}}%
\eqno ({\rm \eqhead\number\equano})}%

\def\Eqno{{\hbox{\global\advance\equano by 1}}%
 ({\rm \eqhead\number\equano})}%

\def\EQN#1{\Eqn\edef\Zwi{\eqhead\number\equano}%
\global\let #1=\Zwi
}

\def\EQNO#1{\Eqno\edef\Zwi{\eqhead\number\equano}%
\global\let #1=\Zwi
}

\def\STM#1{{\global\advance \theono by 1}% 
\eqhead\number\theono
\edef\Zwi{\eqhead\number\theono }
\global\let#1=\Zwi
}

\def\PRB#1{{\global\advance \probno by 1}% 
\eqhead\number\probno
\edef\Zwi{\eqhead\number\probno }
\global\let#1=\Zwi
}

\def\PG#1{\def\Zwi{\number\pageno }
\global\let#1=\Zwi
}

\def\Stm{{\global\advance \theono by 1}% 
\eqhead\number\theono
}

\def\Prb{{\global\advance \probno by 1}% 
\eqhead\number\probno
}

\def\EDEF#1#2{%   Export definition to .txs file
\def\tEmP{#1}\expandafter\gdef\tEmP{#2}
}

%%%%%%%%%%%%%%%%%%%%%%%%%%%%%%%%%%%%%%%%%%%%%%%%%%%%%%%%%%%%%%%%%%%%%%%%%%%%%%%
%%%%%%%%%%%%%   Macros for figure insertion
%%%%%%%%%%%%%%%%%%%%%%%%%%%%%%%%%%%%%%%%%%%%%%%%%%%%%%%%%%%%%%%%%%%%%%%%%%%%%%%
%%%%%%%
%%%%%%%  The two main figure insertion macros are
%%%%%%%
%                \figput{<filename w/o extension>}
%                \figplace{<filename w/o extension>}{<hor shift>}{<vert shift>}
%%%%%%%
%%%%%%%  The first just inserts the figure at the current location. The
%%%%%%%  second inserts the figure at the current location but then shifts 
%%%%%%%  horizontally by the second argument and vertically by the third.
%%%%%%%
%%%%%%%  Some typical TeX commands for inserting figures are
%%%%%%%      \centerline{\figput{<filename w/o extension>}}
%%%%%%%      \vadjust{\centerline{\figput{<filename w/o extension>}}}
%%%%%%%      \midinsert\centerline{\figput{<filename w/o extension>}}\endinsert
%%%%%%%      \topinsert\centerline{\figput{<filename w/o extension>}}\endinsert

%%%%%%%
%%%%%%%   TO SET A FIGURE DIRECTORY INSERT, FOR EXAMPLE,
%%%%%%%                 \def\figdir{figures/}
%%%%%%%   IN YOUR SOURCE FILE. REMEMBER THE TAILING /
%%%%%%%

%%%%%%%
%%%%%%%     SELECT (a) YOUR POSTSCRIPT FILE SUFFIX AND (b) YOUR SYSTEM  NOW!
%%%%%%%
\def\suffix{ps}
\newcount\system
%\global\system=1   % for textures 
%\global\system=2   % for msdos
\global\system=3   % for unix(dvips)
%\global\system=4   % for unix(dvips) scaled by a factor of 1.2
%\global\system=6   % for xdvik

\def\ifundefined#1{\expandafter\ifx\csname#1\endcsname\relax}
\ifundefined{figdir}\def\figdir{}\fi
%
% Now for the definitions and main macro for figure inclusion.
%
\newcount\firstline
\newdimen\pswidth  \newdimen\xleft
\newdimen\psheight \newdimen\ytop \newdimen\ybot
\newcount\justx \newcount\justy
\global\justx=0 \global\justy=0
\newdimen\vpos \newtoks\labeL 
\newread\labeLfile \newdimen\xcoord \newdimen\ycoord
\newif\ifdoit 
\newbox\labox
%  variables for use with xdvik
\newdimen\xdvikwid 
\newdimen\xdvikht
\newdimen\pspoints
\newdimen\rwi
\pspoints=1bp
\newcount\temp
\def\readdim#1{\global\read\labeLfile to \temp
\global #1=\temp pt}
%
% 
%    figcrop{<filename,w/o extension>} treats the first two labels as marking
%    the upper left and lower right corners of the figure. This is for
%    positioning purposes only. The figure may extend beyond the corners.
%    The corner markers are not printed.
%
%
\def\figcrop#1{\par%  #1=filename
\openin\labeLfile=\figdir#1.lbl                                              
\global\read\labeLfile to\firstline\message{#1}               
\global\read\labeLfile to\temp%read overall dimensions                                     
\readdim{\ybot}
\readdim{\xleft}%               read upper left point
\readdim{\ytop}
\global\read\labeLfile to\justx%ignore
\global\read\labeLfile to\justy%ignore
\global\read\labeLfile to\labeL%ignore
\readdim{\pswidth}%            read lower right point
\global\advance\pswidth by -\xleft
\readdim{\psheight}
\global\advance\ybot by -\psheight
\global\advance\psheight by -\ytop
\global\read\labeLfile to\justx%ignore
\global\read\labeLfile to\justy%ignore
\global\read\labeLfile to\labeL%ignore                                    
\vbox to\psheight{\vfill
%%%
%%% NOTE: next line may have to be changed for your DVIPS driver %%%
\ifnum\system=1
\fi %textures
\ifnum\system=2
\fi %msdos
\ifnum\system=3
  %%  \special{" grestore newpath gsave}
                                                 \fi         %%unix:dvips
\ifnum\system=4
\fi         %%unix:dvips,scaled
\ifnum\system=1
\hbox to \pswidth{\kern-\xleft\special{postscriptfile \figdir#1.\suffix }\hfil}\fi
                                                              %textures
\ifnum\system=2
\hbox to \pswidth{\kern-\xleft\special{ps: plotfile \figdir#1.\suffix }\hfil}\fi
                                                              %mdos 
\ifnum\system=3
\hbox to \pswidth{\kern-\xleft\includegraphics{\figdir#1.\suffix}\hfil}\fi
                                                             %unix:dvips 
\ifnum\system=4
\hbox to \pswidth{\kern-\xleft\includegraphics{\figdir#1.\suffix}\hfil}\fi
                                                             %unix:dvips,scaled
\ifnum\system=5
\hbox to \pswidth{\kern-\xleft\includegraphics{\figdir#1.\suffix}\hfil}\fi %orphee
\ifnum\system=6
   \xdvikwid=\pswidth
   \xdvikht=\psheight
   {\global\divide\xdvikwid by \pspoints}
   {\global\divide\xdvikht by \pspoints}
   \rwi=\xdvikwid
    {\global\multiply\rwi by 10}
\hbox to \pswidth{\kern-\xleft\includegraphics{\figdir#1.\suffix\space}\hfil}\fi                   %xdvik
%%%
\vskip -\baselineskip
\vskip -\ybot 
\vskip-\psheight %                                     
\hbox to\pswidth  {\hss}%                                            
\parindent=0pt\offinterlineskip                                       
\vpos=0 pt%                                                              
\loop\readdim{\xcoord}                                 
\ifdim \xcoord < -999pt \doitfalse\else\doittrue\fi                        
\ifdoit \advance \xcoord by -\xleft
\readdim{\ycoord}
\advance \ycoord by -\ytop                              
\global\read\labeLfile to\justx                                       
\global\read\labeLfile to\justy                                       
\global\read\labeLfile to\labeL
\global\setbox\labox=\hbox{\labeL\hskip-0.3em}%    
\advance\vpos by-\ycoord                                              
\vskip-\vpos \vpos=\ycoord                                         
\hbox to\pswidth{\hskip\xcoord %                                 
\hbox to 0pt{\ifnum\justx>0\hss\fi%                                   
\vbox to0pt{%                                                         
\ifnum\justy<2\vss\fi%                                                
\copy\labox\kern0pt%  
\ifnum\justy>0\vss\fi}%                                               
\ifnum\justx<2\hss\fi}%                                               
\hss}%                                                                
\repeat%                                                              
\advance\vpos by-\psheight%                                           
\vskip-\vpos %                                                     
}\closein\labeLfile}
%
%
%     \figplace{<filename w/o extension>}{<hor shift>}{<vert shift>}
%     moves to the right by <hor shift> and down by <vert shift>
%     and then applies \figcrop
% 
\def\figplace#1#2#3{
\openin\labeLfile=\figdir#1.lbl
\ifeof \labeLfile
       \immediate\write16{***Can't find \figdir#1.lbl; Skipping it.***}
\else  \closein\labeLfile
       \null\hskip#2\raise #3 \hbox{\figcrop{#1}}
\fi
}
%
%
%     \figput{<filename w/o extension>}
%     
%     just applies \figcrop
% 

%%%%%%%%%%%%%%%%%%%%%%%%%%%%%%%%%%%%%%%%%%%%%%%%%%%%%%%%%%%%%%%%%%%%%%%
%%%%%%%%    omacros
%%%%%%%%%%%%%%%%%%%%%%%%%%%%%%%%%%%%%%%%%%%%%%%%%%%%%%%%%%%%%%%%%%%%%%%

    \def\squiggle{\raise2pt\hbox{${\scriptstyle\sim}$}}
    \def\stoday{\number\day\space\ifcase\month\or Jan\or Feb\or 
                      Mar\or Apr\or May\or Jun\or Jul\or Aug\or Sep\or 
                      Oct\or Nov\or Dec\fi, \number\year}

    \def\veps{{\varepsilon}}
    \def\smchoose#1#2{{\tst {#1\choose #2}}}
    \def\abcst{{\sst const}}
    \def\cst#1#2{{\rm const}^{#1}_{#2}\,}
    \def\Cont#1#2#3{\mathop{{\rm\ \, {\cal C}on}_{#3}}\limits_{#1\rightarrow#2}}

    \def\abcst{{\sst const}}
    \def\cb{{\frak c}}
    \def\ib{{\rm b}}

    \def\dunion{\cup\kern-0.7em\cdot\kern0.45em}

    \def\cD{{\cal D}}
    \def\rD{{\rm D}}

    \def\cV{{\cal V}}    
    \def\cW{{\cal W}}

    \def\fe{{\frak e}}

    \def\fN{{\frak N}}

    \def\bbe{\pmb{$\be$}}
    \def\bga{\pmb{$\ga$}}
    \def\bde{{\mathchoice{\pmb{$\de$}}{\pmb{$\de$}}
                              {\pmb{$\sst\de$}}{\pmb{$\ssst\de$}}}}

    \def\rw{\mathclose{:}}
    \def\lw{\mathopen{:}}
    \def\lW{\mathopen{{\tst{\hbox{.}\atop\raise 2.5pt\hbox{.}}}}}
    \def\rW{\mathclose{{\tst{{.}\atop\raise 2.5pt\hbox{.}}}}}
    \def\lww{\mathopen{{\tst{\raise 1pt\hbox{.}\atop\raise 1pt\hbox{.}}}}}
    \def\rww{\mathclose{{\tst{\raise 1pt\hbox{.}\atop\raise 1pt\hbox{.}}}}}

    \def\tn{|\kern-1pt|\kern-1pt|}
    \def\TN{\big|\kern-1.5pt\big|\kern-1.5pt\big|}
    \def\TTN{\Big|\kern-2pt\Big|\kern-2pt\Big|}

    \def\cnorm{\kern8pt\check{\kern-8pt\|}}
    \def\Cnorm{\kern8pt\check{\kern-8pt\big\|}}
    \def\CNorm{\kern8pt\check{\kern-8pt\Big\|}}

    \def\tnorm{\kern8pt\tilde{\kern-8pt\|}}
    \def\Tnorm{\kern8pt\tilde{\kern-8pt\big\|}}
    \def\TNorm{\kern8pt\tilde{\kern-8pt\Big\|}}

    \def\tv{\kern8pt\tilde{\kern-8pt\pmb{$\vert$}}}
    \def\tV{\kern8pt\tilde{\kern-8pt\pmb{$\big\vert$}}}
    \def\tVV{\kern8pt\tilde{\kern-8pt\pmb{$\Big\vert$}}}

    \def\jbar{{\mathchoice
                   {{\smash{\lower1ex\hbox{$\mathchar'26$}}\mkern-9mu j}}
                   {{\smash{\lower1ex\hbox{$\mathchar'26$}}\mkern-9mu j}}
                   {{\smash{\lower1.2ex\hbox{$\mathchar'26$}}\mkern-10.2mu j}}
                   {{\smash{\lower1.2ex\hbox{$\mathchar'26$}}\mkern-10.2mu j}}}}

\def\Eqnb{{\hbox{\global\advance\equano by 1}}%
\eqno ({\rm \eqhead\number\equano}}%

\def\EQNB#1{\Eqnb\edef\Zwi{\eqhead\number\equano}%
\global\let #1=\Zwi
}

   \font\sixrm=cmr6   \font\eightrm=cmr8  
   \font\sixi=cmmi6   \font\eighti=cmmi8  
  \font\sixsy=cmsy6  \font\eightsy=cmsy8 
  \font\sixbf=cmbx6  \font\eightbf=cmbx8 
                     \font\eightit=cmti8 
                     \font\eightsl=cmsl8 
                     \font\eighttt=cmtt8 

\font\eightfrak=eufm7 at 8pt

\def\eightpoint{\def\rm{\fam0\eightrm}% switch to 8-point type
 \textfont0=\eightrm \scriptfont0=\sixrm \scriptscriptfont0=\fiverm
 \textfont1=\eighti \scriptfont1=\sixi \scriptscriptfont1=\fivei
 \textfont2=\eightsy \scriptfont2=\sixsy \scriptscriptfont2=\fivesy
 \textfont3=\tenex \scriptfont3=\tenex \scriptscriptfont3=\tenex
 \textfont\itfam=\eightit \def\it{\fam\itfam\eightit}%
 \textfont\slfam=\eightsl \def\sl{\fam\slfam\eightsl}%
 \textfont\ttfam=\eighttt \def\tt{\fam\ttfam\eighttt}%
 \textfont\frakfam=\eightfrak \def\frak{\fam\frakfam\tenfrak}%
 \textfont\bffam=\eightbf \scriptfont\bffam=\sixbf
 \scriptscriptfont\bffam=\fivebf \def\bf{\fam\bffam\eightbf}%
 \normalbaselineskip=9pt
 \setbox\strutbox=\hbox{\vrule height7pt depth2pt width0pt}%
 \let\sc=\sixrm \let\big=\eightbig \normalbaselines\rm}
\catcode`@=11
\def\footnote#1{\edef\@sf{\spacefactor\the\spacefactor}#1\@sf
     \insert\footins\bgroup\eightpoint
     \interlinepenalty100 \let\par=\endgraf
     \leftskip=0pt \rightskip=0pt
     \splittopskip=10pt plus 1pt minus 1pt \floatingpenalty=20000
     \smallskip\item{#1}\bgroup\strut\aftergroup\@foot\let\next}
\skip\footins=12pt plus 2pt minus 4pt
\dimen\footins=30pc
\catcode`@=12

%%%%%%%%%%%%%%%%%%%%%%%%%%%%%%%%%%%%%%%%%%%%%%%%%%%%%%%%%%%%%%%%%%%%%%%
%%%%%%%%    allr.txs - automatic numbering data from FTKr1,2
%%%%%%%%%%%%%%%%%%%%%%%%%%%%%%%%%%%%%%%%%%%%%%%%%%%%%%%%%%%%%%%%%%%%%%%

  \IgNoRe{PG}
  \IgNoRe{STM Assertion }
  \IgNoRe{PG}
  \IgNoRe{PG}
  \IgNoRe{STM Assertion }
  \IgNoRe{PG}
  \IgNoRe{STM Assertion }
  \IgNoRe{STM Assertion }
  \IgNoRe{EQN}
  \IgNoRe{STM Assertion }
  \IgNoRe{STM Assertion }
  \IgNoRe{PG}
  \IgNoRe{STM Assertion }
  \IgNoRe{STM Assertion }
  \IgNoRe{EQN}
  \IgNoRe{STM Assertion }
  \IgNoRe{STM Assertion }
  \IgNoRe{STM Assertion }
  \IgNoRe{STM Assertion }
  \IgNoRe{STM Assertion }
  \IgNoRe{STM Assertion }
  \IgNoRe{PG}
 \def\defFancynorm{\frefwarning II.15} \IgNoRe{STM Assertion }
  \IgNoRe{STM Assertion }
  \IgNoRe{STM Assertion }
 \def\defsymnorm{\frefwarning II.18} \IgNoRe{STM Assertion }
  \IgNoRe{STM Assertion }
  \IgNoRe{STM Assertion }
  \IgNoRe{STM Assertion }
 \def\lemGrasscompatnorm{\frefwarning II.22} \IgNoRe{STM Assertion }
  \IgNoRe{STM Assertion }
  \IgNoRe{STM Assertion }
 \def\defcontractintbound{\frefwarning II.25} \IgNoRe{STM Assertion }
 \def\egIIcompatnorm{\frefwarning II.26} \IgNoRe{STM Assertion }
  \IgNoRe{PG}
  \IgNoRe{EQN}
  \IgNoRe{STM Assertion }
 \def\theorII{\frefwarning II.28} \IgNoRe{STM Assertion }
  \IgNoRe{STM Assertion }
  \IgNoRe{PG}
  \IgNoRe{STM Assertion }
  \IgNoRe{STM Assertion }
 \def\corwicknorm{\frefwarning II.32} \IgNoRe{STM Assertion }
  \IgNoRe{STM Assertion }
  \IgNoRe{EQN}
  \IgNoRe{EQN}
  \IgNoRe{STM Assertion }
  \IgNoRe{PG}
  \IgNoRe{PG}
  \IgNoRe{STM Assertion }
  \IgNoRe{EQN}
  \IgNoRe{STM Assertion }
  \IgNoRe{STM Assertion }
  \IgNoRe{STM Assertion }
  \IgNoRe{EQN}
  \IgNoRe{STM Assertion }
  \IgNoRe{STM Assertion }
  \IgNoRe{STM Assertion }
  \IgNoRe{PG}
  \IgNoRe{STM Assertion }
  \IgNoRe{STM Assertion }
  \IgNoRe{PG}
  \IgNoRe{STM Assertion }
  \IgNoRe{STM Assertion }
  \IgNoRe{STM Assertion }
  \IgNoRe{PG}
  \IgNoRe{STM Assertion }
 \def\theoremIVb{\frefwarning IV.4} \IgNoRe{STM Assertion }
  \IgNoRe{STM Assertion }
  \IgNoRe{STM Assertion }
  \IgNoRe{STM Assertion }
  \IgNoRe{STM Assertion }
  \IgNoRe{STM Assertion }
 \def\propBII{\frefwarning A.2} \IgNoRe{STM Assertion }
  \IgNoRe{PG}
  \IgNoRe{STM Assertion }
  \IgNoRe{STM Assertion }
  \IgNoRe{STM Assertion }
  \IgNoRe{STM Assertion }
  \IgNoRe{STM Assertion }
  \IgNoRe{STM Assertion }
 \def\propGII{\frefwarning B.1} \IgNoRe{STM Assertion }
  \IgNoRe{STM Assertion }
  \IgNoRe{PG}
  \IgNoRe{STM Assertion }
  \IgNoRe{STM Assertion }
  \IgNoRe{PG}
  \IgNoRe{PG}
  \IgNoRe{STM Assertion }
  \IgNoRe{STM Assertion }
  \IgNoRe{PG}
  \IgNoRe{PG}
  \IgNoRe{STM Assertion }
  \IgNoRe{STM Assertion }
  \IgNoRe{EQN}
  \IgNoRe{STM Assertion }
  \IgNoRe{PG}
  \IgNoRe{STM Assertion }
  \IgNoRe{STM Assertion }
  \IgNoRe{STM Assertion }
  \IgNoRe{PG}
  \IgNoRe{STM Assertion }
  \IgNoRe{STM Assertion }
  \IgNoRe{STM Assertion }
  \IgNoRe{EQN}
  \IgNoRe{STM Assertion }
  \IgNoRe{PG}
  \IgNoRe{EQN}
  \IgNoRe{STM Assertion }
  \IgNoRe{STM Assertion }
  \IgNoRe{STM Assertion }
  \IgNoRe{STM Assertion }
  \IgNoRe{EQN}
  \IgNoRe{EQN}
  \IgNoRe{PG}
  \IgNoRe{PG}
  \IgNoRe{STM Assertion }
  \IgNoRe{STM Assertion }
  \IgNoRe{STM Assertion }
  \IgNoRe{STM Assertion }
  \IgNoRe{STM Assertion }
  \IgNoRe{STM Assertion }
  \IgNoRe{STM Assertion }
  \IgNoRe{STM Assertion }
  \IgNoRe{PG}
  \IgNoRe{STM Assertion }
  \IgNoRe{STM Assertion }
  \IgNoRe{STM Assertion }
  \IgNoRe{STM Assertion }
  \IgNoRe{STM Assertion }
  \IgNoRe{STM Assertion }
  \IgNoRe{STM Assertion }
  \IgNoRe{STM Assertion }
  \IgNoRe{STM Assertion }
  \IgNoRe{STM Assertion }
  \IgNoRe{PG}
  \IgNoRe{STM Assertion }
  \IgNoRe{STM Assertion }
  \IgNoRe{STM Assertion }
  \IgNoRe{PG}
  \IgNoRe{STM Assertion }
  \IgNoRe{STM Assertion }
  \IgNoRe{STM Assertion }
  \IgNoRe{STM Assertion }
  \IgNoRe{PG}
  \IgNoRe{STM Assertion }
  \IgNoRe{STM Assertion }
  \IgNoRe{PG}
  \IgNoRe{STM Assertion }
  \IgNoRe{PG}
  \IgNoRe{STM Assertion }
  \IgNoRe{PG}
  \IgNoRe{STM Assertion }
  \IgNoRe{STM Assertion }
  \IgNoRe{STM Assertion }
  \IgNoRe{STM Assertion }
  \IgNoRe{PG}
  \IgNoRe{PG}
  \IgNoRe{STM Assertion }
  \IgNoRe{STM Assertion }
  \IgNoRe{EQN}
  \IgNoRe{EQN}
  \IgNoRe{STM Assertion }
  \IgNoRe{STM Assertion }
  \IgNoRe{STM Assertion }
  \IgNoRe{STM Assertion }
  \IgNoRe{PG}
  \IgNoRe{STM Assertion }
  \IgNoRe{STM Assertion }
  \IgNoRe{STM Assertion }
  \IgNoRe{STM Assertion }
  \IgNoRe{STM Assertion }
  \IgNoRe{STM Assertion }
  \IgNoRe{STM Assertion }
  \IgNoRe{PG}
  \IgNoRe{STM Assertion }
  \IgNoRe{EQN}
  \IgNoRe{EQN}
  \IgNoRe{PG}
  \IgNoRe{STM Assertion }
  \IgNoRe{EQN}
  \IgNoRe{STM Assertion }
  \IgNoRe{STM Assertion }
  \IgNoRe{STM Assertion }
  \IgNoRe{PG}
  \IgNoRe{STM Assertion }
  \IgNoRe{EQN}
  \IgNoRe{STM Assertion }
  \IgNoRe{PG}
  \IgNoRe{PG}

%%%%%%%%%%%%%%%%%%%%%%%%%%%%%%%%%%%%%%%%%%%%%%%%%%%%%%%%%%%%%%%%%%%%%%%
%%%%%%%%    fl-all.txs - automatic numbering data from FTKf1,2,3
%%%%%%%%%%%%%%%%%%%%%%%%%%%%%%%%%%%%%%%%%%%%%%%%%%%%%%%%%%%%%%%%%%%%%%%

  \IgNoRe{STM Assertion }
  \IgNoRe{PG}
  \IgNoRe{EQN}
  \IgNoRe{EQN}
  \IgNoRe{EQN}
  \IgNoRe{EQN}
  \IgNoRe{EQN}
  \IgNoRe{STM Assertion }
  \IgNoRe{STM Assertion }
  \IgNoRe{STM Assertion }
  \IgNoRe{STM Assertion }
  \IgNoRe{STM Assertion }
  \IgNoRe{STM Assertion }
  \IgNoRe{STM Assertion }
  \IgNoRe{STM Assertion }
  \IgNoRe{STM Assertion }
  \IgNoRe{STM Assertion }
  \IgNoRe{STM Assertion }
  \IgNoRe{PG}
  \IgNoRe{PG}
  \IgNoRe{PG}
  \IgNoRe{PG}
  \IgNoRe{EQN}
  \IgNoRe{EQN}
  \IgNoRe{EQN}
  \IgNoRe{EQN}
  \IgNoRe{PG}
  \IgNoRe{PG}
  \IgNoRe{PG}
  \IgNoRe{EQN}
  \IgNoRe{PG}
  \IgNoRe{PG}
  \IgNoRe{EQN}
  \IgNoRe{EQN}
  \IgNoRe{EQN}
  \IgNoRe{EQN}
  \IgNoRe{EQN}
  \IgNoRe{EQN}
  \IgNoRe{EQN}
  \IgNoRe{EQN}
  \IgNoRe{EQN}
  \IgNoRe{EQN}
  \IgNoRe{PG}
  \IgNoRe{PG}
  \IgNoRe{EQN}
  \IgNoRe{EQN}
  \IgNoRe{EQN}
  \IgNoRe{STM Assertion }
  \IgNoRe{PG}
  \IgNoRe{EQN}
  \IgNoRe{EQN}
  \IgNoRe{EQN}
  \IgNoRe{EQN}
  \IgNoRe{STM Assertion }
  \IgNoRe{STM Assertion }
  \IgNoRe{STM Assertion }
  \IgNoRe{STM Assertion }
  \IgNoRe{STM Assertion }
  \IgNoRe{STM Assertion }
  \IgNoRe{STM Assertion }
  \IgNoRe{STM Assertion }
  \IgNoRe{STM Assertion }
  \IgNoRe{EQN}
  \IgNoRe{EQN}
  \IgNoRe{EQN}
  \IgNoRe{EQN}
  \IgNoRe{STM Assertion }
  \IgNoRe{EQN}
  \IgNoRe{STM Assertion }
  \IgNoRe{EQN}
  \IgNoRe{STM Assertion }
  \IgNoRe{PG}
  \IgNoRe{EQN}
  \IgNoRe{STM Assertion }
  \IgNoRe{STM Assertion }
  \IgNoRe{EQN}
  \IgNoRe{PG}
  \IgNoRe{PG}
  \IgNoRe{STM Assertion }
  \IgNoRe{STM Assertion }
  \IgNoRe{PG}
  \IgNoRe{STM Assertion }
  \IgNoRe{STM Assertion }
  \IgNoRe{STM Assertion }
  \IgNoRe{STM Assertion }
  \IgNoRe{STM Assertion }
  \IgNoRe{STM Assertion }
  \IgNoRe{STM Assertion }
  \IgNoRe{STM Assertion }
  \IgNoRe{PG}
  \IgNoRe{STM Assertion }
  \IgNoRe{STM Assertion }
  \IgNoRe{STM Assertion }
  \IgNoRe{STM Assertion }
  \IgNoRe{EQN}
  \IgNoRe{STM Assertion }
  \IgNoRe{STM Assertion }
  \IgNoRe{STM Assertion }
  \IgNoRe{EQN}
  \IgNoRe{STM Assertion }
  \IgNoRe{STM Assertion }
  \IgNoRe{STM Assertion }
  \IgNoRe{STM Assertion }
  \IgNoRe{PG}
  \IgNoRe{STM Assertion }
  \IgNoRe{STM Assertion }
  \IgNoRe{STM Assertion }
  \IgNoRe{STM Assertion }
  \IgNoRe{STM Assertion }
  \IgNoRe{STM Assertion }
  \IgNoRe{STM Assertion }
  \IgNoRe{STM Assertion }
  \IgNoRe{STM Assertion }
  \IgNoRe{EQN}
  \IgNoRe{EQN}
  \IgNoRe{EQN}
  \IgNoRe{STM Assertion }
  \IgNoRe{PG}
  \IgNoRe{STM Assertion }
  \IgNoRe{STM Assertion }
  \IgNoRe{STM Assertion }
  \IgNoRe{EQN}
  \IgNoRe{EQN}
  \IgNoRe{EQN}
  \IgNoRe{STM Assertion }
  \IgNoRe{STM Assertion }
  \IgNoRe{EQN}
  \IgNoRe{EQN}
  \IgNoRe{EQN}
  \IgNoRe{STM Assertion }
  \IgNoRe{PG}
  \IgNoRe{PG}
  \IgNoRe{STM Assertion }
  \IgNoRe{STM Assertion }
  \IgNoRe{PG}
  \IgNoRe{STM Assertion }
  \IgNoRe{STM Assertion }
  \IgNoRe{EQN}
  \IgNoRe{EQN}
  \IgNoRe{EQN}
  \IgNoRe{EQN}
  \IgNoRe{EQN}
  \IgNoRe{EQN}
  \IgNoRe{EQN}
  \IgNoRe{EQN}
  \IgNoRe{EQN}
  \IgNoRe{EQN}
  \IgNoRe{EQN}
  \IgNoRe{EQN}
  \IgNoRe{EQN}
  \IgNoRe{EQN}
  \IgNoRe{STM Assertion }
  \IgNoRe{EQN}
  \IgNoRe{PG}
  \IgNoRe{EQN}
  \IgNoRe{STM Assertion }
  \IgNoRe{EQN}
  \IgNoRe{STM Assertion }
  \IgNoRe{STM Assertion }
  \IgNoRe{EQN}
  \IgNoRe{EQN}
  \IgNoRe{EQN}
  \IgNoRe{STM Assertion }
  \IgNoRe{EQN}
  \IgNoRe{EQN}
  \IgNoRe{EQN}
  \IgNoRe{EQN}
  \IgNoRe{EQN}
  \IgNoRe{EQN}
  \IgNoRe{EQN}
  \IgNoRe{EQN}
  \IgNoRe{EQN}
  \IgNoRe{EQN}
  \IgNoRe{EQN}
  \IgNoRe{EQN}
  \IgNoRe{EQN}
  \IgNoRe{EQN}
  \IgNoRe{EQN}
  \IgNoRe{EQN}
  \IgNoRe{EQN}
  \IgNoRe{EQN}
  \IgNoRe{EQN}
  \IgNoRe{EQN}
  \IgNoRe{EQN}
  \IgNoRe{STM Assertion }
  \IgNoRe{PG}
  \IgNoRe{PG}
  \IgNoRe{STM Assertion }
  \IgNoRe{PG}
  \IgNoRe{EQN}
  \IgNoRe{EQN}
  \IgNoRe{EQN}
  \IgNoRe{EQN}
  \IgNoRe{STM Assertion }
  \IgNoRe{EQN}
  \IgNoRe{PG}
  \IgNoRe{EQN}
  \IgNoRe{EQN}
  \IgNoRe{EQN}
  \IgNoRe{EQN}
  \IgNoRe{EQN}
  \IgNoRe{EQN}
  \IgNoRe{EQN}
  \IgNoRe{EQN}
  \IgNoRe{EQN}
  \IgNoRe{EQN}
  \IgNoRe{EQN}
  \IgNoRe{EQN}
  \IgNoRe{STM Assertion }
  \IgNoRe{STM Assertion }
  \IgNoRe{EQN}
  \IgNoRe{EQN}
  \IgNoRe{PG}
  \IgNoRe{PG}
  \IgNoRe{STM Assertion }
  \IgNoRe{EQN}
  \IgNoRe{STM Assertion }
  \IgNoRe{PG}
  \IgNoRe{EQN}
  \IgNoRe{EQN}
  \IgNoRe{EQN}
  \IgNoRe{STM Assertion }
  \IgNoRe{STM Assertion }
  \IgNoRe{EQN}
  \IgNoRe{EQN}
  \IgNoRe{EQN}
  \IgNoRe{EQN}
  \IgNoRe{EQN}
  \IgNoRe{STM Assertion }
  \IgNoRe{EQN}
  \IgNoRe{EQN}
  \IgNoRe{EQN}
  \IgNoRe{EQN}
  \IgNoRe{STM Assertion }
  \IgNoRe{STM Assertion }
  \IgNoRe{EQN}
  \IgNoRe{STM Assertion }
  \IgNoRe{STM Assertion }
  \IgNoRe{STM Assertion }
  \IgNoRe{STM Assertion }
  \IgNoRe{PG}
  \IgNoRe{STM Assertion }
  \IgNoRe{STM Assertion }
  \IgNoRe{STM Assertion }
  \IgNoRe{STM Assertion }
  \IgNoRe{STM Assertion }
  \IgNoRe{STM Assertion }
  \IgNoRe{STM Assertion }
  \IgNoRe{STM Assertion }
  \IgNoRe{STM Assertion }
  \IgNoRe{STM Assertion }
  \IgNoRe{STM Assertion }
  \IgNoRe{STM Assertion }
  \IgNoRe{STM Assertion }
  \IgNoRe{STM Assertion }
  \IgNoRe{STM Assertion }
  \IgNoRe{PG}
  \IgNoRe{STM Assertion }
  \IgNoRe{STM Assertion }
  \IgNoRe{STM Assertion }
  \IgNoRe{STM Assertion }
  \IgNoRe{STM Assertion }
  \IgNoRe{STM Assertion }
  \IgNoRe{STM Assertion }
  \IgNoRe{STM Assertion }
  \IgNoRe{STM Assertion }
  \IgNoRe{STM Assertion }
  \IgNoRe{STM Assertion }
  \IgNoRe{STM Assertion }
  \IgNoRe{EQN}
  \IgNoRe{STM Assertion }
  \IgNoRe{STM Assertion }
  \IgNoRe{STM Assertion }
  \IgNoRe{STM Assertion }
  \IgNoRe{STM Assertion }
  \IgNoRe{STM Assertion }
  \IgNoRe{EQN}
  \IgNoRe{STM Assertion }
  \IgNoRe{PG}
  \IgNoRe{PG}
  \IgNoRe{STM Assertion }
  \IgNoRe{STM Assertion }
  \IgNoRe{STM Assertion }
  \IgNoRe{EQN}
  \IgNoRe{STM Assertion }
  \IgNoRe{PG}
  \IgNoRe{EQN}
  \IgNoRe{STM Assertion }
  \IgNoRe{STM Assertion }
  \IgNoRe{EQN}
  \IgNoRe{EQN}
  \IgNoRe{EQN}
  \IgNoRe{EQN}
  \IgNoRe{EQN}
  \IgNoRe{EQN}
  \IgNoRe{EQN}
  \IgNoRe{EQN}
  \IgNoRe{EQN}
  \IgNoRe{EQN}
  \IgNoRe{EQN}
  \IgNoRe{EQN}
  \IgNoRe{EQN}
  \IgNoRe{EQN}
  \IgNoRe{EQN}
  \IgNoRe{EQN}
  \IgNoRe{EQN}
  \IgNoRe{EQN}
  \IgNoRe{EQN}
  \IgNoRe{EQN}
  \IgNoRe{STM Assertion }
  \IgNoRe{STM Assertion }
  \IgNoRe{PG}
  \IgNoRe{EQN}
  \IgNoRe{EQN}
  \IgNoRe{EQN}
  \IgNoRe{EQN}
  \IgNoRe{EQN}
  \IgNoRe{EQN}
  \IgNoRe{EQN}
  \IgNoRe{EQN}
  \IgNoRe{EQN}
  \IgNoRe{EQN}
  \IgNoRe{EQN}
  \IgNoRe{EQN}
  \IgNoRe{EQN}
  \IgNoRe{EQN}
  \IgNoRe{EQN}
  \IgNoRe{EQN}
  \IgNoRe{EQN}
  \IgNoRe{EQN}
  \IgNoRe{EQN}
  \IgNoRe{EQN}
  \IgNoRe{EQN}
  \IgNoRe{STM Assertion }
  \IgNoRe{PG}
  \IgNoRe{STM Assertion }
  \IgNoRe{STM Assertion }
  \IgNoRe{EQN}
  \IgNoRe{EQN}
  \IgNoRe{PG}
  \IgNoRe{EQN}
  \IgNoRe{EQN}
  \IgNoRe{EQN}
  \IgNoRe{EQN}
  \IgNoRe{EQN}
  \IgNoRe{EQN}
  \IgNoRe{EQN}
  \IgNoRe{EQN}
  \IgNoRe{STM Assertion }
  \IgNoRe{STM Assertion }
  \IgNoRe{STM Assertion }
  \IgNoRe{STM Assertion }
  \IgNoRe{STM Assertion }
  \IgNoRe{PG}
  \IgNoRe{PG}

%%%%%%%%%%%%%%%%%%%%%%%%%%%%%%%%%%%%%%%%%%%%%%%%%%%%%%%%%%%%%%%%%%%%%%%
%%%%%%%%    fl-chap.tex - chapter numbering data from FTKf1,2,3
%%%%%%%%%%%%%%%%%%%%%%%%%%%%%%%%%%%%%%%%%%%%%%%%%%%%%%%%%%%%%%%%%%%%%%%

\newcount\CHAPNO
\newcount\APPNO
\CHAPNO=0
\APPNO=1
\def\advCHAPNO{\advance\CHAPNO by 1}
\def\advAPPNO{\advance\APPNO by 1}

\def\caproman#1{\ifcase#1\or I\or II\or III\or IV\or V\or VI\or VII\or
VIII\or IX\or X\or XI\or XII\or XIII\or XIV\or XV\or XVI\or XVII\or XVIII\or
XIX\or XX\or XXI\or XXII\or XXIII\or XXIV\or XXV\or XXVI\or XXVII\or XXVIII\or XXIX\or XXX\or XXXI\or XXXII\or XXXIII\or XXXIV\or XXXV\or XXXVI\or XXXVII\or XXXVIII\or XXXIX\fi}%

\def\capletter#1{\ifcase#1\or A\or B\or C\or D\or E\or F\or G\or
H\or I\or J\or K\or L\or M\or N\or O\or P\or Q\or R\or
S\or T\or U\or V\or W\or X\or Y\or Z\fi}%

\newcount\cHintroI \cHintroI=\CHAPNO \advCHAPNO 
                                       %I
\newcount\cHintroOverview  \cHintroOverview=\CHAPNO \advCHAPNO 
                                %II
\newcount\cHrenmap \cHrenmap=\CHAPNO \advCHAPNO 
                                       %III

 \advAPPNO

\newcount\cHintroII \cHintroII=\CHAPNO \advCHAPNO 
                              \edef\CHintroII{\caproman\CHAPNO}
\newcount\cHfirstscale \cHfirstscale=\CHAPNO \advCHAPNO
                              
\newcount\cHnewsectors \cHnewsectors=\CHAPNO \advCHAPNO
                              
\newcount\cHphladders \cHphladders=\CHAPNO \advCHAPNO
                              
\newcount\cHfinitescale \cHfinitescale=\CHAPNO \advCHAPNO
                              
\newcount\cHstep \cHstep=\CHAPNO \advCHAPNO
                              
\newcount\cHrecurs \cHrecurs=\CHAPNO \advCHAPNO
                              
 \advAPPNO

\newcount\cHintroIII \cHintroIII=\CHAPNO \advCHAPNO
                              
\newcount\cHtildefinitescale \cHtildefinitescale=\CHAPNO \advCHAPNO
                              
\newcount\cHtildenewsectors \cHtildenewsectors=\CHAPNO \advCHAPNO
                              
\newcount\cHtildephladders \cHtildephladders=\CHAPNO \advCHAPNO
                              
\newcount\cHtildestep  \cHtildestep=\CHAPNO \advCHAPNO

 \advAPPNO
 \advAPPNO

%%%%%%%%%%%%%%%%%%%%%%%%%%%%%%%%%%%%%%%%%%%%%%%%%%%%%%%%%%%%%%%%%%%%%%%
%%%%%%%%    os-all.txs - automatic numbering data from FTKo1,2,3,4
%%%%%%%%%%%%%%%%%%%%%%%%%%%%%%%%%%%%%%%%%%%%%%%%%%%%%%%%%%%%%%%%%%%%%%%

 \def\pgOSI{\frefwarning 1} \IgNoRe{PG}
 \def\eqnOSintroI{\frefwarning I.1} \IgNoRe{EQN}
 \def\defOSmultideriv{\frefwarning II.1} \IgNoRe{STM Assertion }
 \def\pgOSII{\frefwarning 4} \IgNoRe{PG}
 \def\lemOSleibniz{\frefwarning II.2} \IgNoRe{STM Assertion }
 \def\defOSdecayop{\frefwarning II.3} \IgNoRe{STM Assertion }
 \def\defOSFancynormdomain{\frefwarning II.4} \IgNoRe{STM Assertion }
 \def\defOSFancynorm{\frefwarning II.5} \IgNoRe{STM Assertion }
 \def\exOSSymmNorm{\frefwarning II.6} \IgNoRe{STM Assertion }
 \def\lemOSelloneinfty{\frefwarning II.7} \IgNoRe{STM Assertion }
 \def\eqnOSnormsI{\frefwarning II.1} \IgNoRe{EQN}
 \def\corOSelloneinfty{\frefwarning II.8} \IgNoRe{STM Assertion }
 \def\defOSFmn{\frefwarning II.9} \IgNoRe{STM Assertion }
 \def\defOSSymmNorm{\frefwarning II.10} \IgNoRe{STM Assertion }
 \def\defOScontnorm{\frefwarning III.1} \IgNoRe{STM Assertion }
 \def\defOScontbnd{\frefwarning III.2} \IgNoRe{STM Assertion }
 \def\remOScontbnd{\frefwarning III.3} \IgNoRe{STM Assertion }
 \def\exOSelloneinftycontr{\frefwarning III.4} \IgNoRe{STM Assertion }
 \def\defOSintbnd{\frefwarning III.5} \IgNoRe{STM Assertion }
 \def\pgOSIII{\frefwarning 11} \IgNoRe{PG}
 \def\remOSelloneinftyintbnd{\frefwarning III.6} \IgNoRe{STM Assertion }
 \def\defOShomogGrAlg{\frefwarning III.7} \IgNoRe{STM Assertion }
 \def\remOSminnorm{\frefwarning III.8} \IgNoRe{STM Assertion }
 \def\defOSgrnorm{\frefwarning III.9} \IgNoRe{STM Assertion }
 \def\thmOSroptheorII{\frefwarning III.10} \IgNoRe{STM Assertion }
 \def\thmOSroptheoremIVb{\frefwarning III.11} \IgNoRe{STM Assertion }
 \def\defIntBndsS{\frefwarning IV.1} \IgNoRe{STM Assertion }
 \def\remIntBndsI{\frefwarning IV.2} \IgNoRe{STM Assertion }
 \def\eqnOScovbndsI{\frefwarning IV.1} \IgNoRe{EQN}
 \def\pgOSIV{\frefwarning 15} \IgNoRe{PG}
 \def\pgOSIVa{\frefwarning 15} \IgNoRe{PG}
 \def\propIntBndsII{\frefwarning IV.3} \IgNoRe{STM Assertion }
 \def\lemIntBndsIII{\frefwarning IV.4} \IgNoRe{STM Assertion }
 \def\propIntBndsIV{\frefwarning IV.5} \IgNoRe{STM Assertion }
 \def\defOSderivmom{\frefwarning IV.6} \IgNoRe{STM Assertion }
 \def\remOSloneprod{\frefwarning IV.7} \IgNoRe{STM Assertion }
 \def\propOSpropbnd{\frefwarning IV.8} \IgNoRe{STM Assertion }
 \def\pgOSIVb{\frefwarning 19} \IgNoRe{PG}
 \def\eqnOSsupbnd{\frefwarning IV.2} \IgNoRe{EQN}
 \def\eqnOSoneinftybnd{\frefwarning IV.3} \IgNoRe{EQN}
 \def\eqnOSonsulI{\frefwarning IV.4} \IgNoRe{EQN}
 \def\eqnOSonsulII{\frefwarning IV.5} \IgNoRe{EQN}
 \def\eqnOSonsulIII{\frefwarning IV.6} \IgNoRe{EQN}
 \def\eqnOSonsulIV{\frefwarning IV.7} \IgNoRe{EQN}
 \def\corOSpropbnd{\frefwarning IV.9} \IgNoRe{STM Assertion }
 \def\defOScbzero{\frefwarning IV.10} \IgNoRe{STM Assertion }
 \def\propOSrealfirstpropbound{\frefwarning IV.11} \IgNoRe{STM Assertion }
 \def\lemOSscalednorm{\frefwarning V.1} \IgNoRe{STM Assertion }
 \def\pgOSV{\frefwarning 28} \IgNoRe{PG}
 \def\thmOSinsulators{\frefwarning V.2} \IgNoRe{STM Assertion }
 \def\eqnOSinsI{\frefwarning V.1} \IgNoRe{EQN}
 \def\remOSgamma{\frefwarning V.3} \IgNoRe{STM Assertion }
 \def\defOSsaturated{\frefwarning A.1} \IgNoRe{STM Assertion }
 \def\lemOSappMonoidI{\frefwarning A.2} \IgNoRe{STM Assertion }
 \def\exOSappMonoidI{\frefwarning A.3} \IgNoRe{STM Assertion }
 \def\pgOSA{\frefwarning 34} \IgNoRe{PG}
 \def\lemOSappMonoidIV{\frefwarning A.4} \IgNoRe{STM Assertion }
 \def\corOSappMonoidIV{\frefwarning A.5} \IgNoRe{STM Assertion }
 \def\remOSappMonoidIV{\frefwarning A.6} \IgNoRe{STM Assertion }
 \def\lemOSappMonoidV{\frefwarning A.7} \IgNoRe{STM Assertion }
 \def\pgOSIref{\frefwarning 39} \IgNoRe{PG}
  \IgNoRe{EQN}
  \IgNoRe{EQN}
  \IgNoRe{PG}
  \IgNoRe{STM Assertion }
  \IgNoRe{STM Assertion }
  \IgNoRe{STM Assertion }
  \IgNoRe{EQN}
  \IgNoRe{PG}
  \IgNoRe{STM Assertion }
  \IgNoRe{STM Assertion }
  \IgNoRe{EQN}
  \IgNoRe{STM Assertion }
  \IgNoRe{STM Assertion }
  \IgNoRe{STM Assertion }
  \IgNoRe{STM Assertion }
  \IgNoRe{PG}
  \IgNoRe{STM Assertion }
  \IgNoRe{STM Assertion }
  \IgNoRe{STM Assertion }
  \IgNoRe{STM Assertion }
  \IgNoRe{STM Assertion }
  \IgNoRe{STM Assertion }
  \IgNoRe{STM Assertion }
  \IgNoRe{STM Assertion }
  \IgNoRe{PG}
  \IgNoRe{STM Assertion }
  \IgNoRe{STM Assertion }
  \IgNoRe{STM Assertion }
  \IgNoRe{STM Assertion }
  \IgNoRe{STM Assertion }
  \IgNoRe{EQN}
  \IgNoRe{STM Assertion }
  \IgNoRe{STM Assertion }
  \IgNoRe{PG}
  \IgNoRe{STM Assertion }
  \IgNoRe{STM Assertion }
  \IgNoRe{STM Assertion }
  \IgNoRe{STM Assertion }
  \IgNoRe{STM Assertion }
  \IgNoRe{STM Assertion }
  \IgNoRe{STM Assertion }
  \IgNoRe{STM Assertion }
  \IgNoRe{STM Assertion }
  \IgNoRe{EQN}
  \IgNoRe{STM Assertion }
  \IgNoRe{STM Assertion }
  \IgNoRe{PG}
  \IgNoRe{STM Assertion }
  \IgNoRe{STM Assertion }
  \IgNoRe{STM Assertion }
  \IgNoRe{STM Assertion }
  \IgNoRe{STM Assertion }
  \IgNoRe{STM Assertion }
  \IgNoRe{STM Assertion }
  \IgNoRe{PG}
  \IgNoRe{STM Assertion }
  \IgNoRe{PG}
  \IgNoRe{PG}
  \IgNoRe{STM Assertion }
  \IgNoRe{STM Assertion }
  \IgNoRe{PG}
  \IgNoRe{STM Assertion }
  \IgNoRe{STM Assertion }
  \IgNoRe{STM Assertion }
  \IgNoRe{STM Assertion }
  \IgNoRe{STM Assertion }
  \IgNoRe{STM Assertion }
  \IgNoRe{STM Assertion }
  \IgNoRe{STM Assertion }
  \IgNoRe{STM Assertion }
  \IgNoRe{STM Assertion }
  \IgNoRe{STM Assertion }
  \IgNoRe{STM Assertion }
  \IgNoRe{STM Assertion }
  \IgNoRe{STM Assertion }
  \IgNoRe{STM Assertion }
  \IgNoRe{STM Assertion }
  \IgNoRe{EQN}
  \IgNoRe{STM Assertion }
  \IgNoRe{STM Assertion }
  \IgNoRe{PG}
  \IgNoRe{STM Assertion }
  \IgNoRe{EQN}
  \IgNoRe{EQN}
  \IgNoRe{STM Assertion }
  \IgNoRe{EQN}
  \IgNoRe{EQN}
  \IgNoRe{STM Assertion }
  \IgNoRe{EQN}
  \IgNoRe{STM Assertion }
  \IgNoRe{EQN}
  \IgNoRe{STM Assertion }
  \IgNoRe{STM Assertion }
  \IgNoRe{STM Assertion }
  \IgNoRe{STM Assertion }
  \IgNoRe{PG}
  \IgNoRe{STM Assertion }
  \IgNoRe{STM Assertion }
  \IgNoRe{STM Assertion }
  \IgNoRe{STM Assertion }
  \IgNoRe{EQN}
  \IgNoRe{EQN}
  \IgNoRe{STM Assertion }
  \IgNoRe{STM Assertion }
  \IgNoRe{PG}
  \IgNoRe{STM Assertion }
  \IgNoRe{STM Assertion }
  \IgNoRe{STM Assertion }
  \IgNoRe{EQN}
  \IgNoRe{STM Assertion }
  \IgNoRe{EQN}
  \IgNoRe{STM Assertion }
  \IgNoRe{STM Assertion }
  \IgNoRe{EQN}
  \IgNoRe{STM Assertion }
  \IgNoRe{EQN}
  \IgNoRe{EQN}
  \IgNoRe{EQN}
  \IgNoRe{EQN}
  \IgNoRe{STM Assertion }
  \IgNoRe{STM Assertion }
  \IgNoRe{STM Assertion }
  \IgNoRe{STM Assertion }
  \IgNoRe{PG}
  \IgNoRe{STM Assertion }
  \IgNoRe{STM Assertion }
  \IgNoRe{STM Assertion }
  \IgNoRe{STM Assertion }
  \IgNoRe{STM Assertion }
  \IgNoRe{STM Assertion }
  \IgNoRe{STM Assertion }
  \IgNoRe{STM Assertion }
  \IgNoRe{STM Assertion }
  \IgNoRe{STM Assertion }
  \IgNoRe{EQN}
  \IgNoRe{EQN}
  \IgNoRe{STM Assertion }
  \IgNoRe{PG}
  \IgNoRe{STM Assertion }
  \IgNoRe{STM Assertion }
  \IgNoRe{EQN}
  \IgNoRe{STM Assertion }
  \IgNoRe{STM Assertion }
  \IgNoRe{EQN}
  \IgNoRe{EQN}
  \IgNoRe{EQN}
  \IgNoRe{EQN}
  \IgNoRe{EQN}
  \IgNoRe{STM Assertion }
  \IgNoRe{STM Assertion }
  \IgNoRe{STM Assertion }
  \IgNoRe{EQN}
  \IgNoRe{EQN}
  \IgNoRe{EQN}
  \IgNoRe{EQN}
  \IgNoRe{EQN}
  \IgNoRe{EQN}
  \IgNoRe{STM Assertion }
  \IgNoRe{STM Assertion }
  \IgNoRe{STM Assertion }
  \IgNoRe{PG}
  \IgNoRe{STM Assertion }
  \IgNoRe{STM Assertion }
  \IgNoRe{EQN}
  \IgNoRe{EQN}
  \IgNoRe{STM Assertion }
  \IgNoRe{EQN}
  \IgNoRe{EQN}
  \IgNoRe{STM Assertion }
  \IgNoRe{EQN}
  \IgNoRe{EQN}
  \IgNoRe{STM Assertion }
  \IgNoRe{STM Assertion }
  \IgNoRe{PG}
  \IgNoRe{STM Assertion }
  \IgNoRe{STM Assertion }
  \IgNoRe{STM Assertion }
  \IgNoRe{PG}
  \IgNoRe{STM Assertion }
  \IgNoRe{STM Assertion }
  \IgNoRe{STM Assertion }
  \IgNoRe{STM Assertion }
  \IgNoRe{PG}
  \IgNoRe{STM Assertion }
  \IgNoRe{STM Assertion }
  \IgNoRe{STM Assertion }
  \IgNoRe{STM Assertion }
  \IgNoRe{STM Assertion }
  \IgNoRe{STM Assertion }
  \IgNoRe{STM Assertion }
  \IgNoRe{STM Assertion }
  \IgNoRe{STM Assertion }
  \IgNoRe{STM Assertion }
  \IgNoRe{STM Assertion }
  \IgNoRe{STM Assertion }
  \IgNoRe{STM Assertion }
  \IgNoRe{STM Assertion }
  \IgNoRe{PG}
  \IgNoRe{STM Assertion }
  \IgNoRe{PG}
  \IgNoRe{STM Assertion }
  \IgNoRe{STM Assertion }
  \IgNoRe{PG}
  \IgNoRe{STM Assertion }
  \IgNoRe{STM Assertion }
  \IgNoRe{EQN}
  \IgNoRe{PG}
  \IgNoRe{EQN}
  \IgNoRe{STM Assertion }
  \IgNoRe{STM Assertion }
  \IgNoRe{PG}
  \IgNoRe{EQN}
  \IgNoRe{EQN}
  \IgNoRe{EQN}
  \IgNoRe{EQN}
  \IgNoRe{EQN}
  \IgNoRe{STM Assertion }
  \IgNoRe{STM Assertion }
  \IgNoRe{EQN}
  \IgNoRe{STM Assertion }
  \IgNoRe{PG}
  \IgNoRe{PG}
  \IgNoRe{PG}
  \IgNoRe{STM Assertion }
  \IgNoRe{EQN}
  \IgNoRe{STM Assertion }
  \IgNoRe{PG}
  \IgNoRe{STM Assertion }
  \IgNoRe{EQN}
  \IgNoRe{STM Assertion }
  \IgNoRe{STM Assertion }
  \IgNoRe{PG}
  \IgNoRe{EQN}
  \IgNoRe{EQN}
  \IgNoRe{EQN}
  \IgNoRe{STM Assertion }
  \IgNoRe{STM Assertion }
  \IgNoRe{STM Assertion }
  \IgNoRe{EQN}
  \IgNoRe{STM Assertion }
  \IgNoRe{STM Assertion }
  \IgNoRe{STM Assertion }
  \IgNoRe{STM Assertion }
  \IgNoRe{STM Assertion }
  \IgNoRe{STM Assertion }
  \IgNoRe{PG}
  \IgNoRe{STM Assertion }
  \IgNoRe{STM Assertion }
  \IgNoRe{STM Assertion }
  \IgNoRe{STM Assertion }
  \IgNoRe{STM Assertion }
  \IgNoRe{STM Assertion }
  \IgNoRe{STM Assertion }
  \IgNoRe{PG}
 \def\pgOSInot{\frefwarning 40} \IgNoRe{PG}
  \IgNoRe{PG}
  \IgNoRe{PG}
  \IgNoRe{PG}

%%%%%%%%%%%%%%%%%%%%%%%%%%%%%%%%%%%%%%%%%%%%%%%%%%%%%%%%%%%%%%%%%%%%%%%
%%%%%%%%    os-chap.tex - chapter numbering data from FTKo1,2,3,4
%%%%%%%%%%%%%%%%%%%%%%%%%%%%%%%%%%%%%%%%%%%%%%%%%%%%%%%%%%%%%%%%%%%%%%%

\newcount\CHAPNO
\newcount\APPNO
\CHAPNO=0
\APPNO=1
\def\advCHAPNO{\advance\CHAPNO by 1}
\def\advAPPNO{\advance\APPNO by 1}

\def\caproman#1{\ifcase#1\or I\or II\or III\or IV\or V\or VI\or VII\or
VIII\or IX\or X\or XI\or XII\or XIII\or XIV\or XV\or XVI\or XVII\or XVIII\or
XIX\or XX\or XXI\or XXII\or XXIII\or XXIV\or XXV\or XXVI\or XXVII\or XXVIII\or XXIX\or XXX\or XXXI\or XXXII\or XXXIII\or XXXIV\or XXXV\or XXXVI\or XXXVII\or XXXVIII\or XXXIX\fi}%

\def\capletter#1{\ifcase#1\or A\or B\or C\or D\or E\or F\or G\or
H\or I\or J\or K\or L\or M\or N\or O\or P\or Q\or R\or
S\or T\or U\or V\or W\or X\or Y\or Z\fi}%

\newcount\cHintroI \cHintroI=\CHAPNO \advCHAPNO 
                              
\newcount\cHnorms  \cHnorms=\CHAPNO \advCHAPNO 
                              \edef\CHnorms{\caproman\CHAPNO}
\newcount\cHproprengrp \cHproprengrp=\CHAPNO \advCHAPNO 
                              
\newcount\cHcovbounds  \cHcovbounds=\CHAPNO \advCHAPNO 
                              
\newcount\cHinsulator \cHinsulator=\CHAPNO \advCHAPNO
                              \edef\CHinsulator{\caproman\CHAPNO}

\edef\APappMonoid{\capletter\APPNO} \advAPPNO

\newcount\cHintroII \cHintroII=\CHAPNO \advCHAPNO 
                              \edef\CHintroII{\caproman\CHAPNO}
\newcount\cHamputate \cHamputate=\CHAPNO \advCHAPNO
                              
\newcount\cHscales \cHscales=\CHAPNO \advCHAPNO
                              
\newcount\cHfourier \cHfourier=\CHAPNO \advCHAPNO
                              \edef\CHfourier{\caproman\CHAPNO}
\newcount\cHmomentum \cHmomentum=\CHAPNO \advCHAPNO

 \advAPPNO
 \advAPPNO

\newcount\cHintroIII \cHintroIII=\CHAPNO \advCHAPNO
                              
\newcount\cHsectors \cHsectors=\CHAPNO \advCHAPNO
                              
\newcount\cHsecpropbounds \cHsecpropbounds=\CHAPNO \advCHAPNO
                              
\newcount\cHladdersNotn  \cHladdersNotn=\CHAPNO \advCHAPNO
                              
\newcount\cHestren  \cHestren=\CHAPNO \advCHAPNO
                              
\newcount\cHsecmomnorm \cHsecmomnorm=\CHAPNO \advCHAPNO
                              
\newcount\cHmomestren \cHmomestren=\CHAPNO \advCHAPNO

 \advAPPNO

\newcount\cHintroIV  \cHintroIV=\CHAPNO \advCHAPNO
                              
\newcount\cHcomparison   \cHcomparison=\CHAPNO \advCHAPNO
                              
\newcount\cHsumsmom  \cHsumsmom=\CHAPNO \advCHAPNO
                              
\newcount\cHsectorsmom   \cHsectorsmom=\CHAPNO \advCHAPNO
                              
\newcount\cHppladsect    \cHppladsect=\CHAPNO \advCHAPNO

 \advAPPNO

%=====================================================================
%========================== TITLE PAGE ===============================
%=====================================================================

{\nopagenumbers
\multiply\baselineskip by \spacingDenominator\divide \baselineskip by\spacingNumerator

\null\vskip3truecm

%  
%   (1) Put title in next two lines
%
\centerline{\tafontt Single Scale Analysis of Many Fermion Systems}

\vskip0.1in
\centerline{\tbfontt Part 1: Insulators}

\vskip0.75in
\centerline{Joel Feldman{\parindent=.15in\footnote{$^{*}$}{Research supported 
in part by the
 Natural Sciences and Engineering Research Council of Canada and the Forschungsinstitut f\"ur Mathematik, ETH Z\"urich}}}
\centerline{Department of Mathematics}
\centerline{University of British Columbia}
\centerline{Vancouver, B.C. }
\centerline{CANADA\ \   V6T 1Z2}
\centerline{feldman@math.ubc.ca}
\centerline{http:/\hskip-3pt/www.math.ubc.ca/\squiggle
feldman/}
\vskip0.3in
\centerline{Horst Kn\"orrer, Eugene Trubowitz}
\centerline{Mathematik}
\centerline{ETH-Zentrum}
\centerline{CH-8092 Z\"urich}
\centerline{SWITZERLAND}
\centerline{knoerrer@math.ethz.ch, trub@math.ethz.ch}
\centerline{http:/\hskip-3pt/www.math.ethz.ch/\squiggle
knoerrer/}

\vskip0.75in
\noindent
%
%   (3) Put abstract below here
{\bf Abstract.\ \ \ } 
We construct, using fermionic functional integrals, thermodynamic 
Green's functions for a weakly coupled fermion gas whose Fermi energy 
lies in a gap. Estimates on the Green's functions are obtained that are 
characteristic of the size of the gap. This prepares the way for the 
analysis of single scale renormalization group maps for a system of 
fermions at temperature zero without a gap.

\vfill
\eject

%=====================================================================
%======================= TABLE OF CONTENTS ===========================
%=====================================================================

\titleb{Table of Contents}
\halign{\hfill#\ &\hfill#\ &#\hfill&\ p\ \hfil#&\ p\ \hfil#\cr
\noalign{\vskip0.05in}
\S I&\omit Introduction                             \span&\:\pgOSI&\omit\cr
\noalign{\vskip0.05in}
\S II&\omit Norms                                   \span&\:\pgOSII\cr
\noalign{\vskip0.05in}
\S III&\omit Covariances and the Renormalization Group Map \span&\:\pgOSIII\cr
\noalign{\vskip0.05in}
\S IV&\omit Bounds for Covariances                    \span&\:\pgOSIV\cr
&&Integral Bounds                                      &\omit&\:\pgOSIVa\cr
&&Contraction Bounds                                      &\omit&\:\pgOSIVb\cr
\noalign{\vskip0.05in}
\S V&\omit Insulators                              \span&\:\pgOSV\cr
\noalign{\vskip0.05in}
{\bf Appendices}\span\cr
\noalign{\vskip0.05in}
\S A&\omit Calculations in the Norm Domain            \span&\:\pgOSA\cr
\noalign{\vskip0.05in}
 &\omit References                                    \span&\:\pgOSIref \cr
\noalign{\vskip0.05in}
 &\omit Notation                                      \span&\:\pgOSInot \cr
}
\vfill\eject
\multiply\baselineskip by \spacingNumerator\divide \baselineskip by\spacingDenominator}
\pageno=1

%=====================================================================
%=======================  INTRODUCTION  ==============================
%=====================================================================

\chap{Introduction}\PG\pgOSI

The standard model for a gas of weakly interacting electrons in a $d$-dimensional crystal  is given in terms of
\item{$\bullet$} a single particle dispersion relation (shifted by the chemical
potential) $e(\k)$ on $\bbbr^d$,
\item{$\bullet$} an ultraviolet cutoff $U(\k)$ on $\bbbr^d$,
\item{$\bullet$} the interaction $v$.

\noindent
Here $\k$ is the momentum variable dual to the position variable 
$\x\in \bbbr^d$. The Fermi surface associated to the dispersion relation $e(\k)$ is by definition
$$
F=\set{\k\in\bbbr^d}{ e(\k)=0}
$$
The ultraviolet cutoff is a smooth function with compact support that fulfills
$0\le U(\k)\le 1$ for all $\k\in\bbbr^d$. We assume that it is identically one 
on a neighbourhood of the Fermi surface\footnote{$^{(1)}$}
{In particular, we assume that $F$ is compact.}. The function $v(\x)$ is rapidly decaying; it gives a spin independent translation invariant interaction
$v(\x_1-\x_2)$ between the fermions. 

This situation can be described in field theoretic terms as follows: 
There are anticommuting fields
$\psi_\si(x_0,\x),\,\bar \psi_\si(x_0,\x)$, where $x_0\in \bbbr$ is the temperature (or Euclidean time) argument and $\si \in \{\uparrow,\downarrow\}$ is the spin argument. For $x=(x_0,\x,\si)$ we write 
$\psi(x)= \psi_\si(x_0,\x)$ and $\bar \psi(x)=\bar \psi_\si(x_0,\x)$. The covariance of the Grassmann Gaussian measure, $d\mu_C$,  for these fields 
has Fourier transform
$$
C(k_0,\k) = \frac{U(\k)}{\imath k_0 - e(\k)}
$$
Precisely, for  $x=(x_0,\x,\si),\ x'=(x_0',\x',\si') 
\in \bbbr\times\bbbr^2\times\{\uparrow,\downarrow\}$
$$
C(x,x') = \int \psi(x)\bar\psi(x') d\mu_C(\psi,\bar\psi) 
= \de_{\si,\si'} \int \frac{d^{d+1}k}{(2\pi)^{d+1}} e^{\imath<k,x-x'>_-}C(k)
$$
where $<k,x-x'>_- = -k_0(x_0-x'_0) + \k\cdot(\x-\x')$ for 
$k=(k_0,\k)\in\bbbr\times\bbbr^d$.
The interaction between the fermions is described by an effective potential
$$
\cV(\psi,\bar\psi) = \int_{(\bbbr\times\bbbr^2\times\{\uparrow,\downarrow\})^4}  \hskip-.7in
V_0(x_1,x_2,x_3,x_4)\, \bar\psi(x_1)\psi(x_2)\bar\psi(x_3)\psi(x_4)\
dx_1dx_2dx_3dx_4
$$
with the interaction kernel
$$
V_0\big( (x_{1,0},\x_1,\si_1),\cdots,(x_{4,0},\x_4,\si_4)\big)
= -\half \de(x_1,x_2) \de(x_3,x_4)\de(x_{1,0}-x_{3,0}) v(\x_1-\x_3)
$$
where $\de\big((x_0,\x,\si),(x_0',\x',\si'))=\de(x_0-x_0')\de(\x-\x')\de_{\si,\si'}$.
More generally we will discuss translation invariant and spin independent interaction kernels $V_0(x_1,x_2,x_3,x_4)$. 

Formally, the generating functional for the  connected amputated\footnote{$^{(2)}$}
{with respect to the bare propagator $C(x,y)$} Green's functions is
$$
\cG_{\rm amp}(\phi,\bar\phi) =  \log\sfrac{1}{Z} 
\int  e^{\cV(\psi+\phi,\bar \psi+\bar\phi)}\,d\mu_{C}(\psi,\bar\psi)
$$
where $Z=\int e^{\cV(\psi,\bar\psi)} d\mu_C(\psi,\bar\psi)$.
The connected amputated Green's functions themselves are determined by
$$
\cG_{\rm amp}(\phi,\bar \phi) = 
\sum_{n=1}^\infty \sfrac{1}{(n!)^2} \int\smprod_{i=1}^n dx_idy_i\ 
G_{2n}^{\rm amp}(x_1,y_1,\cdots,x_n,y_n) 
\smprod_{i=1}^n \bar\phi(x_i)\phi(y_i) 
$$
The fields $\phi,\bar\phi$ are called source fields, the fields $\psi,\bar\psi$
internal fields.

In a renormalization group analysis, the covariance $C$ is written as a sum
$$
C=\sum_{j=0}^\infty C^{(j)}
$$
where $C^{(j)}(k)$ is supported on the set $k=(k_0,\k)\in\bbbr\times\bbbr^d$ for which $|C(k)|$ is approximately $M^j$. Here $M>1$ is a scale parameter.
To successively integrate out scales $j=0,1,\cdots$, we use the renormalization group map analyzed in [FKTr1]. For any
covariance $S$ we set
$$
\Om_S(\cW)(\phi,\bar\phi,\psi,\bar\psi)=
 \log\sfrac{1}{Z} 
\int  e^{\cW(\phi,\bar\phi,\psi+\ze,\bar \psi+\bar\ze)}\,d\mu_{S}(\ze,\bar\ze)
$$
Here, $\cW$ is a Grassmann function and the partition function is
$Z=\int  e^{\cW(0,0,\ze,\bar\ze)}\,d\mu_{S}(\ze,\bar\ze)$. $\Om_S$
maps Grassmann functions in the variables $\phi,\bar\phi,\psi,\bar \psi$
to Grassmann functions in the same variables. Clearly
$$
\cG_{\rm amp}(\phi,\bar \phi)=\Om_C(\cV)(0,0,\phi,\bar\phi)
\EQN\eqnOSintroI$$
where we view $\cV(\psi,\bar\psi)$ as a function of the four variables
$\phi,\bar\phi,\psi,\bar\psi$ that happens to be independent of $\phi,\bar\phi$.
Also, $\Om$ obeys the semigroup property
$$
\Om_{S_1+S_2}=\Om_{S_1}\circ\Om_{S_2}
$$
Even for a cutoff covariance $S$, it is not a priori clear that $\Om_S(\cW)$
makes sense for a reasonable set of $\cW$'s. On the other hand, it is easy
to see, using graphs, that each term in the formal Taylor expansion 
of $\Om_S(\cW)$ in powers of $\cW$ is well--defined for a large class 
of $\cW$'s and cutoff $S$'s.  The Taylor expansion of 
$\int  e^{\cW(\phi,\bar\phi,\psi+\ze,\bar \psi+\bar\ze)}\,d\mu_{S}(\ze,\bar\ze)$
is $\sum_{n=1}^\infty\cG_n(\cW,\cdots,\cW)$ where the 
$n^{\rm th}$ term is the multilinear form
$$
\cG_n(\cW_1,\cdots,\cW_n)= \sfrac{1}{n!}
\int  \cW_1(\phi,\bar\phi,\psi+\ze,\bar \psi+\bar\ze)
\cdots\cW_n(\phi,\bar\phi,\psi+\ze,\bar \psi+\bar\ze)\ d\mu_{S}(\ze,\bar\ze)
$$
restricted to the diagonal. Explicit evaluation of the Grassmann 
integral expresses $\cG_n$ as the sum of all graphs with vertices 
$\cW_1,\ \cdots,\ \cW_n$ and lines $S$. The (formal) Taylor coefficient 
$\sfrac{d\hfill}{dt_1}\cdots\sfrac{d\hfill}{dt_n}
\Om_S(t_1\cW_1+\cdots +t_n\cW_n)\Big|_{t_1=\cdots=t_n=0}$ 
of $\Om_S(\cW)$ is similar, but with only connected graphs. We prove in 
[FKTr1] that, under hypotheses that will be satisfied here, the formal 
Taylor series for $\Om_S(\cW)$ 
converges to an analytic\footnote{$^{(3)}$}{For an elementary discussion of analytic maps between Banach spaces see, for example, Appendix A of [PT].}
function of $\cW$.

In part 3 of this paper, we analyze the maps $\Om_{C^{(j)}}$ in great detail.
These results are used in our proof [FKTf1,2,3] that the temperature zero 
renormalized perturbation expansions of a class of interacting 
many--fermion models in two space dimensions have nonzero radius of
convergence.  
In this first part, we apply the general results of
[FKTr1] to  many fermion systems. In particular, we introduce concrete norms that fulfill the conditions of \S II.4 of 
[FKTr1] and develop contraction and integral bounds
for them. Then, we apply Theorem \theorII\ of [FKTr1]
and (\eqnOSintroI) to models for which the dispersion relation is both infrared and ultraviolet finite (insulators). For these models, no scale decomposition is necessary.

In the second part of this paper, we introduce scales and apply the results of part 1 to integrate out the first few scales. It turns out that for higher scales the norms introduced in parts 1 and 2 are inadequate and, in particular, power count poorly. Using sectors, we introduce finer norms that, in dimension 
two, power count appropriately. For these sectorized norms, passing from one 
scale to the next is not completely trivial. This question is dealt with in   
part 4.  Cumulative notation tables are provided at the end of each part 
of this paper.

\vfill\eject
%=====================================================================
%=============================== NORMS ===============================
%=====================================================================

\chap{Norms}\PG\pgOSII

Let $A$ be the Grassmann algebra freely generated by the fields
$\phi(y), \bar\phi(y)$ with $y~\in~\bbbr\times\bbbr^d\times \{\uparrow,\downarrow\}$. The generating functional for the connected Greens functions 
is a Grassmann Gaussian integral in the Grassmann algebra with coefficients in $A$ that is generated by the fields $\psi(x), \bar\psi(x)$ with $x \in \bbbr\times\bbbr^d\times \{\uparrow,\downarrow\}$. We want to apply the results
of [FKTr1] to this situation.

To simplify notation we define, for $\xi=(x_0,\x,\si,a)=(x,a) \in \bbbr \times \bbbr^d \times \{\uparrow,\downarrow\} \times \{0,1\}$, the internal fields
$$
\psi(\xi) = \cases{\psi(x) & if $a=0$\cr
                   \bar\psi(x) & if $a=1$\cr
}$$ 
Similarly, we define for an external variable  
$\eta=(y_0,\y,\tau,b)=(y,b) \in \bbbr \times \bbbr^d \times \{\uparrow,\downarrow\} \times \{0,1\}$, the source fields
$$
\phi(\eta) = \cases{\phi(y) & if $b=0$\cr
                   \bar\phi(y) & if $b=1$\cr
}$$ 
$\cB = \bbbr \times \bbbr^2 \times \{\uparrow,\downarrow\}\times\{0,1\}$ is called the ``base space'' parameterizing the fields. 
The Grassmann algebra $A$ is the direct sum of the vector spaces $A_m$
generated by the products $\phi(\eta_1)\cdots\phi(\eta_m)$.  
Let $V$ be the vector space generated by $\psi(\xi),\ \xi\in\cB$. 
An antisymmetric function $C(\xi,\xi')$ on $\cB\times\cB$ defines a covariance
on $V$ by $C({\sst \psi(\xi),\psi(\xi')})=C(\xi,\xi')$.
The Grassmann Gaussian integral  with this 
covariance, $\int\ \cdot\ d\mu_C(\psi)$, is a linear functional 
on the Grassmann algebra $\bigwedge_AV$ with values in $A$.

We shall define norms on $\bigwedge_AV$ by specifying norms on the spaces of
functions on $\cB^m\times\cB^n,\ m,n\ge 0$. The rudiments of such norms and 
simple examples are discussed in this section. In the next section we recall the results of [FKTr1] in the current concrete situation.

The norms we construct are $(d+1)$--dimensional seminorms in the sense
of Definition \defFancynorm\ of [FKTr1]. They measure the spatial decay of the functions, i.e. derivatives of their Fourier transforms.

\definition{\STM\defOSmultideriv (Multiindices)}{
\Item{i)}
A multiindex is an element 
$\de=(\de_0,\de_1,\cdots,\de_d) \in \bbbn_0 \times \bbbn_0^d$.
The length of a multiindex $\de=(\de_0,\de_1,\cdots\de_d)$ is 
$|\de| =\de_0+\de_1+\cdots+\de_d$ and its factorial is
$\de!=\de_0!\de_1!\cdots\de_d!$. For two multiindices $\de,\de'$ we say that
$\de \le \de'$ if $\de_i \le \de'_i$ for $i=0,1,\cdots,d$.
The spatial part of the multiindex $\de=(\de_0,\de_1,\cdots\de_d)$ is 
$\bde=(\de_1,\cdots\de_d)\in\bbbn_0^d$. It has length
 $|\bde| =\de_1+\cdots+\de_d$.
\Item{ii)} Let $\de,\de^{(1)},\cdots,\de^{(r)}$ be multiindices such that
$ \de=\de^{(1)}+\cdots+\de^{(r)}$. Then by definition
$$
\smchoose{\de}{\de^{(1)},\cdots,\de^{(r)}} =  
\sfrac{\de!}{\de^{(1)}!\cdots \de^{(r)}!}
$$
\Item{iii)}
For a  multiindex $\de$ and 
$x=(x_0,\x,\si),\,x'=(x'_0,\x',\si') \in \bbbr\times\bbbr^d\times \{ \uparrow,\,\downarrow\}$ set
$$
(x-x')^\de = (x_0-x_0')^{\de_0}\,(\x_1-\x'_1)^{\de_1}\cdots
(\x_d-\x_d')^{\de_d}
$$
If  $\xi=(x,a),\,\xi'=(x',a') \in \cB$ we define
$\ (\xi -\xi')^\de = (x-x')^\de$. 
\Item{iv)}
For a function $f(\xi_1,\cdots,\xi_n)$ on $\cB^n$, a multiindex $\de$, and
$1\le i,j\le n;\,i\ne j$ set
$$
\cD_{i,j}^\de f\,(\xi_1,\cdots,\xi_n) 
= (\xi_i-\xi_j)^\de f(\xi_1,\cdots,\xi_n)
$$
}

\lemma{\STM\lemOSleibniz (Leibniz's rule)}{
Let $f(\xi_1,\cdots,\xi_n)$ be a function on $\cB^n$ and $f'(\xi_1,\cdots,\xi_m)$
a function on $\cB^m$. Set 
$$
g(\xi_1,\cdots,\xi_{n+m-2}) 
= \int_\cB d\eta\,f(\xi_1,\cdots,\xi_{n-1},\eta)\,f'(\eta,\xi_n\cdots,\xi_{n+m-2})
$$
Let $\de$ be a multiindex and $1\le i \le n-1,\ n\le j \le n+m-2$. Then
$$
\cD_{i,j}^\de g\,({\sst \xi_1,\cdots,\xi_{n+m-2}})
= \smsum_{\de'\le \de} \smchoose{\de}{\de',\de-\de'}
\int_\cB d\eta\ \cD_{i,n}^{\de'} f({\sst\xi_1,\cdots,\xi_{n-1},\eta})\,
\,\cD_{1,j-n+2}^{\de-\de'}f'\,({\sst \eta,\xi_n\cdots,\xi_{n+m-2}})
$$
} 
\prf
For each $\eta \in \cB$
$$
(\xi_i-\xi_j)^\de\ = \ \big((\xi_i-\eta) + (\eta-\xi_j)\big)^\de
=\smsum_{\de'\le \de} \smchoose{\de}{\de',\de-\de'}
(\xi_i-\eta)^{\de'}(\eta-\xi_j)^{\de-\de'}
$$
\endproof

\definition{\STM\defOSdecayop (Decay operators)}{
Let $n$ be a positive integer. A decay operator $\cD$ on the set of 
functions on $\cB^n$ is an operator of the form
$$
\cD = \cD^{\de^{(1)}}_{u_1,v_1} \cdots 
 \cD^{\de^{(k)}}_{u_k,v_k}
$$
with multiindices $\de^{(1)},\cdots,\de^{(k)}$ and 
$1\le u_j,v_j\le n,\ u_j\ne v_j$. The indices $u_j,v_j$ are called variable
indices.  The total order of derivatives in $\cD$ is 
$$
\de(\cD) = \de^{(1)} + \cdots + \de^{(k)}
$$
In a similar way, we define the action of a decay operator on the set of 
functions on $\big(\bbbr\times\bbbr^d \big)^n$ or on
$\big(\bbbr\times\bbbr^d \times \{\uparrow,\downarrow\} \big)^n$.
}

\definition{\STM\defOSFancynormdomain}{
\Item{i)}
On $\bbbr_+\cup\{\infty\} = \set{x\in\bbbr}{x\ge0}\cup\{+\infty\}$, addition
 and the total ordering $\le$ are defined in the standard way. With the convention that $0\cdot\infty=\infty$, multiplication is also defined in the standard way.
\Item{ii)}
Let $d\ge -1$. For $d\ge 0$, 
the $(d+1)$--dimensional norm domain $\fN_{d+1}$ is the 
set of all formal power series
$$
X = \sum_{\de\in\bbbn_0\times\bbbn_0^d} X_\de \
t_0^{\de_0}t_1^{\de_1}\cdots t_d^{\de_d}
$$
in the variables $t_0,t_1,\cdots,t_d$ with coefficients $X_\de \in \bbbr_+\cup\{\infty\}$. To shorten notation, we set
 $t^\de=t_0^{\de_0}t_1^{\de_1}\cdots t_d^{\de_d} $.
 Addition and partial ordering on $\fN_{d+1}$ are defined componentwise. Multiplication is defined by
$$
(X\cdot X')_\de = \smsum_{\be+\ga=\de} X_\be X'_\ga
$$
The max and min of two elements of $\fN_{d+1}$ are again defined componentwise.

\noindent
The zero--dimensional norm domain $\fN_0$ is defined to be $\bbbr_+\cup\{\infty\}$.
We also identify $\bbbr_+\cup\{\infty\}$ with the set of all $X\in \fN_{d+1}$ with
$X_\de=0$ for all $\de\ne \0=(0,\cdots,0)$.

\noindent
If $a>0$, $X_\0\ne\infty$ and $a-X_\0 >0$ then $(a-X)^{-1}$ is defined as
$$
(a-X)^{-1} = \sfrac{1}{a-X_\0} 
\smsum_{n=0}^\infty \big(\sfrac{X-X_\0}{a-X_\0}\big)^n
$$

\noindent
For an element 
$\ 
X = \sum_{\de\in\bbbn_0\times\bbbn_0^d} X_\de \ t^{\de}
\ $
of $\fN_{d+1}$ and $0\le j\le d$ the formal 
derivative $\sfrac{\partial\hfill}{\partial t_j}X$ is defined as
$$
\sfrac{\partial\hfill}{\partial t_j}X 
= \sum_{\de\in\bbbn_0\times\bbbn_0^d} (\de_j+1)X_{\de+\ep_j} \ t^{\de}
$$
where $\ep_j$ is the $j^{\rm th}$ unit vector.
}

\definition{\STM\defOSFancynorm}{
Let $E$ be a complex vector space. A $(d+1)$--dimensional seminorm on $E$ is a 
map $\ \|\cdot\|:\,E\rightarrow \fN_{d+1}\ $ such that
$$
\|e_1+e_2\| \le \|e_1\|+\|e_2\| \qquad,\qquad \|\la\,e\| = |\la|\,\|e\|
$$
for all $e,e_1,e_2\in E$ and $\la \in \bbbc$.
}

\example{\STM\exOSSymmNorm}{
For a function $f$ on $\cB^m \times \cB^n$ we 
define the (scalar valued) $L_1$--$L_\infty$--norm as
$$
\tn f\tn_{1,\infty} 
= \cases{ 
\max\limits_{1\le j_0 \le n}\ 
\sup\limits_{\xi_{j_0} \in \cB}\  
\int \prod\limits_{j=1,\cdots, n \atop j\ne j_0} d\xi_j\, 
| f( \xi_1,\cdots,\xi_n) | & if \ $m=0$  \cr
\sup\limits_{\eta_1,\cdots,\eta_m \in \cB}
\int \prod\limits_{j=1,\cdots, n}  d\xi_j\  
| f( \eta_1,\cdots,\eta_m;\,\xi_1,\cdots,\xi_n) | & if \ $m\ne0$ 
}$$
and the $(d+1)$--dimensional $L_1$--$L_\infty$ seminorm
$$
\| f\|_{1,\infty}=\cases{
{\dst \sum_{\de\in\bbbn_0\times\bbbn_0^d}}\ \frac{1}{\de!}\ \Big(
\max\limits_{\cD\ {\rm decay\ operator} \atop {\rm with\ } \de(\cD) =\de} 
 \tn\cD\,f\tn_{1,\infty}\Big)\ 
 t_0^{\de_0}t_1^{\de_1}\cdots t_d^{\de_d} & if $m=0$\cr
\noalign{\vskip.1in}
\tn f\tn_{1,\infty}& if $m\ne 0$\cr
}
$$
Here $\tn f\tn_{1,\infty}$ stands for the formal power series with constant coefficient $\tn f\tn_{1,\infty}$ and all other coefficients zero.
}

\lemma{\STM\lemOSelloneinfty}{
Let $f$ be a function on $\cB^m \times \cB^n$
and $f'$ a function on $\cB^{m'} \times \cB^{n'}$.
Let $1\le \mu \le n,\ 1\le \nu \le n'$. 
Define the function $g$ on 
$\cB^{m+m'} \times \cB^{n+n'-2}$ by
$$\eqalign{
&g({\sst \eta_1,\cdots,\eta_{m+m'};
\xi_1,\cdots,\xi_{\mu-1},\,\xi_{\mu+1},\cdots,\xi_n,\,
\xi_{n+1},\cdots,\xi_{n+\nu-1},\,\xi_{n+\nu+1},\cdots,\xi_{n+n'}}) \cr
&\hskip .7cm  =\int_\cB {\sst d\zeta}\, f( {\sst \eta_1,\cdots,\eta_m;\, 
\xi_1,\cdots,\xi_{\mu-1},\,\zeta,\,\xi_{\mu+1},\cdots,\xi_n })\, 
f'( {\sst \eta_{m+1},\cdots,\eta_{m+m'};
\,\xi_{n+1},\cdots,\xi_{n+\nu-1},\,\zeta,\,\xi_{n+\nu+1},\cdots,\xi_{n+n'}})
\cr
}$$
If $m=0$ or $m'=0$
$$\eqalign{
\tn g\tn_{1,\infty} &\le \tn f\tn_{1,\infty}\ \tn f'\tn_{1,\infty}\cr
\|g\|_{1,\infty} &\le \|f\|_{1,\infty}\ \|f'\|_{1,\infty}\cr
}$$
}

\prf We first consider the norm $\tn\ \cdot \ \tn_{1,\infty}$.
In the case $m\ne 0, m'=0$, for all $\eta_1,\cdots,\eta_m \in \cB$
$$\eqalign{
\Big| \int \smprod_{j=1 \atop j \ne \mu,n+\nu}^{n+n'}& d\xi_j\ 
g({\sst \eta_1,\cdots,\eta_{m};
\xi_1,\cdots,\xi_{\mu-1},\,\xi_{\mu+1},\cdots,\xi_n,\,
\xi_{n+1},\cdots,\xi_{n+\nu-1},\,\xi_{n+\nu+1},\cdots,\xi_{n+n'}})  \Big| \cr 
& \le \Big| \int {\sst d\xi_1\cdots d\xi_n}\  
f( {\sst \eta_1,\cdots,\eta_m;\, \xi_1,\cdots,\xi_n}) \Big| 
\ \ \sup_{\zeta \in \cB} 
\Big| \int \smprod_{j=1 \atop j \ne \nu}^{n'} {\sst\,d\xi'_j}\ 
f'( {\sst \xi'_1,\cdots,\xi'_{\nu-1},\,\zeta,\,\xi'_{\nu+1},\cdots,\xi'_{n'}})
\Big|  \cr  
& \le\tn f\tn_{1,\infty}\ \tn f'\tn_{1,\infty}
}$$
The case $m=0,\ m'\ne 0$ is similar. In the case $m=m'=0$ first fix 
$j_0 \in \{1,\cdots,n\}\setminus \{\mu\}$, and fix $\xi_{j_0} \in \cB$. 
As in the case $m\ne 0, m'=0$ one shows that
$$\eqalign{
\Big| &\int \smprod_{j=1 \atop j \ne j_0,\mu,n+\nu}^{n+n'} d\xi_j\ 
g({\sst \xi_1,\cdots,\xi_{\mu-1},\,\xi_{\mu+1},\cdots,\xi_n,\,
\xi_{n+1},\cdots,\xi_{n+\nu-1},\,\xi_{n+\nu+1},\cdots,\xi_{n+n'}})  \Big| \cr 
& \hskip 3cm \le \Big| \int \smprod_{j=1 \atop j\ne j_0}^n{\sst d\xi_j}\  
f( {\sst \xi_1,\cdots,\xi_n}) \Big| 
\ \ \sup_{\zeta \in \cB} 
\Big| \int \smprod_{j=1 \atop j \ne \nu}^{n'} {\sst\,d\xi'_j}\ 
f'( {\sst \xi'_1,\cdots,\xi'_{\nu-1},\,\zeta,\,\xi'_{\nu+1},\cdots,\xi'_{n'}})
\Big|  \cr  
& \hskip 3cm \le\tn f\tn_{1,\infty}\ \tn f'\tn_{1,\infty}
}$$
If one fixes one of the variables $\xi_{j_0}$ with 
$j_0 \in \{n+1,\cdots n+n'\}\setminus \{n+\nu\}$, the argument is similar.

We now consider the norm $\|\ \cdot \ \|_{1,\infty}$.
If $m\ne 0$ or $m'\ne 0$ this follows from the first part of this Lemma and
$$
\|g\|_{1,\infty}=\tn g\tn_{1,\infty} 
\le \tn f\tn_{1,\infty}\ \tn f'\tn_{1,\infty}
 \le \|f\|_{1,\infty}\ \|f'\|_{1,\infty}
$$
So assume that $m=m'=0$. Set
$$\eqalign{
\|f\|_{1,\infty}
&= \smsum_{\de\in\bbbn_0\times\bbbn_0^d}\sfrac{1}{\de!}\,X_\de\,t^\de\quad\qquad
\|f'\|_{1,\infty}
= \smsum_{\de\in\bbbn_0\times\bbbn_0^d}\sfrac{1}{\de!}\,X'_\de\,  t^\de\cr
\|g\|_{1,\infty}
&= \smsum_{\de\in\bbbn_0\times\bbbn_0^d}\sfrac{1}{\de!}\,Y_\de\,t^\de\cr
}$$
with $X_\de,X'_\de,Y_\de\in\bbbr_+\cup\{\infty\}$.
Let $\cD$ be a decay operator of degree $\de$ 
acting on $g$. 
The variable indices for $g$ lie in the set $I\cup I'$, where
$$\eqalign{
I & = \{1,\cdots,\mu-1,\,\mu+1,\cdots,n\} \cr
I' & =\{n+1,\cdots,n+\nu-1,\,n+\nu+1,\cdots,n+n'\} \cr
}$$ 
We can factor the decay operator $\cD$ in the form
$$
\cD = \pm\,\tilde\cD\,\cD_2\,\cD_1
$$ 
where all variable indices of $\cD_1$ lie in $I$, all variable indices of 
$\cD_2$ lie in $I'$, and
$$
\tilde \cD =\cD^{\de^{(1)}}_{u_1,v_1} \cdots \cD^{\de^{(k)}}_{u_k,v_k}
$$
with $u_1,\cdots,u_k \in I,\,v_1,\cdots,v_k \in I'$. Set
$h = \cD_1f $ and $h'=\cD_2 f'$.
By Leibniz's rule 
$$\eqalign{
\cD\,g 
&= \pm\tilde \cD\,\int_\cB {\sst d\zeta}\, 
h({\sst \xi_1,\cdots,\xi_{\mu-1},\,\zeta,\,\xi_{\mu+1},\cdots,\xi_n })\, 
h'( {\sst \xi_{n+1},\cdots,\xi_{n+\nu-1},\,\zeta,\,\xi_{n+\nu+1},\cdots,\xi_{n+n'}})
\cr
&= \pm\sum_{\al^{(i)} + \be^{(i)} = \de^{(i)} \atop {\rm for\ } i=1,\cdots,k}
\Big( \smprod_{i=1}^k \smchoose{\de^{(i)}}{\al^{(i)}, \be^{(i)}} \Big)
\int d\zeta \, \big(  \smprod_{i=1}^k \cD^{\al^{(i)}}_{u_i,\mu} h \big)
   ({\sst \xi_1,\cdots,\xi_{\mu-1},\,\zeta,\,\xi_{\mu+1},\cdots,\xi_n } )\cr
& \hskip 6cm \big(  \smprod_{i=1}^k \cD^{\be^{(i)}}_{\nu,v_i} h' \big) 
({\sst \xi_{n+1},\cdots,\xi_{n+\nu-1},\,\zeta,\,\xi_{n+\nu+1},\cdots,\xi_{n+n'}}) \cr
}$$
By the first part of this Lemma, the $L_1$--$L_\infty$--norm of each integral 
on the right hand side is bounded by
$$
\TN  \smprod_{i=1}^k \cD^{\al^{(i)}}_{u_i,\mu} h \TN_{1,\infty}\,
\TN  \smprod_{i=1}^k \cD^{\be^{(i)}}_{\nu,v_i} h'\TN_{1,\infty}
$$
Therefore, setting $\tilde\de=\de(\tilde\cD)=\de^{(1)}+\cdots+\de^{(k)}$,
$$\eqalignno{
\tn \cD\,g\tn_{1,\infty} t^\de
& \le  \sum_{\al^{(i)} + \be^{(i)} = \de^{(i)} \atop {\rm for\ }
i=1,\cdots,k}
\Big( \smprod_{i=1}^k \smchoose{\de^{(i)}}{\al^{(i)}, \be^{(i)}} \Big)
t^{\de(\cD_1)}\,t^{\al^{(1)}+\cdots+\al^{(k)}}\,
\TN  \smprod_{i=1}^k \cD^{\al^{(i)}}_{u_i,\mu}(\cD_1f) \TN_{1,\infty}
\cr &\hskip 5cm 
t^{\de(\cD_2)}\,t^{\be^{(1)}+\cdots+\be^{(k)}}\,
\TN  \smprod_{i=1}^k \cD^{\be^{(i)}}_{\nu,v_i}(\cD_2f') \TN_{1,\infty} \cr
%%%
& \le  \sum_{\al+\be=\tilde\de}\sum_{\al^{(i)} + \be^{(i)} = \de^{(i)} \atop {\al^{(1)}+\cdots+\al^{(k)} =\al
\atop  \be^{(1)}+\cdots+\be^{(k)} =\be } }
\Big( \smprod_{i=1}^k \smchoose{\de^{(i)}}{\al^{(i)}, \be^{(i)}} \Big)
\ \ X_{\de(\cD_1)+\al}\,t^{\de(\cD_1)+\al}
\ \ X'_{\de(\cD_2)+\be}\,t^{\de(\cD_2)+\be} \cr
%%%
&= \sum_{\al+\be=\tilde\de}
\smchoose{\tilde \de}{\al\,,\, \be}
\ \ X_{\de(\cD_1)+\al}\,t^{\de(\cD_1)+\al}
\ \ X'_{\de(\cD_2)+\be}\,t^{\de(\cD_2)+\be} \cr
%%%
&\le \sum_{\al+\be=\tilde\de}
\smchoose{\de}{\de(\cD_1)+\al\ ,\ \de(\cD_2)+\be}
\ \ X_{\de(\cD_1)+\al}\,t^{\de(\cD_1)+\al}
\ \ X'_{\de(\cD_2)+\be}\,t^{\de(\cD_2)+\be}&\EQNO\eqnOSnormsI \cr
}$$
In the equality, we used the fact that for each pair of multiindices $\al,\be$ 
with $\al+\be=\tilde \de$ and each $k$--tuple of multiindices $\de^{(i)},\ 
1\le i\le k$, with $\smsum_i \de^{(i)}=\tilde\de$
$$
 \sum_{\al^{(i)} + \be^{(i)} = \de^{(i)} \atop {\al^{(1)}+\cdots+\al^{(k)} =\al
\atop  \be^{(1)}+\cdots+\be^{(k)} =\be } }
 \smprod_{i=1}^k \smchoose{\de^{(i)}}{\al^{(i)}, \be^{(i)}} 
=\smchoose{\tilde \de}{\al\,,\, \be}
$$
This standard combinatorial identity follows from 
$$\eqalign{
\sum_{\al+\be=\tilde\de}\smchoose{\tilde \de}{\al\,,\, \be}\, x^\al y^\be
&=(x+y)^{\tilde \de}=\smprod_{i=1}^k(x+y)^{\de^{(i)}}
=\smprod_{i=1}^k\Big(
\sum_{\al^{(i)}+\be^{(i)}=\de^{(i)}}
\smchoose{\de^{(i)}}{\al^{(i)}, \be^{(i)}}\,
 x^{\al^{(i)}} y^{\be^{(i)}} \Big)\cr
&=
\sum_{\al^{(i)}+\be^{(i)}=\de^{(i)}\atop i=1,\cdots,k}
\Big[\smprod_{i=1}^k\smchoose{\de^{(i)}}{\al^{(i)}, \be^{(i)}}\Big]\ 
x^{\al^{(1)}+\cdots+\al^{(k)}} y^{\be^{(1)}+\cdots+\be^{(k)}} \cr
}$$
by matching the coefficients of $x^\al y^\be$.

It follows from (\eqnOSnormsI) that
$$
\sfrac{1}{\de!}\,Y_\de\, t^\de
\le \sum_{\al'+\be'=\de}
\sfrac{1}{\al'!}\, X_{\al'}\,t^{\al'}
\ \ \sfrac{1}{\be'!}\, X'_{\be'}\,t^{\be'}
$$
and
$$
\|g\|_{1,\infty}
\le \sum_\de  \sum_{\al'+\be'=\de}
\sfrac{1}{\al'!}\, X_{\al'}\,t^{\al'}
\ \ \sfrac{1}{\be'!}\, X'_{\be'}\,t^{\be'}
\ =\ \|f\|_{1,\infty}\ \|f'\|_{1,\infty}
$$
\endproof 

\corollary{\STM\corOSelloneinfty}{
Let $f$ be a function on $\cB^n$, $f'$ a function on $\cB^{n'}$ and 
$C_2,C_3$ functions on $\cB^2$. Set
$$
h({\sst \xi_4,\cdots,\xi_n,\,\xi'_4,\cdots,\xi'_{n'}}) 
  =\int{\sst d\zeta\,d\xi_2 d\xi'_2\,d\xi_3 d\xi'_3}\, 
f({\sst \ze,\xi_2,\xi_3,\xi_4,\cdots \xi_n })\
C_2({\sst \xi_2,\xi'_2})\,C_3({\sst \xi_3,\xi'_3})\  
f'({\sst \ze,\xi'_2,\xi'_3,\xi'_4,\cdots \xi'_{n'} })
$$
Then
$$
\|h\|_{1,\infty} \le \sup_{\xi,\xi'} |C_2({\sst \xi,\xi'})|
\ \sup_{\eta,\eta'}|C_3({\sst\eta,\eta'})|\ 
\|f\|_{1,\infty}\ \|f'\|_{1,\infty}
$$
}

\prf
Set
$$
g({\sst \xi_2,\cdots,\xi_n,\,\xi'_2,\cdots,\xi'_{n'}}) 
  =\int{\sst d\zeta}\, 
f({\sst \ze,\xi_2,\xi_3,\xi_4,\cdots \xi_n })\,
f'({\sst \ze,\xi'_2,\xi'_3,\xi'_4,\cdots \xi'_{n'} })
$$ 
Let $\cD$ be a decay operator acting on $h$. Then
$$
\cD h =\int{\sst d\xi_2 d\xi'_2\,d\xi_3 d\xi'_3}\, 
C_2({\sst \xi_2,\xi'_2})\,C_3({\sst \xi_3,\xi'_3}) 
\cD g({\sst \xi_2,\cdots,\xi_n,\,\xi'_2,\cdots,\xi'_{n'}})
$$
Consequently
$$
\tn \cD h \tn_{1,\infty} \le \sup|C_2|\,\sup|C_3|\ \tn\cD g\tn_{1,\infty}
$$
and therefore
$$
\| h \|_{1,\infty} \le \sup|C_2|\,\sup|C_3|\ \| g\|_{1,\infty}
$$
The Corollary now follows from Lemma \lemOSelloneinfty.
\endproof

\definition{\STM\defOSFmn}{ 
 Let $\cF_m(n)$ be the space of all functions 
$f({\sst \eta_1,\cdots,\eta_m;\,\xi_1,\cdots\xi_n})$ on $\cB^m \times \cB^n$
that are antisymmetric in the $\eta$ variables. If
$f({\sst\eta_1,\cdots,\eta_m;\,\xi_1,\cdots\xi_n})$ is any function on 
$\cB^m \times\cB^n$, its antisymmetrization in the external variables is
$$
{\rm Ant_{ext}}f\,({\sst\eta_1,\cdots,\eta_m;\,\xi_1,\cdots\xi_n})
=\sfrac{1}{m!} \smsum_{\pi \in S_m} \sgn(\pi)\,
f({\sst\eta_{\pi(1)},\cdots,\eta_{\pi(m)};\,\xi_1,\cdots\xi_n})
$$ 
For $m,n\ge 0$, the symmetric group $S_n$ acts on
$\cF_m(n)$ from the right by 
$$
f^\pi(\eta_1,\cdots,\eta_m;\,\xi_1,\cdots\xi_n) =
f(\eta_1,\cdots,\eta_m;\,\xi_{\pi(1)},\cdots\xi_{\pi(n)})
\qquad {\rm for\ } \pi \in S_n
$$
}

\definition{\STM\defOSSymmNorm}{  A seminorm
$\|\,\cdot\,\|$ on $\cF_m(n)$ is called symmetric, if for every 
$f\in\cF_m(n)$ and $\pi \in S_n$
$$
\|f^\pi\| = \|f\|
$$
and $\|f\|=0$ if $m=n=0$.
}

\noindent
For example, the seminorms $\|\ \cdot\ \|_{1,\infty}$ of Example
\exOSSymmNorm\ are symmetric.

\vfill\eject
%=====================================================================
%========================== RENORM GROUP MAP =========================
%=====================================================================

\chap{Covariances and the Renormalization Group Map}\PG\pgOSIII

\definition{\STM\defOScontnorm (Contraction)}{ 
Let $C(\xi,\xi')$ be any skew symmetric function on $\cB\times\cB$. Let 
$m,n\ge 0$ and $1\le i < j\le n$. For $f\in \cF_m(n)$ the contraction 
$\Cont{i}{j}{C} f \in \cF_m(n-2) $ is defined as
$$\eqalign{
\Cont{i}{j}{C} f\, &({\sst \eta_1,\cdots,\eta_m;
\xi_1,\cdots,\xi_{i-1},\,\xi_{i+1},\cdots,\xi_{j-1},\,\xi_{j+1},\cdots,\xi_n})
\cr
& \hskip 1.5cm= (-1)^{j-i+1} \int {\sst d\zeta\,d\zeta'}\  C({\sst\zeta,\zeta'})\,
f({\sst \eta_1,\cdots,\eta_m;
\xi_1,\cdots,\xi_{i-1},\,\zeta,\,\xi_{i+1},\cdots,\xi_{j-1},\,\zeta',\,
\xi_{j+1},\cdots,\xi_n}) 
}$$  
}

\definition{\STM\defOScontbnd (Contraction Bound)}{
Let $\|\cdot\|$ be a family of symmetric seminorms on the spaces $\cF_m(n)$.
We say that $\cb\in\fN_{d+1}$ is a contraction bound for the covariance $C$ with respect to 
this family of seminorms, if for all $m,n,m',n'\ge 0$ there exist $i$ and
$j$ with
$1\le i \le n,\ 1\le j \le n'$ such that
$$
\big\| \Cont{i}{n+j}{C} \big({\rm Ant_{ext}} (f\otimes f')\big) \big\|
\le \cb\,\|f\|\,\|f'\|
$$
for all $f \in \cF_m(n),\ f' \in \cF_{m'}(n')$.
Observe that $f\otimes f'$ is a function on
$(\cB^m\times\cB^n)\times(\cB^{m'}\times\cB^{n'})
\cong \cB^{m+m'}\times\cB^{n+n'}$, so that 
${\rm Ant_{ext}} (f\otimes f') \in \cF_{m+m'}(n+n')$.
 }
\remark{\STM\remOScontbnd}{ If $\cb$ is a contraction bound for the 
covariance $C$ with respect to a family of symmetric seminorms, then,
by symmetry,
$$
\big\| \Cont{i}{n+j}{C} \big({\rm Ant_{ext}} (f\otimes f')\big) \big\|
\le \cb\,\|f\|\,\|f'\|
$$
for {\bf all} $1\le i \le n,\ 1\le j \le n'$ and all
$f \in \cF_m(n),\ f' \in \cF_{m'}(n')$.
}

\example{\STM\exOSelloneinftycontr}{ 
The $L_1$--$L_\infty$--norm introduced in Example \exOSSymmNorm\ 
has $\ \max\{ \|C\|_{1,\infty},\tn C\tn_\infty \}$ as a contraction bound  for covariance $C$. Here, $\tn C\tn_\infty$ is the element of $\fN_{d+1}$ whose constant term is $\sup_{\xi,\xi'}|C(\xi,\xi')|$ and is the only nonzero term.
This is easily proven by iterated application of Lemma \lemOSelloneinfty.
See also Example \egIIcompatnorm\ in [FKTr1]. A more general statement will 
be formulated and proven in Lemma \:\lemOSscalednorm.iii. 
}

\definition{\STM\defOSintbnd (Integral Bound)}{  
Let $\|\cdot\|$ be a family of symmetric seminorms on the spaces $\cF_m(n)$.
We say that $\ib\in\bbbr_+$ is an integral bound for the covariance $C$ with respect 
to this family of seminorms, if  the following holds: 

{\parindent=.25in\item{}
Let $m\ge 0$, $1\le n' \le n$. For $f\in \cF_m(n)$ define $f'\in \cF_m(n-n')$ by
$$
f'({\sst \eta_1,\cdots,\eta_m;\,\xi_{n'+1},\cdots,\xi_n})\! = \!\!
\int\!\!\!\!\int_{\cB^{n'}} {\sst d\xi_1\cdots d\xi_{n'}}\ 
f({\sst\eta_1,\cdots,\eta_m;\hskip.5pt\xi_1,\cdots,\xi_{n'},\xi_{n'+1},\cdots,\xi_n})
\,\psi({\sst\xi_1})\cdots\psi({\sst\xi_{n'}})\,d\mu_C(\psi)
$$
Then
$$
\|f'\| \le (\ib/2)^{n'}\,\|f\|
$$

}
}

\remark{\STM\remOSelloneinftyintbnd}{
Suppose that there is a constant $S$ such that
$$
\Big| \int \psi(\xi_1)\cdots\psi(\xi_n)\,d\mu_C(\psi) \Big| \le S^n
$$
for all $\xi_1,\cdots,\xi_n \in \cB$. Then $2S$ is an integral bound for 
$C$ with respect to the $L_1$--$L_\infty$--norm introduced in Example 
\exOSSymmNorm.
}

\definition{\STM\defOShomogGrAlg}{
\Item{i)} 
We define $A_m[n]$ as the subspace of the Grassmann algebra $\bigwedge_AV$ 
that consists of all elements of the form
$$
Gr(f) = \int {\sst d\eta_1\cdots d\eta_m\,d\xi_1\cdots d\xi_n}\ 
f({\sst \eta_1,\cdots,\eta_m;\,\xi_1,\cdots,\xi_n})\
\phi(\eta_1)\cdots\phi(\eta_m)\,\psi(\xi_1)\cdots\psi(\xi_n)
$$
with a function $f$ on $\cB^m\times\cB^n$.
\Item{ii)} 
Every element of $A_m[n]$ has a unique representation of the form $Gr(f)$ with a function 
$f({\sst \eta_1,\cdots,\eta_m;\,\xi_1,\cdots,\xi_n}) \in \cF_m(n)$ that is 
antisymmetric in its $\xi$ variables. Therefore a seminorm $\|\cdot\|$
on $\cF_m(n)$ defines a canonical seminorm on $A_m[n]$, which we denote by the 
same symbol $\|\cdot\|$.
}

\remark{\STM\remOSminnorm}{
For $F\in A_m[n]$
$$
\|F\| \le \|f\|\qquad\qquad\hbox{for all } f\in \cF_m(n)\ {\rm with\ } Gr(f) =F
$$
}

\prf
Let $f\in \cF_m(n)$. Then $f'= \sfrac{1}{n!}\smsum_{\pi\in S_n} \sgn(\pi)\,f^\pi$
is the unique element of $\cF_m(n)$ that is antisymmetric in its $\xi$ variables
such that $Gr(f') = Gr(f)$. Therefore
$$
\|Gr(f)\|\ =\ \|f'\| \le \sfrac{1}{n!}\smsum_{\pi\in S_n} \|f^\pi\| 
\ =\ \sfrac{1}{n!}\smsum_{\pi\in S_n} \|f\|\ =\ \|f\|
$$
since the seminorm is symmetric.
\endproof

\definition{\STM\defOSgrnorm}{Let $\|\,\cdot\,\|$ be a family of symmetric seminorms, and let
$\cW(\phi,\psi)$ be a Grassmann function. Write
$$
\cW = \smsum_{m,n \ge0} \cW_{m,n}
$$
with $\cW_{m,n} \in A_m[n]$.
For any constants
$\cb\in\fN_{d+1},\ \ib >0$ and $ \al \ge 1$ set
$$\eqalign{
N(\cW;\cb,\ib,\al)
&=\sfrac{1}{\ib^2}\,\cb\!\sum_{m,n\ge 0}\,
\al^{n}\,\ib^{n} \,\|\cW_{m,n}\|\cr
}$$
}

In practice, the quantities $\ib,\cb$ will reflect the ``power counting'' of $\cW$
with respect to the covariance $C$ and the number $\al$ is proportional to an
inverse power of the largest allowed modulus of the coupling constant.

\vskip1cm

In this paper, we will derive bounds on the renormalization group map for several kinds of seminorms. The main ingredients from [FKTr1] are

\theorem{\STM\thmOSroptheorII}{
Let $\|\,\cdot\,\|$ be a family of symmetric 
seminorms and let $C$ be a covariance on $V$ with contraction bound 
$\cb$ and integral bound $\ib$. Then the formal Taylor series $\Om_C(\lw\cW\rw)$
converges to an analytic map on $\set{\cW}{\cW\hbox{ even},\  
N\big(\cW;\cb,\ib,  8\al\big)_\0  <\sfrac{\al^2}{4}}$.
Furthermore, if $\cW(\phi,\psi)$ is an even Grassmann function such that
$$
N\big(\cW;\cb,\ib,  8\al\big)_\0 <\sfrac{\al^2}{4} 
$$
then
$$
N\big(\Om_C(\lw \cW\rw)-\cW;\cb,\ib,  \al\big)\ \le\ 
\sfrac{2}{\al^2}\,
     \sfrac{N( \cW;\cb,\ib, 8\al)^2}{1-{4\over\al^2}N(\cW;\cb,\ib, 8\al)}
$$
Here, $\lw \cdot \rw$ denotes Wick ordering with respect to the covariance $C$.

}

In \S\CHinsulator\ we will use this Theorem to discuss the situation of an insulator.
More generally we have

\theorem{\STM\thmOSroptheoremIVb}{
Let, for $\ka$ in a neighbourhood of $0$, $\cW_\ka(\phi,\psi)$ be an even Grassmann 
function and $C_\ka,D_\ka$ be antisymmetric functions on $\cB\times\cB$. Assume that $\al\ge 1$ and 
$$
N(\cW_0;\cb,\ib, 32\al)_\0<\al^2
$$
and that
$$\meqalign{
&C_0\ {\rm has\ contraction\ bound} \ \cb \quad&&\quad
&\sfrac{1}{2}\ib \ {\rm is\ an\ integral\ bound\ for\ } C_0,D_0 \cr
&\sfrac{d\hfill}{d\ka}C_\ka\big|_{\ka=0}\ {\rm has\ contraction\ bound} \ \cb' \quad&&\quad
&\sfrac{1}{2}\ib' \ {\rm is\ an\ integral\ bound\ for\ } \sfrac{d\hfill}{d\ka}D_\ka\big|_{\ka=0} \cr
}$$
and that $\cb\le\sfrac{1}{\mu}\cb^2$.
Set
$$\eqalign{
\lw \tilde \cW_\ka(\phi,\psi)\rw_{\psi,D_\ka}\ 
&=\ \Om_{C_\ka}(\lw \cW_\ka \rw_{\psi,C_\ka+D_\ka})\cr
}$$
Then
$$\eqalign{
N\big(\,\sfrac{d\hfill}{d\ka}[\tilde \cW_\ka-\cW_\ka]_{\ka=0}\,;\cb,\ib, \al\big)
&\le\ \sfrac{1}{2\al^2}\,
     \sfrac{N( \cW_0;\cb,\ib, 32\al)}{1-{1\over\al^2}N( \cW_0;\cb,\ib, 32\al)}
N\big(\,\sfrac{d\hfill}{d\ka}\cW_\ka\big|_{\ka=0}\,;\cb,\ib, 8\al\big) \cr
&\hskip3cm+\sfrac{1}{2\al^2}\,
     \sfrac{N( \cW_0;\cb,\ib, 32\al)^2}{1-{1\over\al^2}N( \cW_0;\cb,\ib, 32\al)}
\Big\{\sfrac{1}{4\mu}\cb'+\big(\sfrac{\ib'}{\ib}\big)^2\Big\}
}$$
}

\proof{ of Theorems \thmOSroptheorII\ and \thmOSroptheoremIVb}

If $f({\sst \eta_1,\cdots,\eta_m;\,\xi_1,\cdots\xi_n})$ is a
function on $\cB^m \times \cB^n$ we define the
corresponding element of $A_m\otimes V^{\otimes n}$ as
$$
{\rm Tens}(f) = \int \smprod_{i=1}^m d\eta_i\, 
\smprod_{j=1}^n d\xi_j\ 
f({\sst\eta_1,\cdots,\eta_m;\, \xi_1,\cdots,\xi_n })\,
 \phi(\eta_1)\cdots\phi(\eta_m)\ 
\psi(\xi_1) \otimes \cdots \otimes \psi(\xi_n)
$$
Each element of $A_m\otimes V^{\otimes n}$ can be uniquely written in the form ${\rm Tens}(f)$ with a function $f\in \cF_m(n)$. Therefore a seminorm
on $ \cF_m(n)$ defines a seminorm on $A_m\otimes V^{\otimes n}$
and conversely. Under this correspondence, symmetric seminorms on $\cF_m(n)$ in the sense of Definition \defOSSymmNorm\ correspond to symmetric seminorms on $A_m\otimes V^{\otimes n}$ in the sense of Definition \defsymnorm\ of [FKTr1], contraction bounds as in Definition \defOScontbnd\ correspond,
by Remark \remOScontbnd,
 to contraction bounds as in Definition \defcontractintbound.i of [FKTr1]
and integral bounds as in Definition \defOSintbnd\ correspond to integral bounds as in
Definition \defcontractintbound.ii of [FKTr1]. Furthermore the norms on the 
spaces $A_m[n]$ defined in Definition \defOShomogGrAlg.ii agrees with those 
of Lemma \lemGrasscompatnorm\ of [FKTr1]. Therefore
Theorem \thmOSroptheorII\ follows directly from Theorem \theorII\ of [FKTr1] 
and Theorem \thmOSroptheoremIVb\ follows from Theorem \theoremIVb\ of [FKTr1].
\endproof

\vfill\eject
%=====================================================================
%========================== COVARIANCE BOUNDS ========================
%=====================================================================

\chap{Bounds for Covariances}\PG\pgOSIV

%%%%%%%%%%%%%%%%%%
\titlec{Integral Bounds}\PG\pgOSIVa
%%%%%%%%%%%%%%%%%%

\definition{\STM\defIntBndsS}{
For any covariance $C=C(\xi,\xi')$ we define
$$
S(C) = \sup_m\sup_{\xi_1,\cdots,\xi_m \in \cB}\
\Big(\ \Big| \int \psi(\xi_1)\cdots\psi(\xi_m)\,d\mu_C(\psi) \Big|\ \Big)^{1/m}
$$

}

\remark{\STM\remIntBndsI}{ 
\Item{i)}
By Remark \remOSelloneinftyintbnd, $2\,S(C)$ is an integral bound for $C$ 
with respect to the $L_1$--$L_\infty$-norms introduced in Example \exOSSymmNorm.
\Item{ii)}
For any two covariances $C,C'$
$$
S(C+C') \le S(C)+S(C')
$$
}
\proof{ of (ii)}
For $\xi_1,\cdots,\xi_m \in \cB$
$$
\int \psi(\xi_1)\cdots\psi(\xi_m)\,d\mu_{(C+C')}(\psi)
= \int \hskip -12pt \int \big(\psi(\xi_1)+\psi'(\xi_1)\big)\cdots
 \big( \psi(\xi_m)+\psi'(\xi_m)\big)\,d\mu_C(\psi)\,d\mu_{C'}(\psi')
$$
Multiplying out one sees that 
$$
\big(\psi(\xi_1)+\psi'(\xi_1)\big)\cdots\big( \psi(\xi_m)+\psi'(\xi_m)\big)
=\smsum_{p=0}^m \smsum_{I\subset \{1,\cdots,m\} \atop |I|=p} {\cal M}(p,I)
$$
with
$$
{\cal M}(p,I)= \pm \smprod_{i\in I}\psi(\xi_i)\ \smprod_{j\notin I}\psi'(\xi_j)
$$
Therefore
$$\eqalign{
\Big| \int \psi(\xi_1)\cdots\psi(\xi_m)\,d\mu_{(C+C')}(\psi) \Big|
&\le \smsum_{p=0}^m \smsum_{I\subset \{1,\cdots,m\} \atop |I|=p}
\Big|\int \hskip -6pt \int {\cal M}(p,I)\,d\mu_C(\psi)\,d\mu_{C'}(\psi') \Big| \cr 
& \le \smsum_{p=0}^m \smsum_{I\subset \{1,\cdots,m\} \atop |I|=p}
  S(C)^p\,S(C')^{m-p} \cr
& = \big(S(C)+S(C') \big)^m
}$$
\endproof

In this Section, we assume that there is a function $C(k)$ such that for
$\xi = (x,a) = (x_0,\x,\si,a),\ \xi' = (x',a') = (x'_0,\x',\si',a') \in \cB$
$$
C(\xi,\xi' )
=\cases{ \de_{\si,\si'} \int\sfrac{d^{d+1}k}{(2\pi)^{d+1}}\, e^{\imath<k,x-x'>_-}C(k)
& if $a=0, a'=1$ \cr
\noalign{\vskip.05in}
-\de_{\si,\si'} \int\sfrac{d^{d+1}k}{(2\pi)^{d+1}}\, e^{\imath<k,x'-x>_-}C(k)
& if $a=1, a'=0$ \cr
\noalign{\vskip.05in}
0 & if $a=a'$\cr
}\EQN\eqnOScovbndsI$$
(as usual, the case $x_0=x'_0=0$ is defined through the limit $x_0-x'_0\rightarrow 0-$)
and derive bounds for $S(C)$ in terms of norms of $C(k)$. 

\proposition{\STM\propIntBndsII (Gram's estimate)}{
\Item{i)}
$$
S(C) \le  \sqrt{\int \sfrac{d^{d+1}k}{(2\pi)^{d+1}} \ |C(k)|}
$$
\Item{ii)} Let, for each $s$ in a finite set $\Si$,
$\,\chi_s(k)$ be a function on $\bbbr\times\bbbr^d$. Set, for $a\in\{0,1\}$,
$$
\hat \chi_s(x-x',a) = \int e^{(-1)^a\imath<k,x-x'>_-}\,\chi_s(k)\,
\sfrac{d^{d+1}k}{(2\pi)^{d+1}}
$$ 
and
$$
\psi_s(x,a) = \int d^{d+1}x'\ \hat\chi_s(x-x',a)\,\psi(x',a)
$$
Then for all $\xi_1,\cdots,\xi_m \in \cB$ and all $s_1,\cdots,s_m\in \Si$
$$
\Big| \int \psi_{s_1}(\xi_1)\cdots\psi_{s_m}(\xi_m)\,d\mu_C(\psi) \Big|
\le \Big[  \max_{s\in\Si}
\int \sfrac{d^{d+1}k}{(2\pi)^{d+1}}\ \big|C(k)\chi_s(k)^2\big|\Big]^{m/2}
$$
}

\prf
Let $\cH$ be the Hilbert space $\cH=L^2(\bbbr\times\bbbr^d)\otimes \bbbc^2$.
For $\si\in \{\uparrow,\downarrow\}$ define the element $\om(\si)\in \bbbc^2$ by
$$
\om(\si) =\cases{ (1,0) & if $\si = \,\uparrow$ \cr
                  (0,1) & if $\si = \,\downarrow$ \cr
}$$
For each $\xi = (x,a)= (x_0,\x,\si,a) \in \cB$ define $w(\xi) \in \cH$ by
$$
w(\xi)=\cases{ \sfrac{e^{-\imath<k,x>_-}}{(2\pi)^{(d+1)/2}} \sqrt{|C(k)|}
     \otimes \om(\si)   & if $a=0$ \cr 
\noalign{\vskip.05in}
              \sfrac{e^{-\imath<k,x>_-}}{(2\pi)^{(d+1)/2}} 
     \sfrac{C(k)}{\sqrt{|C(k)|}}\otimes\om(\si)
                   & if $a=1$ \cr
}$$
Then 
$$
\|w(\xi)\|^2_\cH =  \int \sfrac{d^{d+1}k}{(2\pi)^{d+1}}\ |C(k)| 
\qquad {\rm for \ all\ } \xi \in \cB
$$
and
$$ 
C(\xi,\xi') = \<w(\xi),w(\xi')\>_\cH
$$
if $\xi=(x,\si,0), \xi'=(x',\si',1) \in \cB$.
Part (i) of the Proposition now follows from part (i) of 
Proposition \propGII\ in [FKTr1].
\Item{ii)}
For each $\xi = (x,a)= (x_0,\x,\si,a) \in \cB$ and $s\in \Si$ 
define $w'(\xi,s) \in \cH$ by
$$
w'(\xi,s)=\cases{\sfrac{e^{-\imath<k,x>_-}}{(2\pi)^{(d+1)/2}}\sqrt{|C(k)|}
   \ \,\overline{\chi_s(k)}\otimes\om(\si)   & if $a=0$ \cr 
\noalign{\vskip.05in}
               \sfrac{e^{-\imath<k,x>_-}}{(2\pi)^{(d+1)/2}} 
     \sfrac{C(k)}{\sqrt{|C(k)|}}\,\chi_s(k)\otimes\om(\si)
                   & if $a=1$ \cr
}$$
Then
$$\eqalign{
\|w'(\xi,s)\|^2_\cH &=  \int \sfrac{d^{d+1}k}{(2\pi)^{d+1}}\ |C(k)| |\ch_s(k)|^2\cr
}$$
and
$$
\int \psi_s(\xi) \psi_{s'}(\xi') d\mu_C(\xi) = \<w(\xi,s),w(\xi',s')\>_\cH
$$
if $\xi=(x_0,\x,\si,0), \xi'=(x'_0,\x',\si',1) \in \cB$. 
Part (ii) of the Proposition now follows from part (i) of 
Proposition \propGII\ in [FKTr1], 
applied to the generating system of fields $\psi_s(\xi)$.

\endproof

\lemma{\STM\lemIntBndsIII}{ Let $\La>0$ and $U(\k)$ a function on $\bbbr^d$.
Assume that
$$
C(k) =  \sfrac{U(\k)}{\imath k_0-\La}
$$
Then 
$$
S(C) \le  \sqrt{\int \sfrac{d^d\k}{(2\pi)^d}\ |U(\k)|}
$$
}

\prf 
For $a=0,\ a'=1$
$$\eqalign{
C\big((x_0,\x,\si,a),(x'_0,\x',\si',a') \big) &=\de_{\si,\si'} 
  \int \sfrac{dk_0}{2\pi} \sfrac{e^{-\imath k_0(x_0-x'_0) }}{\imath k_0-\La} 
 \ \int\sfrac{d^d \k}{(2\pi)^d} \, {e^{\imath<\k,\x-\x'>}}\,U(\k)\cr
&=-\de_{\si,\si'} \int\sfrac{d^d \k}{(2\pi)^d} \, {e^{\imath<\k,\x-\x'>}}\,U(\k)
\cases{ e^{-\La(x_0 - x_0')} & if $x_0>x_0'$ \cr
        0 & if $x_0 \le x_0'$ \cr
}\cr
}$$ 
Let $\cH$ be the Hilbert space $\cH=L^2(\bbbr^d)\otimes \bbbc^2$.
For $\si\in \{\uparrow,\downarrow\}$ define the element $\om(\si)\in \bbbc^2$ as
in the proof of Proposition \propIntBndsII, and
for each $\xi = (x_0,\x,\si,a) \in \cB$ define $w(\xi) \in \cH$ by
$$
w(\xi)=\cases{  \sfrac{e^{-\imath<\k,\x>}}{(2\pi)^{d/2}} \sqrt{|U(\k)|}
     \otimes\om(\si)   & if $a=0$ \cr 
\noalign{\vskip.05in}
               -\sfrac{e^{-\imath<\k,\x>}}{(2\pi)^{d/2}} 
     \sfrac{U(\k)}{\sqrt{|U(\k)|}}\otimes\om(\si)
                   & if $a=1$ \cr
}$$
Again 
$$
\|w(\xi)\|^2_\cH = \sfrac{1}{(2\pi)^d} \int d^d\k\ |U(\k)| 
\qquad {\rm for \ all\ } \xi \in \cB
$$
Furthermore set $\tau(x_0,\x,\si,a) = \La x_0$. Then for
$\xi=(x_0,\x,\si,0), \xi'=(x_0',\x',\si',1) \in \cB$
$$ 
C(\xi,\xi') = \cases{ e^{-(\tau(\xi)-\tau(\xi'))}\,\<w(\xi),w(\xi')\>_\cH             
                & if $\tau(\xi) > \tau(\xi')$ \cr
                       0 & if $\tau(\xi) \le \tau(\xi')$
}$$
The Lemma now follows from part(ii) of 
Proposition \propGII\ in [FKTr1].
\endproof

\proposition{\STM\propIntBndsIV}{
 Assume that $C$ is of the form
$$
C(k) = \sfrac{U(\k)-\chi(k)}{\imath k_0-e(\k)}
$$
with real valued measurable functions $U(\k),\,e(\k)$ on $\bbbr^d$ and $\chi(k)$
on $\bbbr\times \bbbr^d$ such that 
$0 \le \chi(k) \le U(\k) \le 1$ for all $k = (k_0,\k) \in \bbbr\times \bbbr^d$. 
Then
$$
S(C)^2\le 9\int  \sfrac{d^d\k}{(2\pi)^d}\ U(\k) 
+\sfrac{3}{E} \int \sfrac{d^{d+1}k}{(2\pi)^{d+1}}\ \chi(k)
+6\int_{|k_0|\le E}  \sfrac{d^{d+1}k}{(2\pi)^{d+1}} 
\ \sfrac{U(\k)-\chi(k)}{|\imath k_0-e(\k)|} 
$$
where $E=\sup\limits_{\k \in {\rm supp U}} |e(\k)|$.
}

\prf
Write
$$
C(k)= \sfrac{U(\k)}{\imath k_0-E} 
      - \sfrac{\chi(k)}{\imath k_0-E}
      + \sfrac{e(\k)-E}{(\imath k_0-e(\k))(\imath k_0-E)}(U(\k)-\chi(k))
$$
By Remark \remIntBndsI, Lemma \lemIntBndsIII\ and Proposition \propIntBndsII.i
$$\eqalign{
\sfrac{1}{3}S(C)^2
&\le \int  \sfrac{d^d\k}{(2\pi)^d}\ |U(\k)| 
     +  \int \sfrac{d^{d+1}k}{(2\pi)^{d+1}}\ 
          \big|\sfrac{\chi(k)}{\imath k_0-E}\big|
     + \int \sfrac{d^{d+1}k}{(2\pi)^{d+1}}\ 
     \big|\sfrac{e(\k)-E}{(\imath k_0-e(\k))(\imath k_0-E)}(U(\k)-\chi(k))\big|\cr
}$$
The first two terms are bounded by
$$
\int  \sfrac{d^d\k}{(2\pi)^d}\ U(\k) 
     +\sfrac{1}{E} \int \sfrac{d^{d+1}k}{(2\pi)^{d+1}}\ \chi(k)
$$
The contribution to the third term having $|k_0|\le E$ is bounded by
$$
\int_{|k_0|\le E} \sfrac{d^{d+1}k}{(2\pi)^{d+1}}\ 
     \big|\sfrac{e(\k)-E}{(\imath k_0-e(\k))(\imath k_0-E)}
\big(U(\k)-\chi(k)\big)\big|
\le
2\int \sfrac{d^{d+1}k}{(2\pi)^{d+1}}\ 
     \sfrac{U(\k)-\chi(k)}{|\imath k_0-e(\k)|}
$$
The contribution to the third term having $|k_0|> E$ is bounded by
$$\eqalign{
\int_{|k_0|> E} \sfrac{d^{d+1}k}{(2\pi)^{d+1}}\ 
     \big|\sfrac{e(\k)-E}{(\imath k_0-e(\k))(\imath k_0-E)}
\big(U(\k)-\chi(k)\big)\big|
&\le 4\int \sfrac{d^{d+1}k}{(2\pi)^{d+1}}\ 
     \sfrac{E}{|\imath k_0-E|^2}U(\k)\cr
&= 2\int \sfrac{d^{d}\k}{(2\pi)^{d}}\ U(\k)\cr
}$$
Hence
$$
\sfrac{1}{3}S(C)^2\le 3\int  \sfrac{d^d\k}{(2\pi)^d}\ U(\k) 
+\sfrac{1}{E} \int \sfrac{d^{d+1}k}{(2\pi)^{d+1}}\ \chi(k)
+2\int \sfrac{d^{d+1}k}{(2\pi)^{d+1}}\ 
     \sfrac{U(\k)-\chi(k)}{|\imath k_0-e(\k)|}
$$
\endproof

%%%%%%%%%%%%%%%%%%
\titlec{Contraction Bounds}\PG\pgOSIVb
%%%%%%%%%%%%%%%%%%

We have observed in Example \exOSelloneinftycontr\ that
the $L_1$--$L_\infty$--norm introduced in Example \exOSSymmNorm\ 
has $\ \max\{ \|C\|_{1,\infty},\tn C\tn_\infty \}$ as a contraction bound  for covariance $C$.
For the propagators of the form (\eqnOScovbndsI), we estimate these  position space quantities by norms of derivatives of $C(k)$ in momentum space.

\definition{\STM\defOSderivmom}{
\Item i)
For a function $f(k)$ on $\bbbr\times \bbbr^d$ and a multiindex $\de$ we set
$$
\rD^\de f\,(k) = \sfrac{\partial^{\de_0}\hfill }{\partial k_0^{\de_0}}\,
\sfrac{\partial^{\de_1}\hfill }{\partial k_1^{\de_1}}\cdots 
\sfrac{\partial^{\de_d}\hfill }{\partial k_d^{\de_d}}\,f\,(k)
$$
and
$$\deqalign{
\| f(k)\cnorm_\infty &= \sum_{\de\in\bbbn_0\times\bbbn_0^d}
\sfrac{1}{\de!}\Big(\sup_k\big|\rD^\de f(k)\big|\Big)\, t^\de&\in\fN_{d+1} \cr
\| f(k)\cnorm_1 &=  \sum_{\de\in\bbbn_0\times\bbbn_0^d}
\sfrac{1}{\de!}\Big(\int \sfrac{d^{d+1}k}{(2\pi)^{d+1}}\,
\big|\rD^\de f(k)\big|\Big)\, t^\de&\in\fN_{d+1}  \cr
}$$
If $B$ is a measurable subset of $\bbbr\times \bbbr^d$,
$$\deqalign{
\| f(k)\cnorm_{\infty,B} &= \sum_{\de\in\bbbn_0\times\bbbn_0^d}
\sfrac{1}{\de!}\Big(\sup_{k\in B}\big|\rD^\de f(k)\big|\Big)\, t^\de&\in\fN_{d+1} \cr
\| f(k)\cnorm_{1,B} &=  \sum_{\de\in\bbbn_0\times\bbbn_0^d}
\sfrac{1}{\de!}\Big(\int_B \sfrac{d^{d+1}k}{(2\pi)^{d+1}}\,
\big|\rD^\de f(k)\big|\Big)\, t^\de&\in\fN_{d+1}  \cr
}$$
\Item ii)
For $\mu>0$ and $X\in\fN_{d+1}$
$$
T_\mu X = \sfrac{1}{\mu^{d+1}}X+\sfrac{\mu}{d+1}
\sum_{j=0}^d\big(
\sfrac{\partial\hfill }{\partial t_0}\cdots
\sfrac{\partial\hfill }{\partial t_d}\big)
\sfrac{\partial\hfill }{\partial t_j}\ X
$$
}
\remark{\STM\remOSloneprod}{
For functions $f(k)$ and $g(k)$ on $B\subset\bbbr\times \bbbr^d$
$$
\| f(k)\,g(k)\cnorm_{1,B}\le 
\| f(k)\cnorm_{1,B} \| g(k)\cnorm_{\infty,B}
$$
by Leibniz's rule for derivatives. The proof is similar to that of
Lemma \lemOSelloneinfty.
 
}

\proposition{\STM\propOSpropbnd}{
Let $d\ge 1$. 
Assume that there is a function $C(k)$ such that for
$\xi = (x,a) = (x_0,\x,\si,a),\ \xi' = (x',a') = (x'_0,\x',\si',a') \in \cB$
$$
C(\xi,\xi' )
=\cases{ \de_{\si,\si'} \int\sfrac{d^{d+1}k}{(2\pi)^{d+1}} \,e^{\imath<k,x-x'>_-}C(k)
& if $a=0, \,a'=1$ \cr
\noalign{\vskip.05in}
0 & if $a=a'$\cr
\noalign{\vskip.05in}
-C\big(\xi',\xi \big)  & if $a=1, \,a'=0$ \cr
}$$
Let $\de$ be a multiindex and $0<\mu \le 1$.
\Item{i)}
$$
\tn\cD^{\de}_{1,2} C\tn_\infty\ \le\  \int\sfrac{d^{d+1}k}{(2\pi)^{d+1}}\,
 |\rD^\de C(k)|
\ \le \ \sfrac{vol}{(2\pi)^{d+1}}\,
 \sup_{k \in \bbbr\times\bbbr^d} |\rD^\de C(k)|
$$
and
%$$\eqalign{
%\tn\cD^{\de}_{1,2} C\tn_{1,\infty}
%&\le \sfrac{\abcst}{\mu^{d+1}}\max_{|\de'|\le d+2}
%\int\sfrac{d^{d+1}k}{(2\pi)^{d+1}}\,\mu^{|\de'|} |\rD^{\de+\de'} C(k)|  \cr 
%&\le \abcst \,\sfrac{vol}{\mu^{d+1}}
%\max_{|\de'|\le d+2}\, \sup_k\, \mu^{|\de'|}|\rD^{\de+\de'} C(k)| \cr
%}$$
$$
\|C\|_{1,\infty}\le\abcst\,T_\mu \| C(k)\cnorm_1
\le\abcst\,\sfrac{vol}{(2\pi)^{d+1}}\,T_\mu \| C(k)\cnorm_\infty
$$
where $vol$ is the volume of the support of $C(k)$ in $\bbbr\times\bbbr^d$
and the constant $\abcst$ depends only on the dimension $d$.
\Item{ii)}
Assume that there is an $r$--times differentiable real valued 
function $e(\k)$ on $\bbbr^d$ such that $|e(\k)| \ge \mu$ for 
all $\,\k\in\bbbr^d$ and a real valued, compactly supported, smooth,
 non negative function $U(\k)$ on $\bbbr^d$ such that
$$
C(k) = \sfrac{U(\k)}{\imath k_0-e(\k)}
$$
Set 
$$
g_1 = \int_{{\rm supp}\,U} d^d\k\ \sfrac{1}{|e(\k)|} \qquad\qquad
g_2 = \int_{{\rm supp}\,U} d^d\k\ \sfrac{\mu}{|e(\k)|^2}
$$
Then there is a constant $\abcst$ such that, for all multiindices $\de$ whose
spatial part $|\bde|\le r-d-1$,
$$
\tn C\tn_\infty\ \le\ 
\abcst\qquad\qquad
\sfrac{1}{\de!}\tn\cD^{\de}_{1,2} C\tn_{1,\infty}\ \le\ 
\sfrac{\abcst}{\mu^{d+|\de|}}\ \cases{
 g_1     & if $|\de|=0$ \cr
\noalign{\vskip.1in}
2^{|\de|} \,g_2     & if $|\de|\ge 1$ \cr
}
$$
The constant $\abcst$ depends only on the dimension $d$, the degree of differentiability $r$,
the ultraviolet cutoff $U(\k)$ and the quantities  $\sup_{\k}|\rD^{\bga} e(\k)|$, 
$\bga \in \bbbn_0^d,\ |\bga| \le r$.
\Item{iii)}
Assume that $C$ is of the form
$$
C(k) = \sfrac{U(\k)-\chi(k)}{\imath k_0-e(\k)}
$$
with real valued functions $U(\k),\,e(\k)$ on $\bbbr^d$ and $\chi(k)$ on 
$\bbbr\times \bbbr^d$ that fulfill the following conditions:

{\parindent=.25in\item{}
The function $e(\k)$ is $r$ times differentiable.
$|\imath k_0-e(\k)| \ge \mu$ for all
$\,k = (k_0,\k)$ in the support of $U(\k) -\chi(k)$. The function $U(\k)$ is smooth and has compact support.  The function $\chi(k)$ is smooth and 
has compact support and 
$0 \le \chi(k) \le U(\k) \le 1$ for all $k = (k_0,\k) \in \bbbr\times \bbbr^d$. 

}
\vskip.1in
\noindent
There is a constant $\const$ such that
$$
\tn C\tn_\infty\ \le\ \const
\EQN\eqnOSsupbnd$$
The constant $\const$ depends on  $d$, $\mu$ and the supports of $U(\k)$ and 
$\chi$.

\noindent
Let $r_0\in\bbbn$.
There is a constant $\const$ such that, for all multiindices $\de$ whose
spatial part $|\bde|\le r-d-1$ and whose temporal part $|\de_0|\le r_0-2$,
$$
\tn\cD^{\de}_{1,2} C\tn_{1,\infty}\ \le\ \const  
\EQN\eqnOSoneinftybnd$$
The constant $\const$ depends on  $d$, $r$, $r_0$, $\mu$, 
 $U(\k)$ and the quantities  $\sup_{\k}|\rD^{\bga} e(\k)|$ with
$\bga \in \bbbn_0^d,\ |\bga| \le r$
and
$\sup_{k}|\rD^{\be} \chi(k)|$ with 
$\be\in\bbbn_0\times\bbbn_0^d,\ \be_0\le r_0,\ 
|\bbe|\le r$.

}
\goodbreak
\prf
\Item{i)}
As the Fourier transform of the operator 
$\rD^{\de'}$ is, up to a sign, multiplication by $[-i(x-x')]^{\de'}$, we have for 
$\xi=(x,\si,a)$ and $\xi'=(x',\si',a')$
$$
\big|(x-x')^{\de'}\big|\,
|\cD^\de_{1,2} C(\xi,\xi')| \le 
\int\sfrac{d^{d+1}k}{(2\pi)^{d+1}}\, |\rD^{\de+\de'} C(k)|
$$ 
In particular
$$
|\cD^\de_{1,2} C(\xi,\xi')| \le 
\int\sfrac{d^{d+1}k}{(2\pi)^{d+1}}\, |\rD^{\de} C(k)|
\EQN\eqnOSonsulI$$ 
and, for $j=0,1,\cdots,d$,
$$
\mu^{d+2}|x_j-x'_j|\smprod_{i=0}^d|x_i-x'_i|\,
|\cD^\de_{1,2} C(\xi,\xi')| \le \mu^{d+2}\, 
\int\sfrac{d^{d+1}k}{(2\pi)^{d+1}}\, |\rD^{\de+\ep+\ep_j} C(k)|
\EQNB\eqnOSonsulII_j)$$ 
where $\ep=(1,1,\cdots,1)$ and $\ep_j$ is the $j^{\rm th}$ unit vector.
Taking the geometric mean of (\eqnOSonsulII$_0$), $\cdots$, (\eqnOSonsulII$_d$) on the left hand side and the arithmetic mean on the right hand side gives 
$$
\mu^{d+2}\smprod_{i=0}^d|x_i-x'_i|^{1+{1\over d+1}}\,
|\cD^\de_{1,2} C(\xi,\xi')| \le 
\sfrac{\mu^{d+2}}{d+1}\,\sum_{j=0}^d 
\int\sfrac{d^{d+1}k}{(2\pi)^{d+1}}\, |\rD^{\de+\ep+\ep_j} C(k)|
\EQN\eqnOSonsulIII$$ 
Adding (\eqnOSonsulI) and (\eqnOSonsulIII) gives
$$\eqalign{
&\Big(1+\mu^{d+2}\smprod_{i=0}^d|x_i-x'_i|^{1+{1\over d+1}}\Big)\,
|\cD^\de_{1,2} C(\xi,\xi')|\cr
&\hskip1.5in\le \int\sfrac{d^{d+1}k}{(2\pi)^{d+1}}\, |\rD^{\de} C(k)|
+\sfrac{\mu^{d+2}}{d+1}\,\sum_{j=0}^d 
\int\sfrac{d^{d+1}k}{(2\pi)^{d+1}}\, |\rD^{\de+\ep+\ep_j} C(k)|
}\EQN\eqnOSonsulIV$$ 
Dividing across and using
$\ 
\int \sfrac{d^{d+1}x}{1+\mu^{d+2}\smprod_{i=0}^d|x_i|^{1+{1\over d+1}}}
\le\abcst\sfrac{1}{\mu^{d+1}}
\ $
we get 
$$
\TN\cD^\de_{1,2} C(\xi,\xi') \TN_{1,\infty}
\le\abcst\Big(\sfrac{1}{\mu^{d+1}}\int\sfrac{d^{d+1}k}{(2\pi)^{d+1}}\, |\rD^{\de} C(k)|
+\sfrac{\mu}{d+1}\,\sum_{j=0}^d 
\int\sfrac{d^{d+1}k}{(2\pi)^{d+1}}\, |\rD^{\de+\ep+\ep_j} C(k)|\Big)
$$
The contents of the bracket on the right hand side are, up to a factor of $\sfrac{1}{\de!}$, the coefficient of $t^\de$ in $T_\mu \| C(k)\cnorm_1$.
%%%%%%%
\Item ii)
Denote by
$$
C(t,\k)=\int\sfrac{dk_0}{2\pi}e^{-\imath k_0 t}\sfrac{U(\k)}{\imath k_0-e(\k)}
=U(\k) e^{-e(\k) t}\cases{-\chi\big(e(\k)>0\big)& if $t>0$\cr
                      \chi\big(e(\k)<0\big) & if $t\le 0$\cr}
$$
the partial Fourier transform of $C(k)$ in the $k_0$ direction.
(As usual, the case $t=0$ is defined through the limit $t\rightarrow 0-$.)
Then, for $|\bde|+|\bde'|\le r$,
$$\eqalign{
\big|(\x-\x')^{\bde'}\big|\,|\cD^\de_{1,2} C(\xi,\xi')| 
&\le  \int\sfrac{d^{d}\k}{(2\pi)^{d}}\, 
|\cD^{\de_0}_{1,2}\rD^{\bde+\bde'} C(t-t',\k)|\cr
&\le  \abcst\int_{{\rm supp\,}U}\kern-20pt d^{d}\k\ \ 
\big(|t-t'|^{\de_0+|\bde|+|\bde'|}+|t-t'|^{\de_0}\big)
e^{-|e(\k)(t-t')|}\cr
&\le  \abcst\int_{{\rm supp\,}U}\kern-20pt d^{d}\k\ \ 
\Big[\sfrac{{({3\over 2})}^{\de_0+|\bde|+|\bde'|}(\de_0+|\bde|+|\bde'|)!}
{{|e(\k)|}^{\de_0+|\bde|+|\bde'|}}
+\sfrac{{({3\over 2})}^{\de_0}\de_0!}{{|e(\k)|}^{\de_0}}
\Big]
e^{-|e(\k)(t-t')/3|}\cr
&\le  \abcst\ 2^{\de_0}\de_0!\int_{{\rm supp\,}U}\kern-20pt d^{d}\k\ \ 
\sfrac{1}{{|e(\k)|}^{|\de|+|\bde'|}}
e^{-|e(\k)(t-t')/3|}\cr
}$$ 
In particular, $\tn C\tn_\infty\le\abcst$ and
$$\eqalign{
 \big|(\x-\x')^{\bde'}\big|\,\int\!\! dt' \ |\cD^\de_{1,2} C(\xi,\xi')| 
&\le\abcst\ 2^{\de_0}\de_0!\int_{{\rm supp\,}U}\kern-20pt d^{d}\k\ \ 
\sfrac{1}{{|e(\k)|}^{|\de|+|\bde'|+1}}  \cr
&\le\abcst\ 2^{\de_0}\de_0!\cases{g_1& if $|\de|+|\bde'|=0$\cr
            \noalign{\vskip.05in}
                  {g_2\over \mu^{|\de|+|\bde'|}}& if $|\de|+|\bde'|>0$\cr}  \cr
&\le\abcst\ \sfrac{2^{\de_0}\de_0!}{\mu^{|\bde'|}}\cases{g_1& if $|\de|=0$\cr
            \noalign{\vskip.05in}
                  {g_2\over \mu^{|\de|}}& if $|\de|>0$\cr}  \cr
}$$
since $g_1\ge g_2$.
As in equations (\eqnOSonsulI) -- (\eqnOSonsulIV), choosing various $\bde'$'s
with $|\bde'|=d+1$,
$$\eqalign{
 \int\!\! dt' \ |\cD^\de_{1,2} C(\xi,\xi')| 
&\le\abcst\ 2^{\de_0}\de_0!
\sfrac{1}{1+\mu^{d+1}\smprod_{i=1}^d|x_i-x_i'|^{1+{1\over d}}}
\cases{g_1& if $|\de|=0$\cr
            \noalign{\vskip.05in}
                  {g_2\over \mu^{|\de|}}& if $|\de|>0$\cr}  \cr
}$$
Integrating $\x'$ gives the desired bound on $\tn\cD^{\de}_{1,2} C\tn_{1,\infty}$.
%%%%%%
\Item{ iii)}
 Write 
$$
C(k) = C_1(k) - C_2(k) +C_3(k)
$$
with
$$\eqalign{
C_1(k) &= \sfrac{U(\k)}{\imath k_0-E} \cr
C_2(k) &= \sfrac{\chi(k)}{\imath k_0-E} \cr
C_3(k) &= \sfrac{e(\k)-E}{(\imath k_0-e(\k))(\imath k_0-E)}(U(\k)-\chi(k)) \cr
}$$
and define the covariances $C_j$ by
$$
C_j(\xi,\xi' )
=\cases{ \de_{\si,\si'} \int\sfrac{d^{d+1}k}{(2\pi)^{d+1}} \,e^{\imath<k,x-x'>_-}C_j(k)
& if $a=0, \,a'=1$ \cr
\noalign{\vskip.05in}
0 & if $a=a'$\cr
\noalign{\vskip.05in}
-C_j\big(\xi',\xi \big)  & if $a=1, \,a'=0$ \cr
}$$
for $j=1,2,3$. 
For $a=0,\ a'=1$
$$
C_1\big((x_0,\x,\si,a),(x'_0,\x',\si',a') \big) 
=-\de_{\si,\si'} \int\sfrac{d^d \k}{(2\pi)^d} \, {e^{\imath<\k,\x-\x'>}}\,U(\k)
\cases{ e^{-E( x_0 -  x_0')} & if $x_0>x_0'$ \cr
        0 & if $x_0 \le x_0'$ \cr
}$$
and, for $|\bde|\le r,\ |\de_0|\le r_0$,
$$
\tn\cD^{\de}_{1,2} C_1\tn_\infty\ 
\le\ \sfrac{\abcst}{E^{\de_0}}\de_0!\ 
\le\ \abcst \qquad\qquad
\tn\cD^{\de}_{1,2} C_1\tn_{1,\infty}\ 
\le\ \sfrac{\abcst}{E^{\de_0+1}}\de_0!\ 
\le\ \abcst    
$$
By Remark \remOSloneprod
$$
\| C_2(k)\cnorm_1
\le \| \chi(k)\cnorm_1 \| \sfrac{1}{\imath k_0-E}\cnorm_\infty
\le \| \chi(k)\cnorm_1 \ \Big(\sum_{n=0}^\infty\sfrac{1}{E^{n+1}}t_0^n\Big)
$$
so that, for $|\de_0|\le r_0-2$ and $|\bde|\le r-d-1$,
$$
\tn\cD^{\de}_{1,2} C_2\tn_\infty\ 
\le\ \abcst\  \qquad\qquad
\tn\cD^{\de}_{1,2} C_2\tn_{1,\infty}\ \le\ \abcst    
$$
by part (i).

%%%%%%%%%%% C_3 
We now bound $C_3$.  Let $B$ be the support of $U(\k)-\chi(k)$. 
On $B$, $|\imath k_0-e(\k)|\ge\mu>0$ and $|e(\k)|\le E$, so we have, 
for $\de=(\de_0,\bde)\ne 0$ with $|\bde|\le r$ and $\de_0\le r_0$,
$$
\Big|\rD^{\de}\sfrac{e(\k)-E}{(\imath k_0-e(\k))(\imath k_0-E)}
\Big|
\le \abcst\sfrac{E}{|\imath k_0-E|}
\Big(\sfrac{1}{|\imath k_0-e(\k)|^{|\de|+1}}
+\sfrac{1}{|\imath k_0-e(\k)|}\Big)
\le \abcst\sfrac{1}{\mu^{|\de|}}\sfrac{E}{|\imath k_0-E||\imath k_0-\mu|}
$$
Integrating
$$\eqalign{
\sfrac{1}{\de!}\int_B \sfrac{d^{d+1}k}{(2\pi)^{d+1}}\ 
\Big|\rD^\de \sfrac{e(\k)-E}{(\imath k_0-e(\k))(\imath k_0-E)}\Big|
&\le\const\cr
}$$
It follows  that
$$
\Big\|\sfrac{e(\k)-E}{(\imath k_0-e(\k))(\imath k_0-E)}\CNorm_{1,B}
\le\const \sum_{|\bde|\le r\atop |\de_0|\le r_0}t^\de
+\sum_{|\bde|> r\atop {\rm or\ }|\de_0|> r_0}\infty\, t^\de
$$
and, by Remark \remOSloneprod, that
$$
\| C_3(k)\cnorm_1
\le \const\Big(
\sum_{|\bde|\le r\atop |\de_0|\le r_0}t^\de
+\sum_{|\bde|> r\atop {\rm or\ }|\de_0|> r_0}\infty\, t^\de\Big)
\Big(\| U(\k)\cnorm_\infty+\| \chi(k)\cnorm_\infty\Big)
$$
By part (i) of this Proposition and the previous bounds on $C_1$ and $C_2$, this concludes the proof of part (iii).
\endproof

\corollary{\STM\corOSpropbnd}{
Under the hypotheses of Proposition \propOSpropbnd.ii, the $(d+1)$--dimensional 
norm
$$\eqalign{
\| C\|_{1,\infty}
&\le \sfrac{\abcst}{\mu^d}\Big(g_1
+g_2\sum_{|\de|\ge 1\atop|\bde|\le r-d-1}\big(\sfrac{2}{\mu}\big)^{|\de|}t^\de 
+\sum_{|\bde|\ge r-d}\infty\, t^\de
\Big)\cr
&\le \sfrac{\abcst\,g_1}{\mu^d}\Big(
\sum_{|\bde|\le r-d-1}\big(\sfrac{2}{\mu}\big)^{|\de|}t^\de 
+\sum_{|\bde|\ge r-d}\infty\, t^\de
\Big)\cr
}$$
Under the hypotheses of Proposition \propOSpropbnd.iii
$$
\| C\|_{1,\infty}\le
\const\Big(
\sum_{|\bde|\le r-d-1\atop |\de_0|\le r_0-2}t^\de
+\sum_{|\bde|> r-d-1\atop {\rm or\ }|\de_0|> r_0-2}\infty\, t^\de\Big)
$$
}

In the renormalization group analysis we shall add a counterterm $\de e(\k)$
to the dispersion relation $e(\k)$. For such a counterterm, we define the Fourier transform\footnote{$^{(1)}$}{A comprehensive set of Fourier transform 
conventions are formulated in \S\CHfourier.}
$$
\de \hat e(\xi,\xi') 
= \de_{\si,\si'}\, \de_{a,a'}\,\de(x_0-x'_0)
\int e^{(-1)^a\imath\,\k\cdot(\x-\x')}\,\de e(\k)\,
\sfrac{d^d\k}{(2\pi)^d} 
$$
for $\xi=(x,a)=(x_0,\x,\si,a),\,\xi'=(x',a') =(x'_0,\x',\si',a')\in \cB$.
%$$
%\de \hat e\big((x_0,\x,\si,a),(x'_0,\x',\si',a') \big)
%=\cases{ \de_{\si,\si'}\de(x_0-x'_0) \int\sfrac{d^d\k}{(2\pi)^d}\, e^{-\imath\k\cdot(\x-\x')}\de e(\k)
%& if $a=0, \,a'=1$ \cr
%\noalign{\vskip.05in}
%0 & if $a=a'$\cr
%\noalign{\vskip.05in}
%-\de \hat e\big((x'_0,\x',\si',a'),(x_0,\x,\si,a) \big)  & if $a=1,\,a'=0$ \cr
%}$$

\definition{\STM\defOScbzero}{
Fix $r_0$ and $r$. Let
$$
\cb_0=\sum_{|\bde|\le r\atop |\de_0|\le r_0}t^\de
+\sum_{|\bde|> r\atop {\rm or\ }|\de_0|> r_0}\infty\, t^\de
\in\fN_{d+1}
$$ 
The map
$
\fe_0(X) = \sfrac{\cb_0}{1- X}
$
from $X \in \fN_{d+1}$ with $X_\0<1$ to $\fN_{d+1}$
is used to implement the differentiability properties of various kernels 
depending on a counterterm whose norm is bounded by $X$.

}

\proposition{\STM\propOSrealfirstpropbound}{
Let
$$
C(k) = \sfrac{U(\k)-\chi(k)}{\imath k_0-e(\k)+\de e(\k)}\qquad\qquad
C_0(k) = \sfrac{U(\k)-\chi(k)}{\imath k_0-e(\k)}
$$
with real valued functions $U(\k),\,e(\k),\,\de e(\k)$ on $\bbbr^d$ 
and $\chi(k)$ on $\bbbr\times \bbbr^d$ that fulfill the following conditions:

{\parindent=.25in\item{}
The function $e(\k)$ is $r+d+1$ times differentiable.
$|\imath k_0-e(\k)| \ge \mu_e>0$ for all
$\,k = (k_0,\k)$ in the support of $U(\k) -\chi(k)$. The function $U(\k)$ is smooth and has compact support.  The function $\chi(k)$ is smooth and 
has compact support and 
$0 \le \chi(k) \le U(\k) \le 1$ for all $k = (k_0,\k) \in \bbbr\times \bbbr^d$.
The function  $\de e(\k)$ obeys
$$ 
\| \de \hat e\|_{1,\infty} < \mu
+\smsum_{\de\ne \0}\infty\, t^\de
$$

}

\noindent
Then, there is a constant $\mu_1>0$ such that if $\mu<\mu_1$, the following
hold
\Item i) $C$ is an analytic function of $\de e$ and 
$$\meqalign{
\hskip0.5in
\tn C\tn_\infty\ &\le\ \const &&
\tn C-C_0\tn_{\infty} &\le \const \tn\de \hat e\tn_{1,\infty}\cr
\noalign{\noindent and}
\|C\|_{1,\infty} &\le \const \fe_0({\sst \|\de \hat e\|_{1,\infty}})\qquad&&
\|C-C_0\|_{1,\infty} &\le \const \fe_0({\sst \|\de \hat e\|_{1,\infty}})
                 \|\de \hat e\|_{1,\infty}
}$$
\Item ii)
Let
$$
C_s(k) = \sfrac{U(\k)-\chi(k)}{\imath k_0-e(\k)+\de e(\k)+s\de e'(\k)}
$$
Then
$$
\TN \sfrac{d\hfill}{ds}C_s\big|_{s=0}\TN_\infty\ 
          \le\ \const\tn \de\hat e'\tn_{1,\infty}\qquad\qquad
\big\| \sfrac{d\hfill}{ds}C_s\big|_{s=0}\big\|_{1,\infty}
\le \const\fe_0({\sst \|\de \hat e\|_{1,\infty}})\,
              \| \de\hat e'\|_{1,\infty}
$$
}
\prf i)
The first bound follows from (\eqnOSsupbnd), by replacing $e$ by $e-\de e$.

Select a smooth, compactly support function $\tilde U(\k)$ and a smooth
compactly supported  function $\tilde\chi(k)$ such that
$0 \le \tilde\chi(k) \le \tilde U(\k) \le 1$ for all 
$k = (k_0,\k) \in \bbbr\times \bbbr^d$, $\tilde U(\k) -\tilde \chi(k)$
is identically 1 on the support of $U(\k) -\chi(k)$
 and $|\imath k_0-e(\k)| \ge \half\mu_e$
 for all $\,k = (k_0,\k)$ in the support of $\tilde U(\k) -\tilde \chi(k)$.
Let
$$
\tilde C_0(k) = \sfrac{\tilde U(\k)-\tilde\chi(k)}{\imath k_0-e(\k)}
$$
Then
$$\eqalign{
C(k) \ &=\  \frac{C_0(k)}{1 + \sfrac{\de e(\k)}{\imath k_0-e(\k)}}
\ =\ \frac{C_0(k)}{1 + \sfrac{\de e(\k)\, (\tilde U(\k)-\tilde \chi(k))}
{\imath k_0-e(\k)}}
\ =\ \frac{C_0(k)}{1 + \de e(\k)\,\tilde C_0(k)} \cr
&=\  C_0(k)\,\smsum_{n=0}^\infty \big(- \de e(\k)\,\tilde C_0(k) \big)^n 
}$$
Then, by  iterated application of Lemma \lemOSelloneinfty\ 
and the second part of Corollary \corOSpropbnd, with $r$ replaced by $r+d+1$ and $r_0$ replaced by $r_0+2$,
$$\eqalign{
\|C \|_{1,\infty} 
\ &\le\  \|C_0 \|_{1,\infty} \
\smsum_{n=0}^\infty \big( \|\de \hat e \|_{1,\infty} \,
   \, \| \tilde C_0 \|_{1,\infty} \big)^n \cr
& \le \ \const \cb_0 \ \smsum_{n=0}^\infty 
   \big( \cst{'}{} \cb_0\,\|\de \hat e \|_{1,\infty} \big)^n \cr
& =\  \const \sfrac{\cb_0}
{1-\cst{'}{}\cb_0\|  \de \hat e \|_{1,\infty}}
}$$
If $\mu_1<\min\{\sfrac{1}{2\cst{'}{}},1\}$, then,
by Corollary \corOSappMonoidIV.i, with 
$\De=\set{\de\in\fN_{d+1}}{|\bde|\le r,\ |\de_0|\le r_0}$, $\mu=\cst{'}{}$,
$\La=1$ and $X= \|\de \hat e \|_{1,\infty}\,$, 
$$
\|C \|_{1,\infty}  \le\  \const \sfrac{\cb_0}{1-\|\de \hat e \|_{1,\infty}}
$$
Similarly
$$\eqalign{
\|C-C_0 \|_{1,\infty} 
\ &\le\  \|C_0 \|_{1,\infty} \
\smsum_{n=1}^\infty \big( \|\de \hat e \|_{1,\infty} \,
   \, \| \tilde C_0 \|_{1,\infty} \big)^n \cr
& \le \ \const \cb_0 \ \smsum_{n=1}^\infty 
   \big( \cst{'}{} \cb_0\,\|\de \hat e \|_{1,\infty} \big)^n \cr
& \le\  \const \sfrac{\cb_0^2\,\|\de \hat e \|_{1,\infty}}
{1-\cst{'}{}\cb_0\|  \de \hat e \|_{1,\infty}}\cr
&\le\  \const \sfrac{\cb_0\|  \de \hat e \|_{1,\infty}}
{1-\|\de \hat e \|_{1,\infty}}
}$$
and
$$\eqalign{
\tn C-C_0 \tn_{\infty}
\ &\le\  \tn C_0 \tn_{\infty} \
\smsum_{n=1}^\infty \big( \tn\de \hat e \tn_{1,\infty} \,
   \, \tn \tilde C_0\tn_{1,\infty} \big)^n \cr
& \le \ \const  \ \smsum_{n=1}^\infty 
   \big( \cst{'}{}\tn\de \hat e \tn_{1,\infty} \big)^n \cr
& \le\  \const \sfrac{\tn\de \hat e \tn_{1,\infty}}
{1-\cst{'}{}\mu}\cr
&\le\  \const\tn \de \hat e \tn_{1,\infty}
}$$

\Item ii) As
$$
\sfrac{d\hfill}{ds}C_s(k)\big|_{s=0}
=-\sfrac{U(\k)-\chi(k)}{[\imath k_0-e(\k)+\de e(\k)]^2}\de e'(\k)
$$
the first bound is a consequence of Proposition \propOSpropbnd.i.

Let $\tilde U(\k)$ and  $\tilde\chi(k)$ be as in part (i) and set
$$
\tilde C(k) = \sfrac{\tilde U(\k)-\tilde\chi(k)}{\imath k_0-e(\k)+\de e(\k)}
$$
Then 
$$
\sfrac{d\hfill}{ds}C_s(k)\big|_{s=0}
=-C(k)\tilde C(k) \de e'(\k)
$$
and
$$\eqalign{
\big\|\sfrac{d\hfill}{ds}C_s(k)\big|_{s=0}\big\|_{1,\infty}
&\le \|C\|_{1,\infty}\|\tilde C\|_{1,\infty}\|\de \hat e'\|_{1,\infty}
\le \const\fe_0({\sst \|\de \hat e\|_{1,\infty}})^2\|\de \hat e'\|_{1,\infty}\cr
&\le \const\fe_0({\sst \|\de \hat e\|_{1,\infty}})\|\de \hat e'\|_{1,\infty}\cr
}$$
by Corollary \corOSappMonoidIV.ii.

\endproof

\vfill\eject
%=====================================================================
%========================== INSULATORS ===============================
%=====================================================================

\chap{Insulators}\PG\pgOSV

An insulator is a many fermion system as described in the introduction, for
which the dispersion relation $e(\k)$ does not have a zero on the support of the
ultraviolet cutoff $U(\k)$. We may assume that there is a constant $\mu>0$ such 
that $e(\k) \ge \mu$ for all $\k\in \bbbr^d$. We shall show in Theorem
\:\thmOSinsulators\  that for sufficiently small coupling constant the Green's
functions for the interacting system exist and differ by very little 
from the Green's functions of the noninteracting system in the supremum norm.

\lemma{\STM\lemOSscalednorm}{
Let  $ \rho_{m;n}$ be a  sequence of nonnegative real numbers such that 
$\rho_{m;n'} \le \rho_{m;n}$ for $n'\le n$. Define
for $f\in\cF_m(n)$
$$
\|f\| =\rho_{m;n} \,\|f\|_{1,\infty}
$$
where $\|f\|_{1,\infty}$ is the $L_1$--$L_\infty$--norm 
introduced in Example \exOSSymmNorm. 
\Item{i)} 
The seminorms $\|\cdot\|$ are symmetric.
\Item{ii)}
For a covariance $C$, let $S(C)$ be the quantity introduced in Definition
\defIntBndsS. Then
$2S(C)$ is an integral bound for the covariance $C$ with 
respect to the family of seminorms $\|\cdot\|$.
\Item{iii)} 
Let $C$ be a covariance. Assume that for all $m\ge 0$ and $n,n'\ge1$
$$
\rho_{m;\,n+n'-2} \le \rho_{m;n}\,\rho_{0;n'} 
$$
Let $\cb\in\fN_{d+1}$ obey 
$$\eqalign{
\cb &\ge \| C\|_{1,\infty}
\cr 
\cb_\0 & \ge 
\sfrac{\rho_{m+m';\,n+n'-2}}{\rho_{m;n}\,\rho_{m';n'}}\,\tn C\tn_\infty  
\qquad {\rm for\ all}\ m,m',n,n'\ge 1
}$$
where $\cb_\0$ is the constant coefficient of the formal power series $\cb$.
Then $\cb$ is a contraction bound for the covariance $C$ with 
respect to the family of seminorms $\|\cdot\|$.

}

\prf
Parts (i) and (ii) are trivial. To prove part (iii), let 
$f\in \cF_m(n),\ f'\in \cF_{m'}(n')$ and $1\le i \le n, \ 1\le j\le n'$.
Set
$$\eqalign{
&g({\sst \eta_1,\cdots,\eta_{m+m'};
\xi_1,\cdots,\xi_{i-1},\,\xi_{i+1},\cdots,\xi_n,\,
\xi_{n+1},\cdots,\xi_{n+j-1},\,\xi_{n+j+1},\cdots,\xi_{n+n'}}) \cr
&\hskip 3cm  =\int {\sst d\zeta\,d\zeta'}\, f( {\sst \eta_1,\cdots,\eta_m;\, 
\xi_1,\cdots,\xi_{i-1},\,\zeta,\,\xi_{i+1},\cdots,\xi_n })\, C({\sst \zeta,\zeta'})
\cr
& \hskip 6cm
f'( {\sst \eta_{m+1},\cdots,\eta_{m+m'};
\,\xi_{n+1},\cdots,\xi_{n+j-1},\,\zeta',\,\xi_{n+j+1},\cdots,\xi_{n+n'}})\cr
}$$
Then
$$
\Cont{i}{n+j}{C}{\rm Ant_{ext}}(f\otimes f') = {\rm Ant_{ext}}\, g
$$
and therefore
$$
\big\| \Cont{i}{n+j}{C}{\rm Ant_{ext}}(f\otimes f') \big\| 
\le \|g\|
$$
If $m,m' \ge 1$
$$
\|g\|_{1,\infty}\ 
\le\ \|f\|_{1,\infty}\,\tn C\tn_\infty\, \|f'\|_{1,\infty}
$$
and consequently
$$\eqalign{
\big\| \Cont{i}{n+j}{C}{\rm Ant_{ext}}(f\otimes f') \big\|
&\le \rho_{m+m';n+n'-2}\,\tn C\tn_\infty\,\|f\|_{1,\infty}\, \|f'\|_{1,\infty} \cr
&\le \cb_\0\, \rho_{m;n}\, \|f\|_{1,\infty}\ \rho_{m';n'}\|f'\|_{1,\infty} \cr 
&\le\, \cb\, \|f\|\, \|f'\|\cr 
}$$
If $m=0$ or $m'=0$, by iterated application of Lemma \lemOSelloneinfty
$$\eqalign{
\|g\|_{1,\infty} 
&\le\ \Big \|  \int_{\cB}  {\sst d\zeta}\   
f(
{\sst\xi_1,\cdots,\xi_m;\,\xi_1,\cdots,\xi_{i-1},\zeta,\xi_{i+1},\cdots,\xi_n}
) 
\,C({\sst\zeta,\zeta'}) \Big\|_{1,\infty}\ \|f'\|_{1,\infty} \cr
& \le  \|f\|_{1,\infty} \,\|C\|_{1,\infty}\,\|f'\|_{1,\infty}  
}$$
and again
$$
\big\| \Cont{i}{n+j}{C}{\rm Ant_{ext}}(f\otimes f') \big\|
\le\, \cb\, \|f\|\, \|f'\|
$$

\endproof

To formulate the result about insulators, we define for a function 
$f(x_1,\cdots,x_n)$, on
$\big( \bbbr \times \bbbr^d \times \{\uparrow,\downarrow\}\big)^n$, the
$L_1$--$L_\infty$--norm as in Example \exOSSymmNorm\ to be
$$
\tn f\tn_{1,\infty} = \max_{1\le j \le n} \sup_{x_j}
 \int\smprod_{i=1 \atop i\ne j}^n dx_i \,|f(x_1,\cdots,x_n)|
$$

%%%%%%% r_0 dependent version
\theorem{\STM\thmOSinsulators(Insulators)}{
Let $r$ and $r_0$ be natural numbers.
Let $e(\k)$ be a dispersion relation on $\bbbr^d$ that is at least $r+d+1$
 times differentiable, and let $U(\k)$ be a compactly supported, 
smooth ultraviolet
cutoff on $\bbbr^d$. Assume that there is a constant $0<\mu<\half$ such that
$$
e(\k) \ge \mu \qquad\qquad {\rm for\ all\ } \k \ {\rm in\ the\ support\ of\ }U
$$
Set 
$$
g = \int_{{\rm supp}\,U} d^d\k\ \sfrac{1}{|e(\k)|} \qquad\qquad
\ga = \max\Big\{ 1, \sqrt{{\tst \int} d^d\k\  U(\k) \log\sfrac{E}{|e(\k)|}}\  \Big\}
$$
where
$E=\max\big\{1,\sup\limits_{\k \in {\rm supp}\,U}|e(\k)|\big\}$.
Let, for $x=(x_0,\x,\si),\,x'=(x'_0,\x',\si') \in \bbbr \times \bbbr^d \times \{\uparrow,\downarrow\}$
$$
C(x,x') = \de_{\si,\si'} \int\sfrac{d^{d+1}k}{(2\pi)^{d+1}} \,e^{\imath<k,x-x'>_-}
 \sfrac{U(\k)}{\imath k_0 -e(\k)}
$$
and set, for $\xi=(x,a),\,\xi'=(x',a') \in \cB$
$$
C(\xi,\xi') = C(x,x')\,\de_{a,0}\de_{a',1} - C(x',x)\,\de_{a,1}\de_{a',0}
$$

\noindent
Furthermore let
$$
\cV(\psi,\bar\psi) =  \int_{(\bbbr\times\bbbr^2\times\{\uparrow,\downarrow\})^4}  
\hskip-.7in dx_1dy_1dx_2dy_2\ 
V_0(x_1,y_1,x_2,y_2)\,\bar\psi(x_1)\psi(y_1)\bar\psi(x_2) \psi(y_2)\
$$
be a two particle interaction with a kernel $V_0$ that is antisymmetric in the variables 
$x_1,x_2$ and $y_1,y_2$ separately. Set
$$
\upsilon = \sup_{\cD\ {\rm decay\ operator} 
                   \atop {\rm with\ } \de_0\le r_0,\ |\bde|\le r}
    \mu^{|\de(\cD)|}\,\tn\cD\ V_0\tn_{1,\infty}
$$ 
Then there exists $\varepsilon >0$ and a constant $\abcst$ such that 
\Item {i)}
If 
$\ 
\tn V_0\tn_{1,\infty} \le \sfrac{ \veps\,\mu^{d}}{g\ga^2 }
\ $, the connected amputated Green's functions 
$
G^{\rm amp}_{2n}(x_1,y_1,\cdots,x_n,y_n) 
$
exist in the space of all functions on 
$\big( \bbbr \times \bbbr^d \times \{\uparrow,\downarrow\}\big)^{2n}$ 
with finite $\tn\ \cdot\ \tn_{1,\infty} $ norms. They are analytic functions 
of $V_0$.
\Item{ii)} 
Suppose that 
$
\upsilon
\le \sfrac{ \veps\,\mu^{d}}{g\ga^2 }
$.
For all decay operators $\cD$ with $\de_0(\cD)\le r_0$ and $|\bde(\cD)|\le r$
$$\eqalign{
\tn\cD\, G_{2n}^{\rm amp}\tn_{1,\infty} 
&\le \sfrac{\abcst^n\,g\ga^{6-2n}}{\mu^{d+|\de(\cD)|}}\ \upsilon^2
\qquad \qquad {\rm if}\ n\ge 3 \cr
\tn\cD( G_4^{\rm amp} -V_0)\tn_{1,\infty} 
&\le \sfrac{\abcst^2\, g\ga^2}{\mu^{d+|\de(\cD)|}}\ \upsilon^2
\cr
\tn\cD( G_2^{\rm amp}  -K)\tn_{1,\infty} 
&\le \sfrac{\abcst\ g\ga^4}{\mu^{d+|\de(\cD)|}}\ \upsilon^2
}$$
where
$$
K(x,y) = 4 \int dx'dy'\  V_0(x,y,x',y')\,C(x',y')
$$

\noindent
The constants $\varepsilon$ and $\abcst$ depend on $r$, $r_0$, $U$, and 
the suprema of the $\k$--derivatives of the dispersion relation $e(\k)$ up 
to order $r+d+1$, but not on $\mu$ or $V_0$.
 }

\prf
By (\eqnOSintroI), the generating functional for the connected 
amputated Green's functions is
$$
\cG_{\rm gen}^{\rm amp}(\phi) =  \Om_C(\cV)(0,\phi)
$$
To estimate it, we use the norms $\|\cdot\|$ of 
Lemma \lemOSscalednorm\ with $\rho_{m;n} =1$.
By part (ii) of  Lemma \lemOSscalednorm\ and Proposition \propIntBndsIV, 
there is a constant $\abcst_0$ such that 
$\ib=\abcst_0\,\ga$ is an integral bound for the covariance $C$ with respect 
to these norms. By part (iii) of Lemma \lemOSscalednorm,
Corollary \corOSpropbnd\ and part (ii) of Proposition \propOSpropbnd, there 
is a constant $\abcst_1$ such that
$$\eqalign{
\cb=  
\sfrac{\abcst_1\,g}{\mu^d}\Big(
\sum_{|\bde|\le r}\big(\sfrac{2}{\mu}\big)^{|\de|}t^\de 
+\sum_{|\bde|> r}\infty\, t^\de
\Big)
 \cr 
}$$
is a contraction bound for $C$ with respect to these norms. Here 
we used that $\sfrac{g}{\mu^d}$ is bounded below by a nonzero
($E$ and $U(\k)$--dependent) constant.  As in Definition \defOSgrnorm, we set for any Grassmann 
function $\cW(\phi,\psi)$ and any $\al >0$
$$\eqalign{
N(\cW;\cb,\ib,\al)
&=\sfrac{1}{\ib^2}\,\cb\!\sum_{m,n\ge 0}\,
\al^{n}\,\ib^{n} \,\|\cW_{m,n}\|\cr
}$$
In particular
$$
N(\cV ;\cb,\ib, \al) =\al^{4}\,\ib^{2}\cb \,\|V_0\|_{1,\infty}
$$
and
$$
N(\cV ;\cb,\ib, 8\al)_\0\ 
\le\ \abcst_3\,\sfrac{8^4\,\al^4\ga^2 g}{\mu^d}
\,\tn V_0\tn_{1,\infty}
\EQN\eqnOSinsI$$
Observe that
$$\eqalign{
\cb \,\|V_0\|_{1,\infty}
&\le \sfrac{\abcst_1\,g}{\mu^d}\Big(
\sum_{|\bde|\le r}\big(\sfrac{2}{\mu}\big)^{|\de|}t^\de 
+\sum_{|\bde|> r}\infty\, t^\de
\Big)
\Big(\sum_{|\bde|\le r\atop |\de_0|\le r_0} \sfrac{1}{\de!}\sfrac{\upsilon}{\mu^{|\de|}}t^\de 
+\sum_{|\bde|> r\atop {\rm or\ }|\de_0|> r_0}\infty\, t^\de\Big)\cr
&\le \abcst_2\,\sfrac{g \upsilon}{\mu^d}\,\Big(
\sum_{|\bde|\le r\atop |\de_0|\le r_0}\sfrac{1 }{\mu^{|\de|}}t^\de
+\sum_{|\bde|> r\atop {\rm or\ }|\de_0|> r_0}\infty\, t^\de
\Big)\cr
}$$
Write $\cV = \lw \cV'\rw_C\ $. By part (i) of Proposition \propBII\ of [FKTr1],
$$
\cV' = \cV + \int dx\,dy\,K(x,y)\,\bar\psi(x) \psi(y) +\const
$$
and by part (i) of Corollary \corwicknorm\ of [FKTr1]
$$\eqalign{
N(\cV' ;\cb,\ib, \al)\ &\le\ N(\cV ;\cb,\ib, 2\al)
\ =\ 16\,\al^{4}\,\ib^{2}\cb \,\|V_0\|_{1,\infty}\cr
\ &\le  \abcst_3\,\al^4\ga^2\sfrac{g \upsilon}{\mu^d}\,\Big(
\sum_{|\bde|\le r\atop |\de_0|\le r_0}\sfrac{1}{\mu^{|\de|}}t^\de 
+\sum_{|\bde|> r\atop {\rm or\ }\de_0|> r_0}\infty\, t^\de
\Big)\cr
}$$
We set $\al =2$ and $\veps = \sfrac{1}{2^{17}\,\abcst_3}$.  Then 
$$
N(\cV' ;\cb,\ib, 16)\ 
\le \sfrac{g\ga^2 \upsilon}{2\veps\mu^d}\,\Big(
\sum_{|\bde|\le r\atop |\de_0|\le r_0}\sfrac{1}{\mu^{|\de|}}t^\de 
+\sum_{|\bde|> r\atop {\rm or\ }\de_0|> r_0}\infty\, t^\de
\Big)
$$
and, by (\eqnOSinsI)
$$
N(\cV' ;\cb,\ib, 16)_\0\ 
\le\ \sfrac{g\ga^2}{2\veps\mu^d}\,\tn V_0\tn_{1,\infty}
$$
Therefore, whenever 
$\tn V_0\tn_{1,\infty} \le \sfrac{ \veps\,\mu^{d}}{g\ga^2 }$,
 $\cV'$ fulfills the hypotheses of Theorem \thmOSroptheorII\ 
and
$\ 
\cG_{\rm gen}^{\rm amp}(\psi) = \Om_C(\lw\cV'\rw)(0,\psi)
\ $
exists. Part (i) follows.

If, in addition, $\upsilon \le \sfrac{ \veps\,\mu^{d}}{g\ga^2 }$, then
$$\eqalign{
N(\cG_{\rm gen}^{\rm amp} -\cV';\cb,\ib, 2)
&\le \sfrac{1}{2}\,\sfrac{N( \cV';\cb,\ib, 16)^2}{1-N(\cV';\cb,\ib, 16)}
\le \half\big(\sfrac{g\ga^2 \upsilon}{2\veps\mu^d}\big)^2
f\big(\sfrac{t}{\mu}\big)\cr
}$$
where
$$
f(t)=\frac{\big(
\sum_{|\bde|\le r\atop |\de_0|\le r_0}t^\de 
+\sum_{|\bde|> r\atop {\rm or\ }\de_0|> r_0}\infty\, t^\de
\big)^2}{1-{1\over 2}\big(
\sum_{|\bde|\le r\atop |\de_0|\le r_0}t^\de 
+\sum_{|\bde|> r\atop {\rm or\ }\de_0|> r_0}\infty\, t^\de
\big)}
=\sum_{|\bde|\le r\atop |\de_0|\le r_0} F_\de\, t^\de 
+\sum_{|\bde|> r\atop {\rm or\ }\de_0|> r_0}\infty\, t^\de
$$
with $F_\de$ finite for all $|\bde|\le r,\  |\de_0|\le r_0$. Hence
$$\eqalign{
N(\cG_{\rm gen}^{\rm amp} -\cV';\cb,\ib, 2)
&\le \abcst_4\,\big(\sfrac{g\ga^2 \upsilon}{\veps\mu^d}\big)^2
\,\Big(
\sum_{|\bde|\le r\atop |\de_0|\le r_0}
\sfrac{1}{\mu^{|\de|}}t^\de 
+\sum_{|\bde|> r\atop {\rm or\ }\de_0|> r_0}\infty\, t^\de
\Big)\cr
}$$
with $\abcst_4=\sfrac{1}{8}\max\limits_{|\bde|\le r\atop |\de_0|\le r_0}F_\de$.
As
$$\eqalign{
&N(\cG_{\rm gen}^{\rm amp} -\cV';\cb,\ib,2)
= \cb\Big( 4\,\|G_2^{\rm amp}-K\| 
+ 16\ib^2\,\|G_4^{\rm amp}-V_0\| +
\smsum_{n=3}^\infty 4(2\ib)^{2n-2}\,\|G_{2n}^{\rm amp}\| \Big)\cr
&
\ge \sfrac{4\abcst_1\,g}{\mu^d}
\Big( \|G_2^{\rm amp}-K\|_{1,\infty}
+ 4\,\abcst_0^2\,\ga^2\,\|G_4^{\rm amp}-V_0\|_{1,\infty} +
\smsum_{n=3}^\infty (\abcst_0\ga)^{2n-2}\,\|G_{2n}^{\rm amp}\|_{1,\infty} \Big)\cr
}$$
we have
$$\eqalign{
& \|G_2^{\rm amp}-K\|_{1,\infty}
+ 4\,\abcst_0^2\,\ga^2\,\|G_4^{\rm amp}-V_0\|_{1,\infty} +
\smsum_{n=3}^\infty (\abcst_0\ga)^{2n-2}\,\|G_{2n}^{\rm amp}\|_{1,\infty}\cr
&\hskip1in\le \abcst_4\,\sfrac{g\ga^4 \upsilon^2}{4\abcst_1\,\veps^2\mu^d}
\,\Big(
\sum_{|\bde|\le r\atop |\de_0|\le r_0}
\sfrac{1}{\mu^{|\de|}}t^\de  
+\sum_{|\bde|> r\atop {\rm or\ }\de_0|> r_0}\infty\, t^\de
\Big)
}$$
The estimates on the amputated Green's functions follow. \endproof

\remark{\STM\remOSgamma}{
\Item i)
In reasonable situations, for example if the gradient of $e(\k)$ is bounded 
below, the constants $\ga$ and $g$ in Theorem \thmOSinsulators\ are of 
order one and $\log\mu$ respectively.

\Item ii) Using Lemma \:\exOSappMonoidI, one may prove an analog
of Theorem \thmOSinsulators\ with the constants $\veps$
and $\abcst$ independent of $r_0$ and
$$\eqalign{
\tn\cD\, G_{2n}^{\rm amp}\tn_{1,\infty} 
&\le \abcst^n\,\de(\cD)!\,g\,\ga^{6-2n}\ 
\big(\sfrac{8(d+1)}{\mu}\big)^{d+|\de(\cD)|}\ \upsilon^2
\qquad \qquad {\rm if}\ n\ge 3 \cr
\tn\cD( G_4^{\rm amp} -V_0)\tn_{1,\infty} 
&\le \abcst^2\,\de(\cD)!\,g\,\ga^{2}\ 
\big(\sfrac{8(d+1)}{\mu}\big)^{d+|\de(\cD)|}\ \upsilon^2
\cr
\tn\cD( G_2^{\rm amp}  -K)\tn_{1,\infty} 
&\le  \abcst\,\de(\cD)!\,g\,\ga^{4}\ 
\big(\sfrac{8(d+1)}{\mu}\big)^{d+|\de(\cD)|}\ \upsilon^2
}$$
\Item iii)
Roughly speaking, the connected Green's function are constructed from the 
connected amputated Green's functions by appending propagators $C$. The details
are given in \S\CHintroII. 
Using Proposition \propOSpropbnd.ii, one sees that,
under the hypotheses of Theorem \thmOSinsulators.i, 
 the connected Green's functions 
exist in the space of all functions on 
$\big( \bbbr \times \bbbr^d \times \{\uparrow,\downarrow\}\big)^{2n}$ 
with finite $\tn\ \cdot\ \tn_{1,\infty} $ 
and $\tn\ \cdot\ \tn_{\infty} $ norms. 

}

\vfill\eject
%=====================================================================
%========================== NORM DOMAIN===============================
%=====================================================================

\appendix{\APappMonoid}{ Calculations in the Norm Domain}\PG\pgOSA
Recall from Definition \defOSFancynormdomain\ that the
$(d+1)$--dimensional norm domain $\fN_{d+1}$ is the 
set of all formal power series
$$
X = \sum_{\de\in\bbbn_0\times\bbbn_0^d} X_\de \
t_0^{\de_0}t_1^{\de_1}\cdots t_d^{\de_d}
= \sum_{\de\in\bbbn_0\times\bbbn_0^d} X_\de\ t^{\de}
$$
in the variables $t_0,t_1,\cdots,t_d$ with coefficients $X_\de \in \bbbr_+\cup\{\infty\}$. 

\definition{\STM\defOSsaturated}{
A nonempty subset $\De$ of $\bbbn_0 \times \bbbn_0^d$ is called saturated if, for every
$\de \in \De$ and every multiindex $\de'$ with $\de'\le \de$, the
multiindex $\de'$ also lies in $\De$. If $\De$ is a finite set, then
$$
N(\De)=\min\set{n\in\bbbn}{n\de\notin\De\hbox{ for all }\0\ne \de\in\De}
$$
is finite.

}

\noindent
For example, if $r,r_0\in \bbbn$ then the set 
$\set{\de\in\bbbn_0 \times \bbbn_0^d }{\de_0\le r_0,\ |\bde|\le r}$
is saturated and $N(\De)=\max\{r_0+1,r+1\}$.
\lemma{\STM\lemOSappMonoidI}{ Let $\De$ be a saturated set of multiindices
and $X,Y\in\fN_{d+1}$. Furthermore, let $f(t_0,\cdots,t_d)$ and 
$g(t_0,\cdots,t_d)$ be analytic functions in a neighbourhood of the origin in 
$\bbbc^{d+1}$ such that, for all $\de\in\De$, the $\de^{\rm th}$ 
Taylor coefficients of $f$ and $g$ at the origin are real and nonnegative.
Assume that $g(0)<1$ and that, for all $\de\in\De$,
$$\eqalign{
X_\de&\le\sfrac{1}{\de!}\Big(\smprod_{i=0}^d
\sfrac{\partial^{\de_i}\hfill}{\partial t_i^{\de_i}}\Big)f(t_0,\cdots,t_d)
\Big|_{t_0=\cdots=t_d=0}\cr
Y_\de&\le\sfrac{1}{\de!}\Big(\smprod_{i=0}^d
\sfrac{\partial^{\de_i}\hfill}{\partial t_i^{\de_i}}\Big)g(t_0,\cdots,t_d)
\Big|_{t_0=\cdots=t_d=0}\cr
}$$
Set $Z=\sfrac{X}{1-Y}$ and $h(t)=\sfrac{f(t)}{1-g(t)}$. Then, for all $\de\in\De$,
$$
Z_\de\le\sfrac{1}{\de!}\Big(\smprod_{i=0}^d
\sfrac{\partial^{\de_i}\hfill}{\partial t_i^{\de_i}}\Big)h(t_0,\cdots,t_d)
\Big|_{t_0=\cdots=t_d=0}
$$

}
\prf Trivial. \endproof
\example{\STM\exOSappMonoidI}{ Let $\De$ be a saturated set and $a\ge0$, 
$0\le\la\le\half$. Then
$$
\frac
{\big(\sum_{\de\in\De}a^{|\de|}t^\de+\sum_{\de\not\in\De}\infty\,t^\de\big)^2}
{1-\la\big(\sum_{\de\in\De}a^{|\de|}t^\de+\sum_{\de\not\in\De}\infty\,t^\de\big)}
\le \sfrac{16}{3}
\smsum_{\de\in\De}\big(4(d+1)\,a\big)^{|\de|}\ t^\de
+\smsum_{\de\not\in\De}\infty\,t^\de
$$
\prf Set
$$\meqalign{
X&=\Big(\smsum_{\de\in\De}a^{|\de|}t^\de+\smsum_{\de\not\in\De}\infty\,t^\de\Big)^2
\qquad&&
f(t)&=\Big(\smsum_{\de}a^{|\de|}t^\de\Big)^2
=\smprod_{i=0}^d \sfrac{1}{(1-at_i)^2}\cr
Y&=\la\Big(\smsum_{\de\in\De}a^{|\de|}t^\de+\smsum_{\de\not\in\De}\infty\,t^\de\Big)
&&
g(t)&=\la\Big(\smsum_{\de}a^{|\de|}t^\de\Big)
=\la\smprod_{i=0}^d \sfrac{1}{1-at_i}\cr
}$$
Set
$$
h(t) = \frac{f(t)}{1-g(t)}
=\frac{1}{\Pi (1-at_i)}\ \frac{1}{\Pi (1-at_i)-\la}
$$
By the Cauchy integral formula, with $\rho=\sfrac{1}{a}\Big(1-\root{d+1}\of{\sfrac{3}{4}}\,\Big)$
$$\eqalign{
\sfrac{1}{\de!}\Big(\smprod_{j=0}^d
\sfrac{\partial^{\de_j}\hfill}{\partial t_j^{\de_j}}\Big)h(t_0,\cdots,t_d)
\Big|_{t_0=\cdots=t_d=0}
\ &=\ \int_{|z_0|=\rho }\cdots \int_{|z_d|=\rho } h(z)\smprod_{j=0}^d\Big(
\sfrac{1}{z_j^{\de_j+1}}\sfrac{dz_j}{2\pi\imath}\Big)\cr
\noalign{\vskip.05in}
&\le\ \sfrac{1}{\rho^{|\de|}}\sup_{|z_0|=\cdots=|z_d|=\rho}|h(z)|\cr
&\le\ \sfrac{1}{\rho^{|\de|}}\sfrac{1}{(1-a\rho)^{d+1}}\ 
\sfrac{1}{ (1-a\rho)^{d+1}-\la}\cr
&\le\ \sfrac{a^{|\de|}}{{(1-(3/4)^{1/(d+1)}\,)}^{|\de|}}\ \sfrac{4}{3}\ 
\sfrac{1}{ 3/4-1/2}\cr
&\le\ \sfrac{16}{3}\ \big(4(d+1)a\big)^{|\de|}\cr
}$$
\endproof
}

\lemma{\STM\lemOSappMonoidIV}{
\Item i)  Let $X,Y\in\fN_{d+1}$ with $X_\0+Y_\0<1$
$$
\sfrac{1}{1-X}\ \sfrac{1}{1-Y}\le\sfrac{1}{1-(X+Y)}
$$
\Item ii) Let $\De$ be a finite saturated set and 
$X,Y\in\fN_{d+1}$ with $X_\0+Y_\0<\half$. There is a constant,
$\abcst$ depending only on $\De$, such that
$$
\sfrac{1}{1-(X+Y)}\le \abcst\ \sfrac{1}{1-X}\ \sfrac{1}{1-Y}
+\sum_{\de\notin \De}\infty\, t^\de\in\fN_{d+1}
$$
}
\prf i)
$$\eqalign{
\sfrac{1}{1-X}\ \sfrac{1}{1-Y}
&\ =\ \sum_{m,n=0}^\infty X^mY^n
\ =\ \sum_{p=0}^\infty\sum_{m=0}^p X^mY^{p-m}
\ \le\ \sum_{p=0}^\infty\sum_{m=0}^p{\textstyle{p\choose m}}X^mY^{p-m}\cr
&\ =\ \sum_{p=0}^\infty(X+Y)^p
\ =\ \sfrac{1}{1-(X+Y)}
}$$
\Item ii) Set $\hat X=X-X_\0$ and $\hat Y=Y-Y_\0$. Then
$$\eqalign{
\sfrac{1}{1-(X+Y)}
&=\sfrac{1}{1-(X_\0+Y_\0)-(\hat X+\hat Y)}
\le\sfrac{1}{{1\over 2}-(\hat X+\hat Y)}\cr
&\le2\sum_{n=0}^{N(\De)-1}(2\hat X+2\hat Y)^n
+\smsum_{\de\notin \De}\infty\, t^\de\cr
&=2\sum_{n=0}^{N(\De)-1}\sum_{m=0}^n2^{n}\smchoose{n}{m}{\hat X}^m
{\hat Y}^{n-m}+\smsum_{\de\notin \De}\infty\, t^\de\cr
&\le 2^{2N(\De)-1}\sum_{n=0}^{N(\De)-1}\sum_{m=0}^n{\hat X}^m
{\hat Y}^{n-m}+\smsum_{\de\notin \De}\infty\, t^\de\cr
&\le 2^{2N(\De)-1}\sfrac{1}{1-\hat X}\sfrac{1}{1-\hat Y}
+\smsum_{\de\notin \De}\infty\, t^\de\cr
&\le 2^{2N(\De)-1}\sfrac{1}{1- X}\sfrac{1}{1- Y}
+\smsum_{\de\notin \De}\infty\, t^\de\cr
}$$

\endproof

\corollary{\STM\corOSappMonoidIV}{
 Let $\De$ be a finite saturated set, $\mu,\La>0$. Set 
$
\cb = \smsum_{\de\in \De}\La^{|\de|}\, t^\de
+\smsum_{\de\notin \De}\infty\, t^\de
$. 
There is a constant, $\abcst$ depending only on $\De$ and $\mu$, such that
the following hold.
\Item i)
For all $X\in\fN_{d+1}$ with $X_\0<\min\{\sfrac{1}{2\mu},1\}$. 
$$
\sfrac{\cb}{1-\mu \cb X}\le \abcst\ \sfrac{\cb}{1-X}
$$
\Item ii)
Set, for $X\in\fN_{d+1}$,
$\ 
\fe(X)=\sfrac{\cb}{1-\La X}
\ $.
If $\mu+\La X_\0<\half$, then
$$
\fe(X)^2\le\abcst\,\fe(X)\qquad\qquad
\sfrac{\fe(X)}{1-\mu \fe(X)}\le \abcst\, \fe(X)
$$

}
\prf i)
Decompose $X=X_\0+\hat X$. Then, by Example \exOSappMonoidI\
  and Lemma \lemOSappMonoidIV,
$$\eqalign{
\sfrac{\cb}{1-\mu \cb X}
&=\sfrac{\cb}{1-\mu X_\0\cb-\mu \cb \hat X}
\le \abcst\ \sfrac{\cb}{1-\mu X_\0\cb}\ \sfrac{1}{1-\mu \cb \hat X}
\le \abcst\ \sfrac{\cb}{1-\cb/2}\ \sfrac{1}{1-\mu \cb \hat X}
\le \abcst\ \sfrac{\cb}{1-\mu \cb \hat X}\cr
}$$
Expanding in a geometric series
$$
\sfrac{\cb}{1-\mu \cb X}
\le \abcst\ \cb\sum_{n=0}^{N(\De)-1}(\mu \cb \hat X)^n
\le \abcst\ \cb^{N(\De)}\big(1+\mu^{N(\De)}\big)\sum_{n=0}^{N(\De)-1} \hat X^n
\le \abcst\ \sfrac{\cb}{1-\hat X}
\le \abcst\ \sfrac{\cb}{1- X}
$$
%%%%%%%%%%
\Item ii)
The first claim follows from the second, by expanding the geometric series.
By Lemma \lemOSappMonoidIV.ii and part (i),
$$\eqalign{
\sfrac{\fe(X)}{1-\mu \fe(X)}
&=\frac{\sfrac{\cb}{1-\La X}}{1-\mu \sfrac{\cb}{1-\La X}}
=\sfrac{\cb}{1-\La X-\mu \cb}
\le\abcst\sfrac{\cb}{1-\mu\cb}\sfrac{1}{1-\La X}
\le\abcst\sfrac{\cb}{1-\La X}
=\abcst\,\fe(X)
}$$
\endproof
\remark{\STM\remOSappMonoidIV}{
The following generalization of Corollary \corOSappMonoidIV\ is proven in the same way.
 Let $\De$ be a finite saturated set, $\mu,\ \la,\ \La>0$. Set 
$
\cb = \smsum_{\de\in \De}\la^{\de_0}\La^{|\bde|}\, t^\de
+\smsum_{\de\notin \De}\infty\, t^\de
$. 
There is a constant, $\abcst$ depending only on $\De$ and $\mu$, such that
the following hold.
\Item i)
For all $X\in\fN_{d+1}$ with $X_\0<\min\{\sfrac{1}{2\mu},1\}$. 
$$
\sfrac{\cb}{1-\mu \cb X}\le \abcst\ \sfrac{\cb}{1-X}
$$
\Item ii)
Set, for $X\in\fN_{d+1}$,
$\ 
\fe(X)=\sfrac{\cb}{1-\La X}
\ $.
If $\mu+\La X_\0<\half$, then
$$
\fe(X)^2\le\abcst\,\fe(X)\qquad\qquad
\sfrac{\fe(X)}{1-\mu \fe(X)}\le \abcst\, \fe(X)
$$

}

\lemma{\STM\lemOSappMonoidV}{Let $\De$ be a finite saturated set and
$$
X=\sum_{\de\in\De}X_\de t^\de+\sum_{\de\notin \De}\infty\, t^\de
\in\fN_{d+1}
$$
Let $f(z)$ be analytic at $X_\0$, with $f^{(n)}(X_\0)\ge 0$ for all $n$,
whose radius of convergence at $X_\0$ is at least $r>0$. 
Let $0<\be<\sfrac{1}{X_\0}$. Then there exists a constant $C$,
depending only on $\De$, $\be$, $r$ and 
$\max\limits_{|z-X_\0|=r}\big|f(z)\big|$ such that
$$
f(X)\le C\sfrac{1}{1-\be X}
$$

}
\prf 
Set $\al=\sfrac{\be}{1-\be X_\0}$ and $\hat X=X-X_\0$. Then
$$\eqalign{
f(X)&=\sum_{n}\sfrac{1}{n!}f^{(n)}(X_\0)\hat X^n\cr
&\le\sum_{n<N(\De)}\sfrac{1}{n!}f^{(n)}(X_\0)\hat X^n
+\sum_{\de\notin \De}\infty\, t^\de\cr
&\le C\sum_{n<N(\De)}\al^n\hat X^n
+\sum_{\de\notin \De}\infty\, t^\de\cr
}$$
where
$$
C=\max_{n<N(\De)}\sfrac{f^{(n)}(X_\0)}{n!\ \be^n}
>\max_{n<N(\De)}\sfrac{f^{(n)}(X_\0)}{n!\ \al^n}
$$
Hence
$$\eqalign{
f(X)&\le \sfrac{C}{1-\al\hat X}+\sum_{\de\notin \De}\infty\, t^\de
= \sfrac{C}{1-\al( X-X_\0)}
= \sfrac{C(1-\be X_\0)}{1-\be X}
\le \sfrac{C}{1-\be X}
}$$

\endproof

\vfill\eject
%=====================================================================
%========================== REFERENCES ===============================
%=====================================================================

\titlea{ References}\PG\pgOSIref

\item{[FKTf1]} J. Feldman, H. Kn\"orrer, E. Trubowitz, 
{\bf A Two Dimensional Fermi Liquid, Part 1: Overview}, preprint.
\smallskip%
\item{[FKTf2]} J. Feldman, H. Kn\"orrer, E. Trubowitz, 
{\bf A Two Dimensional Fermi Liquid, Part 2: Convergence}, preprint.
\smallskip%
\item{[FKTf3]} J. Feldman, H. Kn\"orrer, E. Trubowitz, 
{\bf A Two Dimensional Fermi Liquid, Part 3: The Fermi Surface}, preprint.
\smallskip%
\item{[FKTr1]} J. Feldman, H. Kn\"orrer, E. Trubowitz, 
{\bf Convergence of Perturbation Expansions in Fermionic Models, Part 1: Nonperturbative Bounds}, preprint.
\smallskip%
\item{[FMRT]} J.Feldman, J. Magnen, V. Rivasseau, E. Trubowitz, 
{\bf Two Dimensional Many Fermion Systems as Vector Models},
Europhysics Letters, {\bf 24} (1993) 521-526.
\smallskip%
\item{[PT]}  J. P\"oschel, E. Trubowitz, {\it Inverse Spectral Theory}, 
Academic Press (1987).
\smallskip%

\vfill\eject
%=====================================================================
%========================== NOTATION   ===============================
%=====================================================================

%\vsize = 9.3truein
%\hoffset=-0.2in
\titlea{Notation }\PG\pgOSInot
\null\vskip-0.8in

\titleb{Norms}
\centerline{
\vbox{\offinterlineskip
\hrule
\halign{\vrule#&
         \strut\hskip0.05in\hfil#\hfil&
         \hskip0.05in\vrule#\hskip0.05in&
          #\hfil\hfil&
         \hskip0.05in\vrule#\hskip0.05in&
          #\hfil\hfil&
           \hskip0.05in\vrule#\cr
height2pt&\omit&&\omit&&\omit&\cr
&Norm&&Characteristics&&Reference&\cr
height2pt&\omit&&\omit&&\omit&\cr
\noalign{\hrule}
height2pt&\omit&&\omit&&\omit&\cr
&$\tn\ \cdot\ \tn_{1,\infty}$&&no derivatives, external positions, acts on functions&&Example \exOSSymmNorm&\cr
height4pt&\omit&&\omit&&\omit&\cr
&$\|\ \cdot\ \|_{1,\infty}$&&derivatives, external positions, acts on functions&&Example \exOSSymmNorm&\cr
height4pt&\omit&&\omit&&\omit&\cr
&$\|\ \cdot\ \cnorm_\infty$&&derivatives, external momenta, acts on functions
&&Definition \defOSderivmom&\cr
height4pt&\omit&&\omit&&\omit&\cr
&$\tn\ \cdot\ \tn_{\infty}$&&no derivatives, external positions, acts on functions&&Example \exOSelloneinftycontr&\cr
height4pt&\omit&&\omit&&\omit&\cr
&$\|\ \cdot\ \cnorm_1$&&derivatives, external momenta, acts on functions
&&Definition \defOSderivmom&\cr
height4pt&\omit&&\omit&&\omit&\cr
&$\|\ \cdot\ \cnorm_{\infty,B}$&&derivatives, external momenta, $B\subset\bbbr\times\bbbr^d$
&&Definition \defOSderivmom&\cr
height4pt&\omit&&\omit&&\omit&\cr
&$\|\ \cdot\ \cnorm_{1,B}$&&derivatives, external momenta, $B\subset\bbbr\times\bbbr^d$
&&Definition \defOSderivmom&\cr
height4pt&\omit&&\omit&&\omit&\cr
&$\|\ \cdot\ \|$&&$\rho_{m;n}\|\ \cdot\ \|_{1,\infty}$&&Lemma \lemOSscalednorm&\cr
height4pt&\omit&&\omit&&\omit&\cr
&$N(\cW;\cb,\ib,\al)$&&$\sfrac{1}{\ib^2}\,\cb\!\sum_{m,n\ge 0}\,
\al^{n}\,\ib^{n} \,\|\cW_{m,n}\|$&&Definition \defOSgrnorm&\cr
height4pt&\omit&&\omit&&\omit&\cr
& && &&Theorem  \thmOSinsulators&\cr
height4pt&\omit&&\omit&&\omit&\cr
}\hrule}}

\vfill
\titleb{Other Notation}
\centerline{
\vbox{\offinterlineskip
\hrule
\halign{\vrule#&
         \strut\hskip0.05in\hfil#\hfil&
         \hskip0.05in\vrule#\hskip0.05in&
          #\hfil\hfil&
         \hskip0.05in\vrule#\hskip0.05in&
          #\hfil\hfil&
           \hskip0.05in\vrule#\cr
height2pt&\omit&&\omit&&\omit&\cr
&Not'n&&Description&&Reference&\cr
height2pt&\omit&&\omit&&\omit&\cr
\noalign{\hrule}
height2pt&\omit&&\omit&&\omit&\cr
&$\Om_S(\cW)(\phi,\psi)$
&&$\log\sfrac{1}{Z} \int  e^{\cW(\phi,\psi+\ze)}\,d\mu_{S}(\ze)$
&&before (\eqnOSintroI)&\cr
height2pt&\omit&&\omit&&\omit&\cr
&$S(C)$&&$\sup_m\sup_{\xi_1,\cdots,\xi_m \in \cB}\
\Big(\ \Big| \int \psi(\xi_1)\cdots\psi(\xi_m)\,d\mu_C(\psi) \Big|\ \Big)^{1/m}$&&Definition \defIntBndsS&\cr
height2pt&\omit&&\omit&&\omit&\cr
&$\cB$&&$\bbbr \times \bbbr^d \times \{\uparrow, \downarrow\}\times\{0,1\}$ 
viewed as position space&&beginning of \S\CHnorms&\cr
height2pt&\omit&&\omit&&\omit&\cr
&$\cF_m(n)$&&functions on $\cB^m \times \cB^n$, antisymmetric in $\cB^m$
arguments&&Definition \defOSFmn&\cr
height2pt&\omit&&\omit&&\omit&\cr
}\hrule}}

\vfill

\end